\documentclass[11pt, oneside]{article} 
\pdfoutput=1

\usepackage{jheppub}
\usepackage{amsmath,amssymb,amsfonts,amsthm}
\usepackage{color}
\usepackage{booktabs}
\definecolor{darkblue}{rgb}{0.1,0.1,.7}
\usepackage[]{amsmath}
\usepackage{graphicx} 

\usepackage[]{latexsym}
\usepackage{amscd}
\usepackage[all,cmtip]{xy}
\usepackage{mathrsfs}
\usepackage{bbold}
\usepackage[margin=10pt,font=small,labelfont=bf]{caption}
\usepackage{subcaption} 
\usepackage{simplewick}
\usepackage{changepage}
\usepackage[]{algorithm2e}
\usepackage{booktabs,multirow}
\usepackage{float}
 
\usepackage{hyperref}

\usepackage[OT2,OT1]{fontenc}
\usepackage{braket}
\usepackage{tikz}
\usepackage{pgfplots}
\pgfplotsset{compat=1.10}
\usepgfplotslibrary{fillbetween}
\usetikzlibrary{patterns}

\usepackage{tikz-feynman}

\usepackage[normalem]{ulem}

\newcommand{\abs}[1]{\left\lvert#1\right\rvert}

\def\d {\partial} 

\newcommand{\hyperF}[4]{{}_2F_1\left(#1,#2;#3;#4\right)}

\newcommand{\threej}[6]{
	\begin{pmatrix}
		#1 & #2 & #3 \\
		#4 & #5 & #6
	\end{pmatrix}
}

\theoremstyle{remark}

\def\XXint#1#2#3{{\setbox0=\hbox{$#1{#2#3}{\int}$}
     \vcenter{\hbox{$#2#3$}}\kern-.5\wd0}}

\makeatletter
\def\@fpheader{\ }
\makeatother

\title{Yang-Mills Flux Tube in AdS}
\author{Barak Gabai, Victor Gorbenko, Jiaxin Qiao}
\affiliation{Laboratory for Theoretical Fundamental Physics, Institute of Physics, École Polytechnique Fédérale de Lausanne (EPFL), CH-1015 Lausanne, Switzerland\\ }

\abstract{ 
We initiate the study of flux tubes in confining gauge theories placed in a rigid AdS background, which serves as an infrared regulator. Varying the AdS radius from large to small allows us to interpolate between the flat space confining string, and a weakly coupled string-like object which is held together by the AdS gravitational potential. At any radius, the string preserves a subgroup of AdS isometries equivalent to the one-dimensional conformal group and hence, from the boundary point of view, can be thought of as a conformal defect. The defect hosts a protected operator, called displacement, which nonlinearly realizes the broken AdS isometries. At small radius the displacement corresponds to the gauge field strength inserted at the boundary, while at large radius it is mapped to the Goldstone mode living on the string worldsheet. This relates gauge field and worldsheet degrees of freedom. We propose a hypothesis according to which the large and small radius perturbative expansions can be smoothly matched with each other. As a test, we calculate the leading order corrections to the scaling dimensions and OPE coefficients of a set of defect operators at weak coupling in planar 3D Yang-Mills.}

\begin{document}
	
\maketitle

\section{Introduction}
In this paper we begin our detailed study of confining flux tubes in asymptotically free gauge theories using Anti-de Sitter space as an infrared regulator. The idea of placing QCD in Anti-de Sitter space to control its strong infrared effects dates back to \cite{Callan:1989em}. Indeed, in flat space theories like QCD have an effective expansion parameter that depends on the energy scale of the process, crudely given by $\Lambda_{\text{QCD}}/E$. As a result, perturbation theory breaks down when the relevant distances become comparable to or exceed the confinement scale. Since QCD is a Lorentz-invariant and renormalizable theory, it can be placed on a rigid AdS space without introducing any ambiguities, aside from the choice of boundary conditions, to which we will return momentarily. This introduces a dimensionless coupling into the theory
\begin{equation}\label{def:lambda}
\lambda\equiv R_\text{AdS}\Lambda_\text{QCD}\,,
\end{equation}
where $\Lambda_\text{QCD}$ is the confinement scale in flat space. 
When $\lambda$ is small, all observables are, in principle, calculable within perturbation theory. The AdS radius effectively acts as an infrared regulator, halting the running of the gauge coupling at a scale comparable to $R_{\text{AdS}}$. The dynamical reasons for this are explained in  \cite{Callan:1989em}. On the other hand, when $\Lambda_\text{QCD}$ is large, the physics on sub-AdS scales is identical to that of flat space for obvious reasons. Let us now imagine keeping $\Lambda_\text{QCD}$ fixed and changing $R_\text{AdS}$ from zero to infinity. In this way, we interpolate between a perturbative theory and our main interest -- the strongly coupled theory in flat space. A crucial question is whether this interpolation is smooth or involves some form of discontinuity along the way. We have strong reasons to believe that, if Neumann boundary conditions are imposed on the gauge fields at the AdS boundary, the interpolation is indeed smooth. Assuming this is the case, QCD on AdS space with small radius is a perfect starting point for the perturbative expansion, at least for the following reasons:
\begin{itemize}
    \item AdS space has as many isometries as flat space, so despite regulating IR divergences it does not break any of space-time symmetries of QCD but rather deforms them.
    \item AdS space has on-shell (asymptotic) observables, which are gauge and field-redefinition invariant. These observables are boundary correlation functions. Moreover, they are well-defined in the Euclidean theory. In this sense, AdS is even superior to Minkowski space which only has Lorentzian on-shell observables.
\end{itemize}
These features of quantum field theories on AdS space, not limited to confining theories, were emphasized in \cite{Paulos:2016fap}, which essentially provoked the renewed interest in this subject. As far as the Dirichlet boundary conditions are concerned, even though they are more standard, for gauge theories the interpolation between weak and strong coupling appears to be less continuous, as was first discussed in \cite{Aharony:2012jf}, and studied in detail more recently in \cite{Ciccone:2024guw}.

Boundary correlation functions of a $d+1$-dimensional QFT in AdS correspond to a $d$-dimensional conformal theory which has all the usual properties, like a discrete spectrum of dimensions, convergent OPE expansion, etc., but lacks a stress tensor \cite{Paulos:2016fap}. In this sense, many technical statements happen to be the same as in AdS/CFT, however, the present setup is not holographic, in the sense that the CFT ``dual'' to QCD in AdS does not have an independent microscopic definition. Nevertheless, many powerful techniques developed recently in the context of AdS/CFT, as well as more broadly in the studies of conformal field theories, can be applied in our setup. In particular, instead of computing directly multi-point correlation functions, it is sufficient to determine operator dimensions and OPE coefficients (the so-called CFT data), which in principle fix all the correlators.

Our program is thus conceptually straightforward: we perform the calculations of the boundary CFT data, corresponding to QCD in AdS at small $\lambda$ and extrapolate it to large $\lambda$. A crucial question is then how well this extrapolation works. In this paper, we provide some evidence that this expansion may work well, at least for a particular sector of large-$N$ QCD. A useful analogy to keep in mind is the $\epsilon$-expansion of Wilson and Fisher. Even for small $\epsilon$, one obtains a good understanding of the operator spectrum in the 3D Ising model, while higher-order expansions combined with resummation techniques can yield very precise determinations—at least for some observables—see, for example, \cite{Henriksson:2025hwi}.

Needless to say, there is {\it{a priori} }no guarantee that the expansion we are proposing will work well in QCD or other confining gauge theories. Let us start collecting some evidence in favor of this. An important qualitative point is that, in AdS, confinement occurs already at weak coupling. First of all, with Neumann boundary conditions, all physical states are singlets under the gauge group. If two colored objects are separated by a large distance inside AdS—for example, pushed toward the boundary in different directions—a string-like flux tube will get stretched between them. Unlike in flat space, where at weak coupling the field lines spread out to create a Coulomb-like potential, in AdS the field lines are pushed together by the AdS potential well. This intuition was also used in \cite{Alday:2007mf} to calculate the anomalous dimensions of double-trace operators with large spin in gauge theories.

As a more quantitative property of QCD, let us briefly discuss chiral symmetry breaking. Consider QCD with massless quarks. As shown in \cite{Rattazzi:2009ux}, there are no boundary conditions for fermions in AdS$_4$ that preserve chiral symmetry; consequently, it must be broken already at the level of the free theory, that is, when $\lambda$ is taken to be very small. Even though the breaking is ``explicit'' on the boundary, it is spontaneous in the bulk, since the chiral current remains conserved there, with the conservation-breaking terms localized on the boundary. This implies that pions are already present in our setup at the level of the free theory. They are massless bosons in AdS$_4$, corresponding to the dimension 3 fermion bilinears. Since the current in the bulk is conserved, their dimension remains protected as we increase $\lambda$. Thus, we have protected boundary operators corresponding to pions even when $\lambda$ is large, that is, in the flat-space limit. 

Chiral perturbation theory, valid at low energies, provides another perturbative expansion at large $\lambda$. It is obtained by placing the chiral Lagrangian in a weakly curved AdS background. The derivative expansion then produces a power series in $1/\lambda$, since typical energies are of order $1/R_\text{AdS}$. Since the boundary correlation functions of pions are unambiguously defined at any coupling, they can be matched between the weak- and strong-coupling expansions.

\begin{figure}[t]
    \centering
    \includegraphics[width=0.75\linewidth]{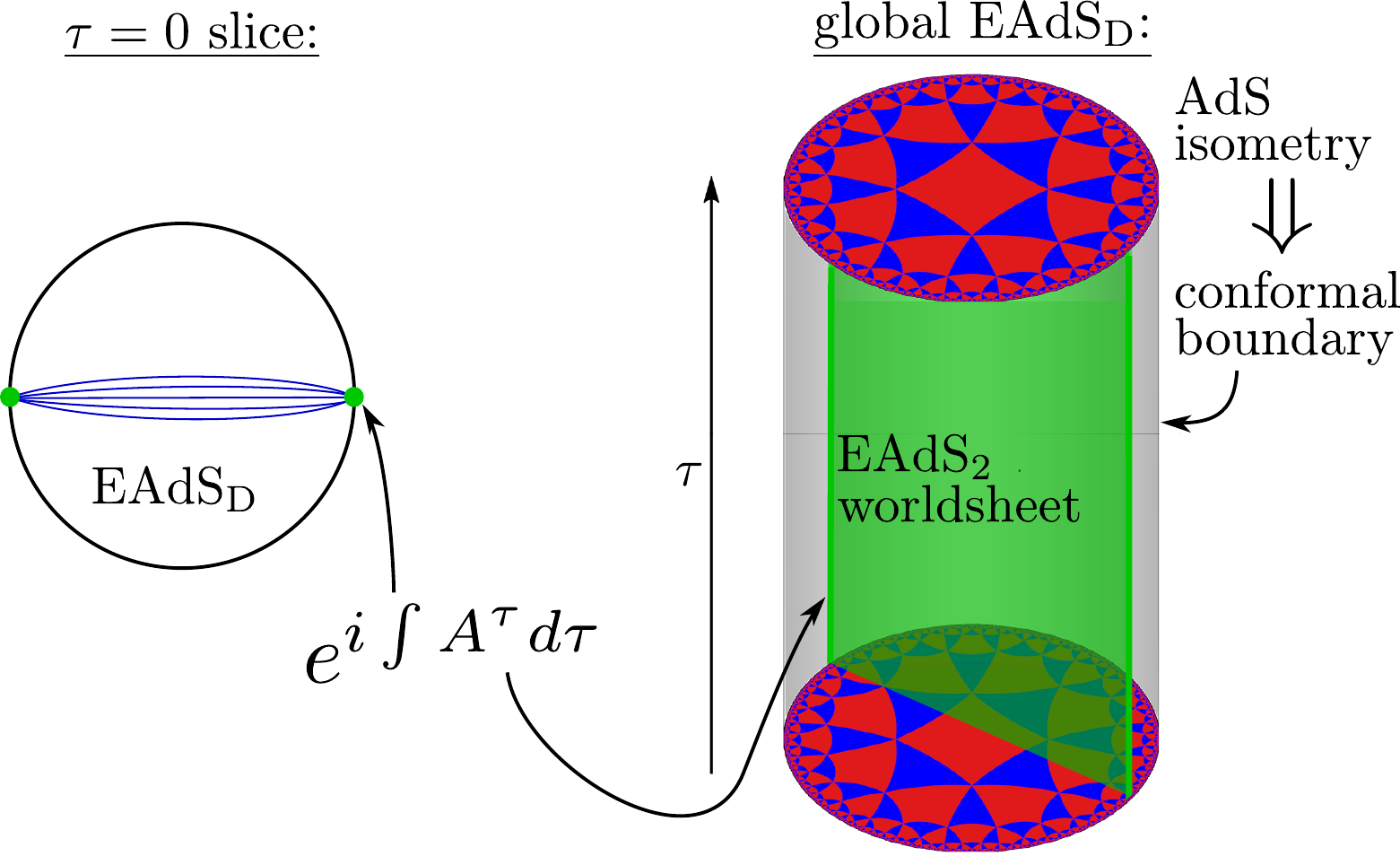}
    \caption{
        Flux tube in AdS generated by a pair of Wilson lines on the boundary. 
        Left: constant-time slice. 
        Right: global EAdS.
    }
    \label{fig:AdS_Flux}
\end{figure}

While the study of pion correlators is certainly an interesting direction to pursue, in this paper we limit our discussion to what appears to be a simpler set of observables. First, we restrict our study to pure large-$N$ Yang-Mills theory in $d+1$ dimensions. In fact, our main goal is to understand the dynamics of confinement, and in this theory confinement is in a sense the most robust, since string-like objects are all stable and cannot split or break. In this theory, we now consider a pair of non-dynamical quarks—or, equivalently, a Wilson line—inserted at the boundary. Following the discussion above, these quarks create a flux tube stretching radially through AdS. Because of large $N$, excitations of this flux tube are decoupled from the rest of the YM degrees of freedom, such as the glueballs that live in the bulk. If the flux tube is straight, it preserves a $SO(1,2)\times SO(d-1)$ subgroup of the $SO(1,d+1)$ isometry group of AdS$_{d+1}$. The center of the flux tube spans an AdS$_{2}$ slice inside AdS$_{d+1}$, as shown in figure \ref{fig:AdS_Flux}. At fixed $\Lambda_{\text{QCD}}$ the width of the flux tube changes with the radius. When $\lambda$ is large, it is of order $\Lambda_{\text{QCD}}^{-1}$, and the string tension is $\Lambda_{\text{QCD}}^{2}$, as in flat space. In this regime, the flux tube is essentially equivalent to an infinitely long confining string in the flat-space pure Yang–Mills theory—an object of many recent studies, which we briefly summarize in the next section. At the same time, for small $\lambda$ the string width is of order $R_{\text{AdS}}$, and, in $D=4$, the string tension is approximately $R_\text{AdS}^{-2} {g}_\text{YM}^2 N$, where ${g}_\text{YM}$ is evaluated at the AdS scale. It is thus consistent that, at $\lambda \sim 1$, the string parameters smoothly interpolate between the two regimes.

From the boundary point of view, at any radius, the flux tube corresponds to a one-dimensional conformal defect. This defect hosts a set of protected operators corresponding to Goldstone bosons of spontaneously broken AdS isometries, similarly to pions discussed above. From the defect point of view, these are $d-1$ displacement operators with protected dimension equal to 2. We can thus study a weak-to-strong coupling interpolation in a one-dimensional CFT with a set of protected operators that can be unambiguously identified at strong and weak coupling. At weak coupling, they correspond to insertions of the gauge field strength into the boundary Wilson line. At strong coupling (large $\lambda$), they are described by the effective theory of the confining string, which we review below. If we are able to establish a matching between these two descriptions, we will obtain a first-principles quantitative understanding of how string-like objects emerge from gauge field degrees of freedom. In this article, we take the first steps in this direction. A crucial question is whether the Goldstones correspond to an emergent collective degree of freedom or whether they roughly correspond to individual gluons. We would like to advocate the point of view that, at least in our present setup, the latter picture is the correct one. While our results are consistent with the statement that all CFT observables are functions of $\lambda$ that are continuous together with all their derivatives, we will not commit to this statement. Most important for us is the ``smoothness'' in a practical sense, namely that by computing to higher and higher order in perturabtion theory we can get better and better agreement between the two descriptions without the need of adding some qualitatively new ingredients. We hope to produce a more precise condition in the future work.

Let us emphasize that, even though we will be dealing with AdS space, our setup is not holographic in any sense and we are not relying on the existence of any kind of holographic dual of QCD. Nevertheless, we will comment on an exciting relation between our proposal and holographic duality for $\mathcal{N}=4$ SYM in the conclusions.

\section{Short Review: the Flux Tube in Flat Space}
\label{sec:flat}
Let us briefly review some recent developments in the studies of confining flux tubes in flat space that will be relevant for our discussion below. Our review is meant to capture only the big picture, avoiding mentioning many interesting subtleties. Two main tools are the lattice field theory and the effective string theory. In both cases we are interested in excited levels of a single long flux tube in a pure large-$N$ Yang-Mills theory in 3 or 4 dimensions. By a {\it{long}} flux tube we mean one of the following: either an infinitely long flux tube stretching along one of the spatial directions, a flux tube winding a compact spatial direction, a folded rotating flux tube at a large angular momentum, or a flux tube with static endpoints fixed by non-dynamical quarks. In all cases, lattice results show that $N=3$ is already very close to $N=\infty$, and that the string can be just a few QCD scales long in order to be described well by the effective theory.\footnote{In this sense rotating strings appear to be more subtle than the other cases, see \cite{Cuomo:2024gek}.}

The effective theory of long strings is a two-dimensional relativistic QFT of $D-2$ massless bosons which parametrize the transverse deformations of the string. They are Goldstone bosons which nonlinearly realize the broken symmetries of the coset 
\begin{equation}
ISO(1,D-1)/O(D-2)\times ISO(1,1)\,.
\end{equation}
There are two equivalent approaches to constructing the effective theory. One is the static gauge approach, which includes only the physical degrees of freedom from the start \cite{Dubovsky:2012sh}. The second approach uses the conformal gauge \cite{Polchinski:1991ax} and includes also the longitudinal string coordinates as auxiliary fields, we will refer to it as the conformal formalism, see \cite{Aharony:2013ipa} for a review. There is also a more modern version of this formalism which uses the Polyakov approach \cite{Hellerman:2014cba}. In either case the derivative expansion goes in the powers of $p \ell_s$, where $p$ is the typical momentum of Goldstones and $\ell_s\sim \Lambda_\text{QCD}^{-1}$ is the string width. The static gauge Lagrangian reads
\begin{equation}
-{1\over 2}\left(\d_\alpha X^i\right )^2 -{\ell_s^2\over 8}\left (\d_\alpha X^i\d^\alpha X^i\right)^2+{\ell_s^2\over 4}\d_\alpha X^i\d_\beta X^j\d^\alpha X^i\d^\beta X^j+O \left (\ell_s^4\d^6 {X^i}^6,\ell_s^6\d^8 {X^i}^4\right)\,,
\label{eq:Static}
\end{equation}
here $i,j=2\ldots D-1$. The leading terms written here come from the expansion of the Nambu-Goto term.
In the PS formalism, the Lagrangian is 
\begin{equation}
-{1\over {2 \ell_s^2}}\left(\d_\alpha X^\mu\right )^2-\frac{(26-D)}{24 \pi}\frac{\left(\d_\beta\d_\gamma X^\nu \d^\gamma X_\nu\right)^2}{\left(\d_\alpha X^{\mu}\right)^2}+ O \left (\ell_s^2 \d^8 {X^{\mu}}^4\right)\,.
\label{eq:PS}
\end{equation}
This theory is conformal and it has to be supplemented with the Virasoro constraints which remove the unphysical degrees of freedom. The second term in the action ensures that the central charge is equal to 26 as necessary for the consistency of this formulation. We refer to this term as the PS-term. It appears non-local, however, around a long-string background $\left(\d_\alpha X^\mu\right )^2$ gets a vev so the action for perturbation around such a background is local. In our loop-counting we treat the PS term as a one-loop term, as its effects are reproduced by one-loop diagrams in the formulation \eqref{eq:Static}. Note that the theory is not free at the tree level, as might appear from \eqref{eq:PS}, because of the constraints. Also the dynamical fields in \eqref{eq:PS} are not canonically normalized, as in \eqref{eq:Static}.

Importantly, non-universal corrections in the effective theory appear only at a two-loop order, that is they are suppressed at least by $(p\ell_s)^4$ as compared to the tree-level contribution to an observable in question.  In case of open or folded rotating strings boundary terms have to be added to the action \cite{Hellerman:2016hnf,Cuomo:2024gek}. For simplicity we discuss winding or infinitely long strings for which such terms are not needed. Either action can be used to compute observables like the S-matrix of goldstones or a finite volume spectrum of excited string states\footnote{The ground state energy is also an interesting observable, but excited states contain significantly more dynamical information.}. It turns out that it is much more beneficial to first calculate an infinite volume S-matrix and then obtain the finite volume spectrum from it, rather than to compute it directly using perturbation theory in the finite volume. This procedure is facilitated by the fact that at tree level the theory is integrable and the scattering is factorized \cite{Dubovsky:2013gi,Caselle:2013dra,Dubovsky:2014fma}. Integrability is, however, broken in 4D by universal one-loop effects \cite{Cooper:2014noa}. In 3D integrability persists at one-loop order and, as proven in \cite{Dubovsky:2015zey}, there is a unique integrable theory consistent with nonlinearly realized Poincare invariance. Lattice data, that we review below, excludes integrability in pure 3D Yang-Mills theory at any $N$.

Let us now briefly describe the lattice methodology \cite{Athenodorou:2010cs,Athenodorou:2011rx}. See \cite{Athenodorou:2024loq} for the most up to date results, as well as \cite{Sharifian:2025fyl} for the related results on open flux tubes. One treats one lattice direction, say $x$ as a compact circle of physical length $L$, and one as a Euclidean time $t$. Then by inserting a pair of Wilson lines winding along $x$ and separated in $t$ one can extract the spectrum of winding strings contributing to the correlation function:
\begin{equation}
\langle W(0) W^\dagger(T)\rangle\sim\sum_n e^{-TE_n(L)}\,.
\end{equation}
By changing the shapes of the Wilson lines one can change the quantum numbers of states contributing.

The most interesting results are obtained by comparing the calculation of functions $E_n(L)$ from the effective theory and from the lattice. Crucially this allows to constrain the non-universal EFT data, like massive particles that propagate on the string worldsheet. These particles are not stable since they can always decay in some number of Goldstones, however, they appear as resonances in the worldsheet S-matrix, or as approximately flat energy levels\footnote{More precisely the difference $E_n(L)-E_0(L)$ is approximately independent of $L$ for a state corresponding to a massive particle at rest on the worldsheet.}. Result of this matching is the following: no massive excitations were detected in 3D, while in 4D a single massive excitation is detected with a very high significance \cite{Dubovsky:2013gi,Dubovsky:2014fma,Athenodorou:2024loq}. This excitation is odd under both parities (transverse and longitudinal to the worldsheet) and has spin zero. It is referred to as the worldsheet axion. The axion can be incorporated in the worldsheet effective theory as follows:
\begin{equation}
S_{a}=\int d^2\sigma\left[-{1\over 2}\left(\d_\alpha a \right)^2-{1\over 2} m_a a^2-\ell_s^2 Q_a a\varepsilon^{ij}\varepsilon^{\alpha\beta}\d_\alpha \d_\gamma X^i\d_\beta \d^\gamma X^j\ldots \right]
\label{ASA}
\end{equation}
with the two new parameters measured from the lattice to be $m_a\sim 1.8 - 1.5 \ell_s^{-1}$, as $N$ changes from 3 to $\infty$ and coupling $Q_a$ being close to 0.37, see \cite{Dubovsky:2013gi,Dubovsky:2015zey,Athenodorou:2017cmw,Athenodorou:2024loq}.  

Note that the axion itself is not parametrically light, its mass is of order of the cutoff of the effective theory, however, the lattice data shows that EFT can still be consistently used at the energies of order of its mass. This, in particular, includes two-axion states. Another curious observation is that the axion coupling happens to be such that the tree-level axion exchange largely cancels the effect of the PS term, or one-loop Goldstone self-interactions \cite{Dubovsky:2015zey}. In this sense the axion unitarizes this very fast growing contribution to the worldsheet amplitudes.

This effect can be studied more carefully using the worldsheet S-matrix bootstrap, pioneered in \cite{EliasMiro:2019kyf} and developed further in \cite{EliasMiro:2021nul,Gaikwad:2023hof,Guerrieri:2024ckc}. Applied to the worldsheet scattering this approach shows that there are two regions in the space of allowed effective string theories in $D>3$: the one where the low energy theory is dominated by the axion, and the one where it is dominated by the dilaton. While at the current level of development this approach cannot rigorously exclude theories where both or additional resonances are present, it at least suggests that the worldsheet theory with axion being the only meta-stable particle is non-perturbatively consistent. It also explains why the value of the axion coupling close to the measured from the lattice is special: it saturates certain dispersion relations, see \cite{Gaikwad:2023hof} for details.

One could expect a presence of another particle, corresponding to a breathing mode of the flux tube, or to a radial coordinate of AdS, if one imagines a holographic description of QCD in the spirit of \cite{Polyakov:1998ju}. There is no such particle seen in the spectrum, which means that there isn't a light and narrow resonance with such properties.

Lattice data on excited flux tubes in three dimensions is even more precise \cite{Athenodorou:2011rx,Athenodorou:2016kpd}, however, no massive particle is observed in the spectrum, which is consistent with the idea of axionic string, according to which the field antisymmetric in transverse indices should be the only massive particle on the worldsheet \cite{Dubovsky:2016cog}, hence no such particles are present in 3D. In the absence of massive particles, the remaining worldsheet parameters are the coefficients of non-universal terms in the effective action, which are being computed using ground state energy \cite{Caselle:2024zoh}, bootstrap \cite{Guerrieri:2024ckc}, as well as excited energy levels \cite{Chen:2018keo}.

Let us conclude the summary with a few non-rigorous observations. Even though there is a single parameter in the effective string action, it appears that the tree-level part in the static gauge formulation, or the piece coming from the constraints in the conformal formulation plays a special role. There are several indications that it is useful to treat this piece independently, and non-perturbatively, while treating perturbatively the rest of the interactions \cite{Chen:2018keo}. The intuition behind this is that this procedure resums large semiclassical effects of the finite string tension and captures the fact that the string becomes longer when a high energy excitation is present on it. In this sense, it is an infrared effect, despite the fact that it grows at high energies. There is a consistent procedure which allows us to separate this piece directly at the level of the worldsheet S-matrix, or the finite volume spectrum, which goes under the name of gravitational dressing, or $T\bar T$ deformation. It results in the following representation for the energy levels:
\begin{equation}
\label{TTbar}
E_n(L)=\sqrt{L^2 \ell_s^{-4}+2 L\ell_s^{-2} E_n^0(L)}-L \ell_s^{-2}\,,
\end{equation}
where $E_n^0(L)$ refers to the spectrum of a putative theory without the semiclassical string tension effects.\footnote{We subtracted the classical contribution proportional to the string length to define the spectrum.} In particular, low-energy scattering in this theory starts with the one-loop contribution. Available data hints that \eqref{TTbar} is indeed a better fit than its Taylor expansion. 

Another interesting observation is that the coefficient of the PS term in $D=4$ appears to be related to the one-loop beta-function in the YM theory, including the one with adjoint matter \cite{Dubovsky:2018vde}. Together with the fact that the worldsheet degrees of freedom can be mapped to the gauge theory degrees of freedom by inserting corresponding components of $F^{\mu \nu}$ in the Wilson line that creates a flux tube,\footnote{For a Wilson line that goes in the $x$ direction insertion of $F^{xy}$ and $F^{xz}$ creates the Goldstones, while $F^{yz}$ has the same quantum numbers as the axion.} this observation hints to the fact that the worldsheet theory at the energies around $\Lambda_\text{QCD}$ can already be described by the gauge theory.

Unfortunately, it is hard to turn this quantitative intuition into a qualitative argument. One obstacle is that it is hard to find any observables that are well-defined both in the UV and IR. For example, if we attempt to take the compactification radius $L$ to be small in order to probe the UV, the theory deconfines. If, instead, one considers a high energy scattering of excitations on a long string, there is always an irreducible IR effect due to the formation of an additional string segment produced by energetic excitations.

This is where placing confining theory in AdS appears to be very beneficial. Spectrum of operator dimensions $\Delta_n(R_\text{AdS})$ is an analog of $E_n(L)$, however, the transition to weakly coupled gauge theory regime at small $R_\text{AdS}$ is much smoother, and the same set of observables exists both at large and small radius.

It is also important to emphasize that our weakly coupled small $R_\text{AdS}$ limit is non-Abelian, in a sense that it perfectly commutes with the large-$N$ limit. This is usually not the case in small-volume compactifications which avoid the deconfinement phase transition by twisting the boundary conditions of gauge or matter fields, see for example \cite{Unsal:2008ch,Tanizaki:2022ngt,Bonanno:2025hzr}. It would be interesting to understand the relation between these compactifications and small-radius AdS.

\section{Two Descriptions: Large and Small AdS Radius}
Let us now introduce in more detail two descriptions of the confining flux tube that are perturbative at small and large AdS radii $R_\text{AdS}$, which we will simply call $R$ from now on. In this section, we will present the arguments for the existence of the protected displacement operator and compare the free theory flux tube spectra in both descriptions.  

\subsection{Yang-Mills in AdS and boundary Wilson loop defect}\label{subsec:WilsonLoopDefect}
Let us first introduce the AdS geometry:
\begin{equation}\label{def:AdSgeometry}
	\begin{split}
		ds^2_{\rm AdS_{d+1}} &= R^2\left[\cosh^2 z\,\frac{d\tau^2 + dx^2}{x^2} + dz^2+\sinh^2z\,d\Omega_{d-2}^2\right] \\
		&\equiv \cosh^2 z\,ds^2_{\rm AdS_{2}} + R^2 dz^2+R^2\sinh^2z\,ds^2_{S^{d-2}}.
	\end{split}
\end{equation}
For $d \geqslant 3$ we take $z \geqslant 0$, but for $d = 2$ we allow $z \in \mathbb{R}$ and drop the last term.

In principle, one could instead use the standard Poincar\'e half--space coordinates commonly employed in the context of the AdS/CFT correspondence. In this work, however, we choose to adopt an AdS$_2$ slicing of AdS$_{d+1}$, which will prove convenient when we later introduce an effective string description of the flux tube. AdS$_2$-sliced coordinates have also been used to describe the fundamental string dual to a half--BPS Wilson line in $\mathcal{N}=4$ super Yang--Mills theory \cite{Giombi:2017cqn}, providing further motivation for this choice. 

The main advantage of our choice is that the transverse coordinates $(z,\Omega)$ do not induce geometric fluctuations at the boundary $x=0$. A detailed discussion of the relation between the two coordinate systems is deferred to appendix~\ref{app:AdS3}. The qualitative difference between them is illustrated in figure~\ref{fig:ads_slicing}.

\begin{figure}
	\centering
	\vspace{-20em} 
	\begin{tikzpicture}[scale=1.0, line cap=round, line join=round]
		\def\R{2.2}
		\begin{scope}[shift={(-0.3,0)}]
			\draw[black, line width=1.1pt] (0,0) circle (\R);
			
			\draw[red!90!black, line width=1pt] (-\R,0) -- (\R,0);
			
			\foreach \a in {-10,-3,-1.5,-0.7,-0.3,0.3,0.7,1.5,3,10} {
				\pgfmathsetmacro{\centery}{\R/\a}
				\pgfmathsetmacro{\radius}{\R *sqrt(1 + 1/(\a*\a))}
				
				\begin{scope}
					\clip (0,0) circle (\R);
					\draw[red!90!black, line width=1.0pt] 
					(0, \centery) circle (\radius);
				\end{scope}
			}
			
			\fill[black] (-\R, 0) circle (2.5pt);
			\node[left] at (-\R-0.1, 0) {$+$};
			\fill[black] (\R, 0) circle (2.5pt);
			\node[right] at (\R+0.1, 0) {$-$};
			
			\node[below] at (0,-\R-0.4) {\large (a) AdS$_2$ slicing};
		\end{scope}
		
		\begin{scope}[shift={(6.5,0)}]
			\draw[black, line width=1.1pt] (0,0) circle (\R);
			
			\draw[blue!80!black, line width=1pt] (-\R,0) -- (\R,0);
			
			\foreach \theta in {15,30,45,60,75} {
				\pgfmathsetmacro{\centery}{\R/cos(\theta)}
				\pgfmathsetmacro{\radius}{\R*tan(\theta)}
				\pgfmathsetmacro{\startangle}{180+\theta}
				\pgfmathsetmacro{\endangle}{360-\theta}
				
				\draw[blue!80!black, line width=1.0pt] 
				(0, \centery) ++(\startangle:\radius) arc (\startangle:\endangle:\radius);
			}
			
			\foreach \theta in {15,30,45,60,75} {
				\pgfmathsetmacro{\centery}{-\R/cos(\theta)}
				\pgfmathsetmacro{\radius}{\R*tan(\theta)}
				\pgfmathsetmacro{\startangle}{\theta}
				\pgfmathsetmacro{\endangle}{180-\theta}
				
				\draw[blue!80!black, line width=1.0pt] 
				(0, \centery) ++(\startangle:\radius) arc (\startangle:\endangle:\radius);
			}
			
			\fill[black] (-\R, 0) circle (2.5pt);
			\node[left] at (-\R-0.1, 0) {$+$};
			\fill[black] (\R, 0) circle (2.5pt);
			\node[right] at (\R+0.1, 0) {$-$};
			
			\node[below] at (0,-\R-0.4) {\large (b) Poincar\'e slicing};
		\end{scope}
	\end{tikzpicture}
	\vspace{-13em}  
	\caption{Foliations of AdS$_3$ in a $\tau=\text{const}$ slice. (a) AdS$_2$ slicing with $z=\text{const}$ arcs (red) connecting antipodal points. (b) Poincaré slicing with arcs (blue) perpendicular to the boundary.}
	\label{fig:ads_slicing}
\end{figure}

We consider the Yang–Mills action in Euclidean AdS$_{d+1}$:
\begin{equation}\label{YM:convention}
	\begin{split}
		S_{\rm YM}[A] &= \frac{1}{4 R^{d - 3}} \int_{{\rm AdS}_{d+1}} d^{d+1}x\,\sqrt{g_\text{AdS}}\,{\rm Tr}\left(F_{\mu\nu}F^{\mu\nu}\right), \\
		F_{\mu\nu} &= \partial_\mu A_\nu - \partial_\nu A_\mu + i g [A_\mu, A_\nu].
	\end{split}
\end{equation}
Here, the coupling $g$ is dimensionless and is related to the conventional Yang–Mills coupling $g_\text{YM}$ via\footnote{A common convention in gauge theory is
	\begin{equation*}
		\begin{split}
			S_{\rm YM}[A] = \frac{1}{4\,g^2_{\rm YM}} \int_{{\rm AdS}_{d+1}} d^{d+1}x\,\sqrt{g_\text{AdS}}\,{\rm Tr}\left(F_{\mu\nu}F^{\mu\nu}\right),
		\end{split}
	\end{equation*}
	where $F_{\mu\nu}=\partial_\mu A_\nu - \partial_\nu A_\mu + i [A_\mu, A_\nu]$ is independent of $g_\text{YM}$. This convention is convenient for saddle-point analysis. However, we wish to rewrite the action so that both the coupling and curvature dependence appear only through the dimensionless parameter $g$ defined in \eqref{def:g}. To this end, we rescale the fields as
	\begin{equation*}
		\begin{split}
			A_\mu &= g_{\rm YM}\,R^{\frac{3 - d}{2}}\,\tilde{A}_\mu, \quad F_{\mu\nu} = g_{\rm YM}\,R^{\frac{3 - d}{2}}\,\tilde{F}_{\mu\nu}, \quad\tilde{F}_{\mu\nu} = \partial_\mu \tilde{A}_\nu - \partial_\nu \tilde{A}_\mu + i g_{\rm YM}\,R^{\frac{3 - d}{2}}[\tilde{A}_\mu, \tilde{A}_\nu].
		\end{split}
	\end{equation*}
	Substituting this and dropping the tildes, we obtain the convention used in \eqref{YM:convention}.
}
\begin{equation}\label{def:g}
	g := g_{\rm YM}\,R^{\frac{3 - d}{2}},
\end{equation}
where $R$ is the AdS radius appearing in the metric \eqref{def:AdSgeometry}.


In this convention, gauge transformations act as
\begin{equation}
	\begin{split}
		A_\mu \quad \longrightarrow \quad A^U_\mu = U A_\mu U^{-1} - \frac{i}{g} U \partial_\mu U^{-1},
	\end{split}
\end{equation}
and the gauge-covariant Wilson line along a path $\gamma$ is defined by
\begin{equation}
\label{eq:defWL}
	\begin{split}
		L[\gamma] = \mathcal{P} \exp\left(-i g \int_\gamma dx^\mu A_\mu(x)\right).
	\end{split}
\end{equation}

At first glance, the factor $R^{3 - d}$ in \eqref{YM:convention} might appear to serve as a saddle-point parameter in the small-radius limit $R \to 0$. However, this factor is canceled by the $R^{d+1}$ coming from the volume element $\sqrt{g_\text{AdS}}$ and $R^{-4}$ from raising the indices on $F_{\mu\nu}$. As a result, the dependence on $R$ cancels in the free (quadratic) part of the action, and the AdS radius enters only through the dimensionless coupling $g \equiv g_{\rm YM} R^{\frac{3 - d}{2}}$ in the interaction terms.

The AdS background thus introduces a length scale $R$, which combines with the confinement scale to form a dimensionless parameter in pure Yang–Mills theory. Recall that the coupling constant of $(d+1)$-dimensional Yang–Mills theory has mass dimension $[g_\text{YM}] = \frac{3 - d}{2}$. When $d < 3$, the coupling $g_\text{YM}$ has positive mass dimension and itself provides a scale, making $g = R^{(3 - d)/2} g_\text{YM}$ a natural expansion parameter in the small-$R$ limit. However, in four dimensions ($d = 3$), one must instead introduce the non-perturbative confinement scale $\Lambda_\text{QCD}$, as discussed in \eqref{def:lambda}.

As mentioned in the introduction, to allow for a smooth interpolation to large $R$ we impose Neumann boundary conditions $F_{ni} = 0$, where $n$ denotes the normal direction and $i$ denotes directions tangent to the boundary. Under Neumann boundary conditions, the gauge field remains dynamical at the boundary, enabling us to define an infinitely long Wilson line on the conformal boundary $\mathbb{R}^d$:
\begin{equation}
	\begin{split}
		W = {\rm Tr}\left(\mathcal{P}e^{-ig\int_\gamma A}\right),
	\end{split}
\end{equation}
where $\gamma$, in the coordinates of \eqref{def:AdSgeometry}, is specified by
\begin{equation}\label{Wilsonloop:position}
	\begin{split}
		\tau(s) = s,\quad x(s) = z(s) = 0\quad(-\infty < s < +\infty).
	\end{split}
\end{equation}
Note that when $z = 0$, the angular coordinates $\Omega_{d-2}$ in \eqref{def:AdSgeometry} become irrelevant.

From the boundary perspective, the Wilson loop defines a conformal line defect in the boundary conformal theory (CT$_{d}$), which is described by a one-dimensional conformal theory (CFT$_1$). Correlation functions in CFT$_1$ are given by expectation values of Wilson loops with insertions of gauge-covariant local operators:
\begin{equation}
	\begin{split}
		\braket{\mathcal{O}_1(\tau_1)\ldots\mathcal{O}_n(\tau_n)}_W := \frac{\braket{{\rm Tr}\left(\mathcal{P}\mathcal{O}_1(\tau_1)\ldots\mathcal{O}_n(\tau_n)e^{-ig\int_\gamma A}\right)}}{\braket{W}}.
	\end{split}
\end{equation}
Here, under a gauge transformation, $\mathcal{O}_i(\tau_i) \rightarrow U(\tau_i)\mathcal{O}_i(\tau_i)U(\tau_i)^{-1}$.

In our case, operators $\mathcal{O}_i$ are built from the field strength and its derivatives along the boundary, both along and perpendicular to the defect. At large $N$, the CFT$_1$ forms a closed unitary theory, since all local gauge-invariant operators are decoupled from the Wilson line. 

For a straight conformal line defect in a \emph{local} CFT$_d$, the Ward identity associated with translation symmetry guarantees the existence of displacement operators $\mathbb{D}^i$ on the defect line $\mathcal{D}$, which generate infinitesimal translations of the defect line \cite{Billo:2016cpy},
\begin{equation}
	\begin{split}
		\partial_{\mu} T^{\mu i}(\tau, x)\, \mathcal{D} = \delta^{(d-1)}(x)\, \mathbb{D}^i(\tau)\, \mathcal{D}.
	\end{split}
\end{equation}
This relation holds at the level of correlation functions when both sides are evaluated away from other operator insertions. Since the stress tensor has protected scaling dimension $\Delta_T = d$, the displacement operators also have a protected scaling dimension:
\begin{equation}
	\begin{split}
		\Delta_{\mathbb{D}^i} = 2.
	\end{split}
\end{equation}

In contrast, for Yang-Mills theory in rigid AdS$_{d+1}$ (no dynamical gravity), the boundary dual is a \emph{non-local} conformal theory, meaning in particular that it lacks a conserved stress tensor. As a result, the above Ward identity argument no longer applies. However, we can still analyze the variation of the Wilson line under infinitesimal deformations of its contour. For a small variation of the path $\gamma$:
\begin{equation}
	\begin{split}
		\gamma \rightarrow \gamma + \delta \gamma,
	\end{split}
\end{equation}
the variation of the Wilson line is given by
\begin{equation}
	\begin{split}
		\delta W = \int_{-\infty}^{+\infty} d\tau\, {\rm Tr} \left( \mathcal{P}\left( ig F_{\tau i}(\tau)\, \delta x^i(\tau) \right) e^{-ig \int_\gamma A} \right).
	\end{split}
\end{equation}
Here, $i$ labels the transverse directions. If we restrict the path variation to lie within the boundary, then by dimensional analysis, $ig F_{\tau i}$ has scaling dimension $2$. It is therefore natural to identify it with the displacement operator:
\begin{equation}
\label{eq:dispWeak}
	\begin{split}
		\mathbb{D}_i \quad \text{``="} \quad ig F_{\tau i}.
	\end{split}
\end{equation}
The quotation marks indicate that this is not a strict operator identity: the left-hand side is an operator in the defect CFT, while the right-hand side appears as an insertion under the path-ordered trace. For example, the two-point function of the displacement operators is given by
\begin{equation}
	\begin{split}
		\braket{\mathbb{D}_i(\tau_1)\mathbb{D}_j(\tau_2)}=-g^2\frac{\braket{{\rm Tr}\left(\mathcal{P}F_{\tau i}(\tau_1)F_{\tau j}(\tau_2)e^{-ig\int_\gamma A}\right)}}{\braket{W}} = \frac{\Lambda\delta_{ij}}{(\tau_2 - \tau_1)^4}\ .
	\end{split}
\end{equation}

Moreover, $F_{\tau i}$ transforms as a vector under the transverse rotation group $SO(d-1)$ (i.e. transverse spin 1), which is the correct transformation property for the displacement operator. $\Lambda$ is a meaningful physical quantity due to the fact that the displacement operator comes with a natural normalization. We refer to it as the two-point function of the displacement operator. The two-point function of $F_{\tau j}$ under Neumann boundary conditions was computed to leading order in \cite{Ciccone:2024guw}, 
\begin{equation}
\label{2pfDisp}
    \Lambda = \frac{2g^2(N^2-1)}{\pi^{d-1} N} + O(g^4)
\end{equation}
for $SU(N)$ fundamental representation in $d=2,3$.

Thus, the Yang-Mills theory predicts the existence of $d-1$ dimension-2, transverse spin-1 operators in CFT$_1$. We will see later that this prediction is consistent with what we expect from the effective string theory.


\subsection{Operator spectrum and correlators at small radius}
Let us now discuss the structure of the theory in the strict weak coupling limit. Specifically, we will comment on the spectrum and $0$-th order OPE coefficients. We will focus on $D=3$ and comment on $D=4$ afterwards.

In the YM description, operators that can be inserted on the Wilson line in a gauge-invariant way are adjoint operators. Since we are pushing the line to the boundary, they should also respect the boundary condition,
\begin{equation}
	F_{ni} = 0\ .
\end{equation}
Here $n$ denotes the radial direction, and other roman letters denote directions parallel to the boundary. When counting such operators, we should also take into account the equations of motion
\begin{equation}
	\partial^\mu F_{\mu \nu} = 0\ ,
\end{equation}
and the Bianchi identity,
\begin{equation}
	\partial_{[\mu} F_{\nu \rho]} = 0\ .
\end{equation}
Finally, $\partial_{\tau}$ is an unbroken generator, and hence operators that are total derivatives with respect to time derivatives are descendants and will not be included. The lightest operators, together with their quantum numbers, are listed in table \ref{tab:examples}. There are two global $\mathbb{Z}_2$ symmetries: transverse parity which we call $\bf{R}$, and the parity along the line which also involves charge conjugation and hence we call it $\bf{CT}$. In the next section \ref{sec:dimreduction}, we introduce a way of keeping track of different operators in terms of bulk ``KK mode" decomposition. Up to overall normalization, the KK mode number is mapped to the number of transverse derivatives acting on the operator (see the last column in \ref{tab:examples}). A comprehensive counting of primary operators is performed in appendix \ref{app:WeakCounting}.  The result for the total  number of primaries is quite simple, 
\begin{equation}
    \mathcal{N}^\text{p,Weak}_{\Delta}=\begin{cases}
        1, & \Delta=0,2 \\
        2^{\Delta-3}, & \Delta=3,4,5,\ldots, \\
        0, & \text{otherwise}. \\
    \end{cases}
\end{equation}
\begin{table}[h]
\centering
\begin{tabular}{|c|c|c|c|c|}
\hline
\textbf{Operator} & \textbf{Dimension} & \textbf{R} & \textbf{CT} &
\textbf{KK mode notation}\\
\hline
$\mathbb{D} = F_{\tau 1}$ & 2 & - & + & $a^{(1)}$ \\
$\partial_1 F_{\tau 1}$ & 3 & + & + & $a^{(2)}$ \\
$\partial^2_1 F_{\tau 1}$ & 4 & - & + & $a^{(3)}$ \\
$F_{\tau 1}F_{\tau 1}$ & 4 & + & + & $a^{(1)}a^{(1)}$ \\
$\partial^3_1 F_{\tau 1}$ & 5 & + & + & $a^{(4)}$ \\
$[\partial_1 F_{\tau 1}F_{\tau 1}]^{-} = \partial_1 F_{\tau 1}F_{\tau 1} - F_{\tau 1}\partial_1F_{\tau 1}$ & 5 & - & - & $a^{(2)}a^{(1)} - a^{(1)}a^{(2)}$ \\
$[\partial_1 F_{\tau 1}F_{\tau 1}]^{+} = \partial_1 F_{\tau 1}F_{\tau 1} + F_{\tau 1}\partial_1F_{\tau 1}$ & 5 & - & + & $a^{(2)}a^{(1)} + a^{(1)}a^{(2)}$ \\
$[F_{\tau 1}F_{\tau 1}]_{1} = \partial_\tau F_{\tau 1}F_{\tau 1} - F_{\tau 1}\partial_\tau F_{\tau 1}$ & 5 & + & + & $\partial_\tau a^{(1)} a^{(1)} - a^{(1)} \partial_\tau a^{(1)}$ \\
\hline
\end{tabular}
\caption{The lowest lying defect operators that can be inserted on the Wilson line. }
\label{tab:examples}
\end{table}
Due to color ordering, the free theory correlators are not the usual Generalized Free Field correlators. For example,
\begin{equation}
\label{eq:0thWeak}
    \Lambda^{-2}\langle D(x_1) D(x_2) D(x_3) D(x_4)\rangle = \frac{1}{(x_2 - x_1)^4(x_4 - x_3)^4} + \frac{1}{(x_3 - x_2)^4(x_4 - x_1)^4} + O(\lambda)
\end{equation}
Here, $\Lambda$ is the unambiguously defined normalization squared of the displacement operator, computed to leading order in \eqref{2pfDisp}. From this we can extract the $0$-th order OPE coefficients \cite{Gaiotto:2013nva},
\begin{equation}
    C^2_{DD[F_{\tau 1} F_{\tau 1} ]_n} = \frac{(4)_n^2}{n!(8+n-1)_n} + O(\lambda)\ ,
\end{equation}
where $[F_{\tau 1} F_{\tau 1} ]_n$ is the unique primary with two $F$'s and $n$ time derivatives, and $(x)_n$ is the Pochhammer symbol.

\paragraph{Comments on 4D.} Weak coupling description in 4D is very similar with the exception that there are three different components of $F$ that can be inserted in the Wilson line and two different transverse derivatives. As a consequence operators come in representations of transverse $O(2)$, rather than just $\mathbb{Z}_2$. Total number of operators still grows exponentially, as it is generic of any large-$N$ gauge theory. Out of three operators made out of components of $F$ two correspond to displacements and are protected, while the third one, with both indices transverse, is not protected. It's classical dimension is also two, however, we expect it to get a growing dimension when $R$ is increased.

%

\subsection{Large radius and effective string}
In the bulk of AdS, the flux tube generated by a boundary Wilson loop can be described by an effective string theory (EST), which is valid when the AdS radius is large. Since EST is formulated in a covariant way in flat space, it can be unambiguously placed in the embedding AdS space. Conformal formulation \eqref{eq:PS} has to be slightly modified since the sigma model on AdS is not conformal at one loop. We believe that the composite dilaton method, used in \cite{Hellerman:2014cba} to fix the central charge, can also be used to reintroduce conformal invariance. However, we leave this for future work and proceed with the static gauge approach generalizing \eqref{eq:Static}.

At leading order, the Nambu-Goto action before gauge fixing reads 
\begin{equation}\label{def:EST}
	\begin{split}
		S_\text{EST} = \frac{1}{\ell_s^2} \int d^2\sigma \sqrt{\det\left(g_{\mu\nu}(X)\,\partial_\alpha X^{\mu} \partial_\beta X^{\nu}\right)} + \ldots
	\end{split}
\end{equation}
Here, $\sigma^\alpha$ ($\alpha = 0,1$) are the string worldsheet coordinates, $X^\mu = X^\mu(\sigma)$ are the target-space AdS coordinates, $g_{\mu\nu}$ is the AdS metric given in \eqref{def:AdSgeometry}, and the ellipsis denotes the other higher-derivative terms allowed by diffeomorphism invariance.

Since the Wilson loop sources the flux tube, it defines the boundary of the worldsheet in EST: 
\begin{equation}
	\begin{split}
		\tau \in \mathbb{R}, \quad x = z = 0.
	\end{split}
\end{equation}
The minimal surface attaching to this boundary is a ``flat plane" given by\footnote{One can view it the minimal surface attaching to two parallel boundary lines with separation distance $L$, and then take the limit $L\rightarrow\infty$. When $L$ is finite, the minimal surface was obtained in \cite{Maldacena:1998im,Rey:1998ik} (see also \cite{Drukker:1999zq} for a nice summary).}
\begin{equation}
	\begin{split}
		\tau \in \mathbb{R}, \quad x \in \mathbb{R}_+, \quad z = 0,
	\end{split}
\end{equation}
so it is natural to fix the worldsheet gauge as
\begin{equation}\label{gaugefix}
	\begin{split}
		X^0(\sigma)=\sigma^0 = \tau, \quad X^1(\sigma)=\sigma^1 = x.
	\end{split}
\end{equation}
Under this gauge choice, the worldsheet theory becomes a theory of $d-1$ scalar fields in AdS$_2$, corresponding to the $d-1$ transverse directions in AdS$_{d+1}$:
\begin{equation}
	\begin{split}
		S_\text{EST}=\frac{R^2}{\ell_s^2}\int\frac{d\tau dx}{x^2}\mathcal{L}(z,\Omega),
	\end{split}
\end{equation}
where $z$ and $\Omega$ are the radial and angular variables for the transverse directions in the AdS geometry \eqref{def:AdSgeometry}.
\begin{equation}\label{NGleading}
	\begin{split}
		\mathcal{L}(z,\Omega)&=1+\frac{1}{2}R^2(\partial z)^2+\frac{1}{2}R^2z^2(\partial\Omega)^2+z^2 \\
		&+\frac{1}{3}z^4+\frac{1}{6}R^2z^4(\partial\Omega)^2+\frac{1}{2}z^4\mathcal{A}+\frac{1}{2}z^2\mathcal{V}^i\mathcal{V}^i -\frac{1}{8}R^4\left((\partial z)^2+z^2(\partial\Omega)^2\right)^2\\
		&+\ldots \\
		\mathcal{A}&:=x^4\left(\left(\frac{\partial\Omega}{\partial\tau}\right)^2\left(\frac{\partial\Omega}{\partial x}\right)^2-\left(\frac{\partial\Omega}{\partial\tau}\cdot\frac{\partial\Omega}{\partial x}\right)^2\right),\quad \mathcal{V}^i:=\epsilon^{\alpha\beta}\partial_\alpha z\partial_\beta \Omega^i. \\
	\end{split}
\end{equation}
Here all the index contractions are with respect to the AdS$_2$ metric
\begin{equation}
	\begin{split}
		ds^2_{\text{AdS}_2}=R^2\frac{d\tau^2+dx^2}{x^2}\,.
	\end{split}
\end{equation}
One can check that the right-hand side of \eqref{NGleading} does not depend on $R$ term by term.

When \( R \gg \ell_s \), the worldsheet theory becomes weakly coupled. After performing the field redefinition \( z\Omega^i := \frac{\ell_s}{R} y^i \), equation \eqref{NGleading} can be expanded as a power series in the small parameter \( \left(\frac{\ell_s}{R}\right)^2 \), leading to the action:
\begin{equation}\label{NGleadingInY}
	\begin{split}
		S_\text{NG}[y]&=\int\frac{d\tau dx}{x^2}R^2\bigg{[}\frac{1}{\ell_s^2}+\frac{1}{2}\partial^\alpha y^i \partial_\alpha y^i+\frac{1}{R^2}y^2 \\
		&\qquad\qquad+\frac{1}{8}\ell_s^2(\partial^\alpha y^i \partial_\alpha y^i)^2   - \frac{1}{4}\ell_s^2\partial_\alpha y^i \partial^\alpha y^j \partial_\beta y^i \partial^\beta y^j  \\
		&\qquad\qquad+ \frac{1}{3}\frac{\ell_s^2}{R^4} (y^2)^2 + \frac{1}{6}\frac{\ell_s^2}{R^2}\left(y^2 \partial^\alpha y^i \partial_\alpha y^i - y^i y^j \partial^\alpha y^i \partial_\alpha y^j\right)+\ldots\bigg{]}.
	\end{split}
\end{equation}
Low-lying operators produce excitations with energies of order $R^{-1}$, and the dimensional coupling constant in the theory is $\lambda_s^{-1}$, where \
\begin{align}
 \lambda_s\equiv R/\ell_s \,.
 \label{lambdas}
\end{align}
$\lambda_s$ differs from $\lambda$, defined in \eqref{def:lambda} only by an order-one rescaling, so the expansion at large radius is likewise a large-$\lambda$ expansion.

The EST in the flat-space limit corresponds to the intermediate length scale \( R \gg L \gg \ell_s \), where $L$ is the length scale of the observables. In this regime, we choose an arbitrary base point \( (\tau_*, x_*) \) in AdS\(_2\) and define the rescaled coordinates as
\begin{equation}
	\begin{split}
		\tau := \tau_*+ \frac{\tau_\text{phys}}{R}x_*, \qquad x := x_* \left(1+ \frac{x_\text{phys}}{R}\right).
	\end{split}
\end{equation}
Recall that \( \tau \) and \( x \) are dimensionless in our earlier parametrization \eqref{def:AdSgeometry}, so \( \tau_\text{phys} \) and \( x_\text{phys} \) have dimensions of length and are of the same order as physical distances.\footnote{In terms of $\tau_\text{phys}$ and $x_\text{phys}$, the AdS$_2$ geometry is given by
	\begin{equation*}
		ds^2_{\text{AdS}_2}=\left(d\tau_\text{phys}^2+d x_\text{phys}^2\right)\left[1+O\left(\tfrac{\abs{x_\text{phys}}}{R}\right)\right].
	\end{equation*}
}
One can then show that when \( \tau_\text{phys}, x_\text{phys} = O(L) \), all non-derivative terms (such as \( y^2 \), \( y^4 \), and \( y^2 (\partial y)^2 \)) are suppressed by powers of \( \left( \tfrac{L}{R} \right)^2 \), while the remaining derivative terms reduce to those in equation~\eqref{eq:Static}.\footnote{In the power counting, we treat derivatives as \( \partial = O(L^{-1}) \) and the transverse field (in string units) as \( y = O(1) \).}

Recall that for a long string in flat spacetime \( \mathbb{R}^{1,d} \), the \( d-1 \) transverse modes are massless due to the spontaneous breaking of Poincaré symmetry:
\begin{equation}
	ISO(1,d) \longrightarrow ISO(1,1) \times O(d-1).
\end{equation}
In contrast, for a long string in AdS\(_{d+1}\), the transverse modes acquire a protected mass \( m^2 = \tfrac{2}{R^2} \) from the AdS\(_2\) point of view, as seen in the quadratic part of \eqref{NGleadingInY}. We will argue that this mass remains protected in the presence of interaction terms, based on a similar spontaneous-symmetry-breaking argument, which we briefly review below.\footnote{We thank Mehrdad Mirbabayi for his notes \cite{Mirbabayi2019unpublished} on this argument.}

Since the theory depends only on the dimensionless parameter \( \ell_s / R \) (and possibly other order-one couplings parameters  appearing in non-universal terms in the action), we set $$R = 1$$ for convenience. We also ignore the presence of the additional couplings, as the argument below is not affected by them.

Since in general spacetime dimensions the Goldstone modes we are discussing transform under the vector representation of \( SO(d-1) \), it suffices to demonstrate the existence of a single such mode. In this sense, it is enough to present the argument in AdS\(_3\), from which the generalization to AdS\(_{d+1}\) is straightforward (we will comment on this at the end). 

The worldsheet action \eqref{def:EST} is invariant under the AdS spacetime symmetry:
\begin{equation}
	\begin{split}
		X^M(\sigma) \rightarrow \Lambda^M{}_N\,X^N(\sigma),
	\end{split}
\end{equation}
where \( X^M \) (\( M = 0,1,\ldots,d+1 \)) are the embedding coordinates in the ambient Lorentzian space \( \mathbb{R}^{1,d+1} \), satisfying \( X \cdot X = -1 \), and \( \Lambda \in SO(1,d+1) \) is a Lorentz transformation. However, generic Lorentz transformations do not preserve the gauge choice \eqref{gaugefix}. To restore the gauge, one must perform a compensating worldsheet diffeomorphism. As a result, the Lorentz group is realized nonlinearly on the worldsheet degrees of freedom. 

For AdS$_3$, the infinitesimal transformation generated by \( L_{03} \) acts on the field \( z \) as
\begin{equation}\label{InfiTransf}
	\begin{split}
		\delta z &= \epsilon\, L_{0,3} \cdot z, \\
		L_{0,3} \cdot z &= \frac{1 + \tau^2 + x^2}{2x} + \tau x \tanh z\, \partial_\tau z + \frac{1}{2}(x^2 - \tau^2 - 1)\tanh z\, \partial_x z,
	\end{split}
\end{equation}
where \( \epsilon \) is an infinitesimal parameter. We provide a detailed derivation of \eqref{InfiTransf} as well as its higher-dimensional generalization in appendix~\ref{app:NLRAdSIso}.

Since this is a symmetry transformation, the variation of the Lagrangian density must be a total derivative:
\begin{equation}
	\delta\mathcal{L} = \epsilon\, \nabla_\alpha j^\alpha.
\end{equation}
Then, by the standard Noether procedure, the associated conserved current is
\begin{equation}\label{def:NoetherCurrent}
	J^\alpha = \frac{\partial \mathcal{L}}{\partial (\partial_\alpha z)}\,(L_{0,3} \cdot z) - j^\alpha.
\end{equation}

Consider now the two-point function \( \braket{J^\alpha(\tau,x)\,z(\tau',x')} \). The conservation equation reads
\begin{equation}
	\braket{\nabla_\alpha J^\alpha(\tau,x)\, z(\tau',x')} = - \braket{(L_{0,3} \cdot z)(\tau',x')}\, \frac{\delta(\tau - \tau')\delta(x - x')}{\sqrt{g}}.
\end{equation}
In coordinates \( (\tau,x) \), this becomes
\begin{equation}\label{Lorentz:conservation}
	\partial_\tau \braket{J_\tau(\tau,x)\, z(\tau',x')} + \partial_x \braket{J_x(\tau,x)\, z(\tau',x')} = - \braket{(L_{0,3} \cdot z)(\tau',x')}\, \delta(\tau - \tau')\delta(x - x').
\end{equation}

By integrating both sides over a region enclosing \( (\tau',x') \) and applying Stokes' theorem, we obtain
\begin{equation}\label{JnIntegral}
	\oint_\Sigma d\Sigma^n\, \braket{J_n(\tau,x)\, z(\tau',x')} = - \braket{(L_{0,3} \cdot z)(\tau',x')},
\end{equation}
where \( \Sigma \) is an arbitrary closed surface surrounding the point \( (\tau',x') \).

In the spirit of Goldstone's theorem in flat space~\cite{Nambu:1960tm,Goldstone:1961eq,Goldstone:1962es}, we postulate that
\begin{itemize}
	\item The vacuum expectation value \( \braket{(L_{0,3} \cdot z)(\tau',x')} \) is \emph{not} identically zero as a function of \( \tau' \) and \( x' \).
\end{itemize}
We will later verify that this is indeed the case in the limit \( \ell_s = 0 \). When the string coupling is turned on, the postulate would fail if and only if
\begin{equation*}
	\frac{1 + \tau^2 + x^2}{2x} + \tau x \braket{\tanh z\, \partial_\tau z} + \frac{1}{2}(x^2 - \tau^2 - 1)\braket{\tanh z\, \partial_x z} = 0
\end{equation*}
holds at every point \( (\tau, x) \) in AdS$_2$. We regard this possibility as highly unlikely, essentially it would correspond to the symmetry restoration.

We now argue that when \( \ell_s \ll R \), the surface integral on the left-hand side of \eqref{JnIntegral} reduces to a single integral along a constant-\( x \) slice:
\begin{equation}\label{IntCurrent}
	\int_{-\infty}^{+\infty} d\tau\, \braket{J_x(\tau,x)\, z(\tau',x')} = \braket{(L_{0,3} \cdot z)(\tau',x')}, \qquad (0 < x < x').
\end{equation}
To see this, we first examine the zero-coupling limit \( \ell_s = 0 \). As discussed below \eqref{NGleading}, in this limit the worldsheet theory \eqref{NGleading} reduces to a free massive scalar theory with a shift symmetry:
\begin{equation}\label{freescalar}
	\begin{split}
		\text{Theory:} \quad &\mathcal{L} = \tfrac{1}{2}(\partial z)^2 + z^2, \\
		\text{Symmetry:} \quad &(L_{0,3} \cdot z)(\tau,x) = \frac{1 + \tau^2 + x^2}{2x}.
	\end{split}
\end{equation}
Here the scalar mass is \( m^2 = 2 \) (in units of \( 1/R^2 \)), protected by the shift symmetry~\cite{Bonifacio:2018zex,Blauvelt:2022wwa}. Imposing Dirichlet boundary conditions \( z|_{\partial} = 0 \), the scaling dimension of the corresponding boundary operator is
\begin{equation}
	\Delta_z = 2 \qquad \text{when } \ell_s = 0.
\end{equation}

The Noether current for this symmetry takes the form
\begin{equation}
	\begin{split}
		J_\tau &= \frac{1 + \tau^2 + x^2}{2x} \partial_\tau z - \frac{\tau}{x} z, \\
		J_x &= \frac{1 + \tau^2 + x^2}{2x} \partial_x z + \frac{1 + \tau^2 - x^2}{2x^2} z,
	\end{split}
\end{equation}
where the second term in each line comes from the \( j^\alpha \) contribution in \eqref{def:NoetherCurrent}.

A direct computation gives
\begin{equation}\label{free:integral}
	\begin{split}
		\int_{-\infty}^{+\infty} d\tau\, \frac{1 + \tau^2 + x^2}{2x} \braket{\partial_x z(\tau,x)\, z(\tau',x')} &= \frac{1 + \tau'^2 + x'^2}{3x'}, \\
		\int_{-\infty}^{+\infty} d\tau\, \frac{1 + \tau^2 - x^2}{2x^2} \braket{z(\tau,x)\, z(\tau',x')} &= \frac{1 + \tau'^2 + x'^2}{6x'},
	\end{split}
\end{equation}
which together reproduce the right-hand side of \eqref{IntCurrent}, in agreement with \eqref{freescalar}.

Finally, one can verify by explicit computation that the remaining contributions to the surface integral vanish due to the rapid decay of the integrands at infinity:
\begin{equation}
	\int_{-\infty}^{+\infty} d\tau\, \braket{J_x(\tau,\infty)\, z(\tau',x')} = 
	\int_{x}^{\infty} dy\, \braket{J_\tau(+\infty,y)\, z(\tau',x')} = 
	\int_{x}^{\infty} dy\, \braket{J_\tau(-\infty,y)\, z(\tau',x')} = 0.
\end{equation}

Now we turn on the coupling \( \ell_s \). When the coupling is sufficiently small, the asymptotic behavior of the two-point function remains largely unchanged. Consequently, the other contributions to the surface integral still vanish due to sufficiently fast power-law decay at infinity. Thus, equation~\eqref{IntCurrent} continues to hold for small coupling.

Since the left-hand side of~\eqref{IntCurrent} is independent of \( x \), we can take the limit \( x \to 0 \). From equations~\eqref{InfiTransf} and~\eqref{def:NoetherCurrent}, the Noether current \( J_x \) takes the form
\begin{equation}
	J_x = \left[\frac{1+\tau^2+x^2}{2x} + \tau x \tanh z\, \partial_\tau z + \frac{1}{2}(x^2 - \tau^2 - 1)\tanh z\, \partial_x z\right] V_x - j_x,
\end{equation}
where
\begin{equation}\label{def:Vx}
	V_x \equiv \frac{\partial \mathcal{L}}{\partial (\partial^x z)} = \partial_x z + \ldots
\end{equation}
is the \( x \)-component of a vector operator. Near the boundary \( x \to 0 \), we have the estimates
\begin{equation}
	\tanh z\, \partial_\tau z = O(x^{4+\gamma}), \qquad \tanh z\, \partial_x z = O(x^{3+\gamma}),
\end{equation}
where \( \gamma = \gamma(\ell_s/R) \) is a small coupling-dependent correction to the scaling. For sufficiently small coupling, these terms vanish as \( x \to 0 \). Therefore,
\begin{equation}
	\begin{split}
		\lim_{x \to 0} \int_{-\infty}^{+\infty} d\tau\, \braket{J_x(\tau,x)\, z(\tau',x')} 
		= \lim_{x \to 0} \int_{-\infty}^{+\infty} d\tau\, \frac{1 + \tau^2 + x^2}{2x}\, \braket{V_x(\tau,x)\, z(\tau',x')} \\
		\qquad - \lim_{x \to 0} \int_{-\infty}^{+\infty} d\tau\, \braket{j_x(\tau,x)\, z(\tau',x')}.
	\end{split}
\end{equation}
Each of the integrals on the right-hand side reduces to its counterpart in equation~\eqref{free:integral} in the free theory. When the coupling is small, we expect both to receive small corrections and remain non-zero. 

From the definition~\eqref{def:Vx}, the near-boundary scaling of \( V_x \) is
\begin{equation}\label{Vx:order}
	V_x(\tau,x) = O(x^{1+\gamma_z}),
\end{equation}
where \( \gamma_z \) denotes the anomalous dimension of the lightest boundary operator appearing in the bulk-to-boundary expansion of \( z \). Since the coupling is small, other operators have larger scaling dimensions and are more suppressed in the small-\( x \) limit. In order for the integral
\begin{equation}\label{Intlim:xtoboundary}
	\lim_{x \to 0} \int_{-\infty}^{+\infty} d\tau\, \frac{1 + \tau^2 + x^2}{2x}\, \braket{V_x(\tau,x)\, z(\tau',x')}
\end{equation}
to equal a finite non-zero number, by \eqref{Vx:order} we must have
\begin{equation}
	\gamma_z = 0.
\end{equation}
Therefore, the boundary scaling dimension of the transverse modes is protected and remains equal to two, at least in the regime of small coupling.

One may then ask whether the transverse modes continue to satisfy \( \Delta_z = 2 \) when the coupling is not small. We expect this to be the case. At zero coupling, the transverse modes are the lightest operators transforming in the vector representation of the transverse rotation group \( SO(d-1) \). All other operators in this representation satisfy \( \Delta - \Delta_z \geqslant 1 \). As the coupling is increased continuously from zero, and assuming no level crossing, these heavier modes should remain above \( \Delta_z \). Consequently, \( \Delta_z = 2 \) must continue to hold in order to ensure a non-vanishing value of the integral~\eqref{IntCurrent}.

The generalization to AdS$_{d+1}$ with $d \geqslant 2$ is straightforward. In this case, one replaces $z$ in AdS$_3$ by $z\,\Omega^i$, and the relevant Noether current is associated with the generator $L_{0,i}$.

Let us comment on the Coleman--Mermin--Wagner--type theorems \cite{Coleman:1973ci,Mermin:1966fe}, which forbid the existence of Goldstone modes associated with global symmetry breaking in two-dimensional flat space. Our argument does not contradict these results. First, even in flat space the implication of this theorem is different for spontaneous breaking of spacetime symmetries. As the flat space flux tube reviewed in section \ref{sec:flat} demonstrates, spontaneous symmetry breaking still happens, even though branons themselves are not well-defined operators. Nevertheless, their derivatives are, and this is enough to define physical observables like the worldsheet $S$-matrix, and to write an effective action. Second, breaking of internal symmetries is also sometimes possible in AdS$_2$, as discussed in \cite{Copetti:2023sya}. Finally, in our case of spacetime symmetry in AdS the Goldstone modes are massive, hence the infrared divergence present in flat space is regulated.

\subsection{Operator spectrum and correlators at large radius}
In the strict large $\lambda_s$ limit we have a $1$ dimensional GFF of operators corresponding to $m^2=2R^{-2}$ Goldstone bosons, which in this limit we denote by $X$. It is the same displacement operator $\mathbb{D}$ which at weak gauge coupling was equal to $F_{\tau 1}$. There is no color ordering, and the lowest lying operators in the spectrum are summarized in table \ref{tab:examplesStr}. 
\begin{table}[h]
\centering
\begin{tabular}{|c|c|c|c|}
\hline
\textbf{Operator} & \textbf{Dimension} & \textbf{R} & \textbf{CT}\\
\hline
$\mathbb{D} = X$ & 2 & - & +  \\
$X^2$ & 4 & + & + \\
$X^3$ & 6 & - & + \\
$[X^2]_{2} \propto 4\,\partial_{\tau}^2 X X - 5\,(\partial_{\tau} X)^2$ & 6 & + & + \\
$[X^3]_{2} \propto 5 X^2 \partial_{\tau}^2 X-4\, X  (\partial_{\tau} X)^2$ & 8 & - & + \\
$[X^2]_{4} \propto 10\,\partial_{\tau}^4 X X - 70 \,\partial_{\tau} X \partial_{\tau}^3 X +63 \,(\partial^2_\tau X)^2$ & 8 & + & + \\
$X^4$ & 8 & + & + \\
$[X^3]_{3} \propto 4\,X^2\partial_{\tau}^3 X  - 18X\partial_{\tau} X \partial_{\tau}^2X + 15 \,(\partial_{\tau} X)^3$ & 9 & - & - \\
\hline
\end{tabular}
\caption{The lowest lying defect in the line theory dual to the effective string.}
\label{tab:examplesStr}
\end{table}
We defer the full counting to future work \cite{paper2}. However, due to the free-boson nature of the theory, the large-dimension asymptotics of the density of states is of the Cardy form. Specifically,
\begin{equation}
    \mathcal{N}^\text{p,Strong}_{\Delta} \sim  \text{exp}\left(\pi \sqrt{\frac{2\Delta}{3}}\right)
\end{equation}
This formula is expected to hold at arbitrarily high dimensions in the strict large radius (strong coupling) limit.\footnote{In principle, the presence of resonances could modify the  exponent at some fixed energy scale. However, to date, there is no evidence or theoretical necessity for the existence of such resonances in 3D YM, see section \ref{sec:flat} for more details and references.} At any finite (but large) $\lambda_s$ the asymptotic behavior above is valid up to roughly $\Delta \sim \lambda_s^2=R^2/\ell_s^2$ when the interactions become large.


The leading order theory is a Generalized Free Field (Mean Field Theory). The computation of the tree level correction was performed in \cite{Giombi:2017cqn}. Overall, we get,
\begin{equation}
\label{eq:01thStrong}
\begin{split}
    G(x) &= 1+\frac{1}{t^4} + \frac{1}{(1+t)^4} \\
    &\hspace{15pt}+\frac{1}{2\pi\lambda_s^2} \frac{2}{3(1+t)^5t^5}\Bigg[-(12 t^{11}+66 t^{10}+150 t^9+180 t^8+120 t^7+36 t^6+12 t^5) \log (t) \\
    &\hspace{15pt}-12 t^{10}-60 t^9-121 t^8-124 t^7-91 t^6-91 t^5-124 t^4-121 t^3-60 t^2-12 t\\
    &\hspace{15pt}+6 (t+1)^5 \left(2 t^6+t^5+t+2\right) \log (t+1)\Bigg]
     + O(\lambda_s^{-4})
\end{split}
\end{equation}
with,
\begin{equation}
    G(t) = \frac{\langle D(x_1) D(x_2) D(x_3) D(x_4)\rangle}{\langle D(x_2) D(x_3)\rangle\langle D(x_1) D(x_4)\rangle}\quad,\quad t=\frac{(x_2-x_1)(x_4-x_3)}{(x_3-x_2)(x_4-x_1)}
\end{equation}
From this we can extract the conformal data of two-letter operators up to tree level following \cite{Giombi:2017cqn}:
\begin{equation}
\label{eq:AnomStr}
    \gamma_{2n} =-\frac{2n^2+7n+4}{ 2\pi \lambda_s^2} + \dots 
\end{equation}
and, 
\begin{equation}
\label{eq:OPEStr}
    C^2_{XX[XX]_{2n}} = \frac{\Gamma^2((n+2)2)\Gamma(2n+7)}{18\Gamma(4n+7)\Gamma(2n+1)} + \frac{1}{2}\partial_{2n}\left(C^2_{XX[XX]_{2n}}\gamma_{2n}\right) + \dots \ ,
\end{equation}
where $[XX]_{2n}$ is the unique primary with two $X$'s and $2n$ time derivatives.

\subsection{Interpolation}
\begin{figure}[h]
    \centering
    \includegraphics[width=1.\linewidth]{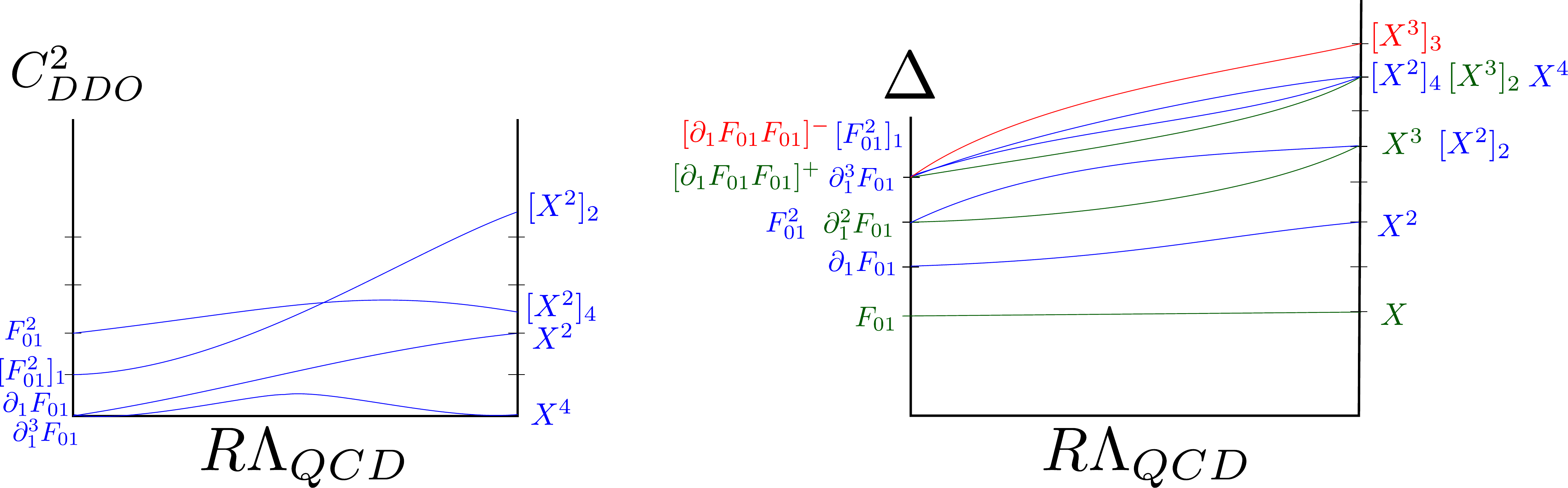}
    \caption{Rough expectation for the behavior of OPE coefficients (left) and dimensions (right) as a function of $\lambda = R \Lambda_\text{QCD}$. The absence of level crossing leads to a prediction for which operator is continuously connected to which operator, for the first few low lying operators. By studying the magnitude of the leading perturbative corrections, we can extend the predictions to higher operators. ++ operators are in blue, -+ in green and -- in red.}
    \label{fig:Inter}
\end{figure}

Now that we understand the large and small radius limits of the theory, we can compare the two on the qualitative level. As demonstrated in figure \ref{fig:Inter}, it is straightforward to imagine how low lying can operators will map to each other. 
{The first ambiguity appears at level 5 at weak coupling when we have two ++ operators that should get dimension 8 at strong coupling. We chose the interpolation $[F_{\tau 1}F_{\tau 1}]_{1}\to [X^2]_4$ and $\d_1^3F_{\tau 1}\to X^4$, as can be seen on the left panel with OPE coefficients. We hope to be able to confirm whether this is the right choice by further calculations in the future.}

In fact, the signs of all strong coupling anomalous dimensions \eqref{eq:AnomStr} are compatible with this scenario. In the next $2$ sections we will compute the leading nontrivial conformal data in the weak coupling limit for single-letter operators, and see that the signs of anomalous dimensions there are also compatible with a smooth monotonic interpolation. {We did not show any +- operators simply because the first one appears at level 6 and interpolates to level 11 at strong coupling}

Another regime we can examine is the asymptotic high energy regime. We see that the asymptotic scaling at weak and strong couplings are
\begin{equation}
    \mathcal{N}^\text{p,Weak}_{\Delta} = 2^\Delta\,, \quad \mathcal{N}^\text{p,Strong}_{\Delta} \sim \text{exp}\left(\pi \sqrt{\frac{2\Delta}{3}}\right)\,.
\end{equation}
In a single sector of the $\mathbb{Z}_2 \times \mathbb{Z}_2$ symmetry, there should be no level crossing, unless there are additional undetected global symmetries. Thus, assuming a smooth interpolation from small to large  $\lambda$ leads to a square root type formula for heavy operators, namely,
\begin{equation}
    \Delta_{n}(0) \sim \sqrt{\Delta_n(\infty)}
    \label{sqrtformula}\,.
\end{equation}
One can try to interpolate these two asymptotics with a formula similar to \eqref{TTbar}, for example using
\begin{align}   \Delta_n(\lambda)\stackrel{?}{=}\sqrt{\lambda_s^4+2 \lambda_s^2 \Delta_n^0(\lambda_s)}-\lambda_s^2\,,
\label{TTbarAdS}
\end{align}
where we imply that $\lambda=0$ corresponds to $\lambda_s$ equal to some order-one number, and $\Delta_n^0(\lambda_s)$ refers to the spectrum of a putative theory without the semiclassical string tension effects and whose anomalous dimensions are not as large. 

A not unrelated phenomenon is the appearance of color ordering. While the Goldstone bosons at large radius are free bosons, the small radius limit is a theory of ordered free fields. This can be seen, for example, from the missing $u$-channel in the $0$-th order four-point function \eqref{eq:0thWeak}, compared to \eqref{eq:01thStrong}. It is in fact the same ordering that is responsible for the faster proliferation of states in the small radius limit, giving rise to the square root relation above. The transition from one limit to the other can be viewed as a change of statistics of the particles. Let us note that this phenomenon is completely universal among all confining large-$N$ gauge theories. It may be easiest to understand it in the context of 2D QCD following the ideas of \cite{Dubovsky:2018dlk,Donahue:2019adv}. 

Coming back to 4D YM, we anticipate that smooth interpolation will require adding the axion on the EST side. Since the axion is massive, it would have infinite dimension at the strict $\lambda_s=\infty$ limit. However, flat space results confirm that the axion can be thought of as a light particle, compared to the string tension scale. We thus believe that it is still possible to keep it in the effective theory by formulating it at finite $\lambda_s$.

\section{Dimensional Reduction and Perturbative Calculations}
\label{sec:dimreduction}
Having discussed the two descriptions of the confining flux tube at small and large AdS radius, we now turn to perturbative calculations. In this work, we restrict our attention to calculations at weak gauge coupling in three dimensions. Calculations of the strong coupling anomalous dimensions and OPE coefficients using the Nambu-Goto action will be presented in \cite{paper2}. We will perform the weak coupling calculation in two distinct ways. In this section, we carry out a direct perturbative analysis, employing the dimensional reduction of Yang–Mills theory from AdS$_3$ to AdS$_2$. In contrast, in section \ref{sec:Nonlin}, we exploit the nonlinearly realized two-dimensional global conformal symmetry to compute observables. Each method accesses a somewhat different subset of observables, with agreement in their overlap providing a nontrivial consistency check.

Let us begin with the dimensional reduction of Yang–Mills theory from AdS$_3$ to AdS$_2$. The key observation is that a gauge field in AdS$_3$ can be recast as a gauge field plus an infinite tower of massive adjoint vector fields in AdS$_2$. This is a special case of the result in \cite{Artsukevich:2008vy}. We present this analysis for two main reasons. First, from a computational standpoint, integrals are more easily evaluated in AdS$_2$. Second, as we reviewed above, at large AdS radius a Wilson line on the boundary of AdS$_3$ is described by EST which is a QFT of a scalar field living on the AdS$_2$ background. Since our goal is to match with that description, it is also natural to start from AdS$_2$ on the weakly coupled Yang-Mills side as well. Using coordinates \eqref{def:AdSgeometry} the Yang-Mills action \eqref{YM:convention} 
is given explicitly by
\begin{equation}\label{YM:AdS2}
	\begin{split}
		S_{\rm YM}[A]&=\frac{1}{2}\int \frac{d\tau\,dx\,dz}{x^{2}}\left[\frac{x^{4}}{\cosh^{2}z}\,{\rm Tr}\left(F_{\tau x}^2\right)+x^2\,{\rm Tr}\left(F_{z\tau}^2\right)+x^2\,{\rm Tr}\left(F_{zx}^2\right)\right]. \\
	\end{split}
\end{equation}
We emphasize that there is no contraction of tensor indices here.

For the reduced modes in AdS$_2$, we will somewhat abusively refer to them as “KK modes”. We remark that the terms “KK reduction” and “KK modes” typically corresponds to the dimensional reduction along a compact extra dimension, e.g.\ $\mathcal{M} \times S^1 \rightarrow \mathcal{M}$. In our case, the extra dimension is $\mathbb{R}$, which is non-compact. Nevertheless, the KK-like modes still acquire discrete masses due to the presence of the warp factor $\cosh^2 z$ and the imposed boundary conditions.

\subsection{KK modes}
\label{sec:KKmodes}
For convenience, we choose the axial gauge:
\begin{equation}\label{AdS3:axialgauge}
	\begin{split}
		A_z = 0.
	\end{split}
\end{equation}
Then, the free part of the Yang-Mills action \eqref{YM:AdS2} becomes
\begin{equation}\label{YM:freepart}
	\begin{split}
		S_{\rm YM}^\text{free}[A] &= \frac{1}{2}\int \frac{d\tau\,dx}{x^{2}}\,\int dz\,{\rm Tr}\left[\frac{x^{4}}{\cosh^2z}\,\left(\partial_\tau A_x - \partial_x A_\tau\right)^2 + x^2\,\left((\partial_z A_{\tau})^2 + (\partial_z A_{x})^2\right)\right]. \\
	\end{split}
\end{equation}
We now decompose the gauge field $A_\mu$ in AdS$_3$ into KK modes in AdS$_2$:
\begin{equation}\label{A:KKexpansion}
	\begin{split}
		A_\mu(\tau, x, z) = \sum_{n} A_\mu^{(n)}(\tau, x)\,\Omega_n(z).
	\end{split}
\end{equation}
Plugging \eqref{A:KKexpansion} into \eqref{YM:freepart}, our basic requirement is that the mode functions $\Omega_n(z)$ decouple after integrating over $z$:
\begin{equation}\label{AdS3:KKintegral}
	\begin{split}
		\int \frac{dz}{\cosh^2z}\,\Omega_m(z)\,\Omega_n(z), \quad \int dz\,\partial_z \Omega_m\,\partial_z \Omega_n \ \propto \ \delta_{m,n}.
	\end{split}
\end{equation}
To achieve this, the standard approach is to let them satisfy the differential equation
\begin{equation}\label{AdS3:KKeq}
	\begin{split}
		\frac{d^2 \Omega}{dz^2} + \frac{\lambda}{\cosh^2z}\,\Omega = 0.
	\end{split}
\end{equation}
After changing the variable to $w \equiv \tanh z$ and setting $\lambda = \nu(\nu + 1)$, equation \eqref{AdS3:KKeq} becomes the Legendre equation with parameter $\nu$:
\begin{equation}\label{AdS3:Legendre}
	\begin{split}
		\frac{d}{dw}\left[(1 - w^2)\frac{d\Omega}{dw}\right] + \nu(\nu + 1)\,\Omega = 0.
	\end{split}
\end{equation}
For a complete basis with respect to the $L^2$ norm
\begin{equation}
	\begin{split}
		\int \frac{dz}{\cosh^2z}\,\abs{f(z)}^2 \ \equiv \ \int dw\,\abs{\tilde{f}(w)}^2, \quad w = \tanh z,\ \tilde{f}(w) = f(z),
	\end{split}
\end{equation}
and satisfying \eqref{AdS3:KKintegral}, it suffices to choose the Legendre polynomials of $w$:
\begin{equation}\label{basis:LegendreP}
	\begin{split}
		\Omega_n(z) = \sqrt{\frac{2n + 1}{2}}\,P_n(\tanh z), \quad n = 0,1,2,\ldots
	\end{split}
\end{equation}
The orthonormality condition of the Legendre polynomials on the interval $[-1,1]$ is equivalent to
\begin{equation}\label{LegendreP:orthonormal}
	\begin{split}
		\int_{-\infty}^{+\infty} \frac{dz}{\cosh^2z}\,\Omega_n(z)\,\Omega_m(z) = \delta_{n,m}, \\
		\sum_{n=0}^{\infty} \Omega_n(z)\,\Omega_n(z') = \cosh^2z\,\delta(z - z').
	\end{split}
\end{equation}

\subsection{Reduced Yang-Mills action in AdS$_2$}
We now rewrite the Yang-Mills action \eqref{YM:AdS2} in terms of KK modes:
\begin{equation}\label{action:KKreduced}
	\begin{split}
		S_{\rm YM}[A] &= S_{\rm YM}^\text{free}[A] + S_{\rm YM}^\text{int}[A], \\
		S_{\rm YM}^\text{free}[A] &= \frac{1}{2}\sum_{n=0}^{\infty}\int \frac{d\tau\,dx}{x^{2}}\,{\rm Tr}\left[x^{4}\,\left(\partial_\tau A^{(n)}_x - \partial_x A^{(n)}_\tau\right)^2 + x^2\,m_n^2\left(\bigl(A^{(n)}_{\tau}\bigr)^2 + \bigl(A^{(n)}_{x}\bigr)^2\right)\right], \\
		S_{\rm YM}^\text{int}[A] &= ig\sum_{n_1,n_2,n_3=0}^{\infty} I^{3pt}_{n_1,n_2,n_3} \int \frac{d\tau\,dx}{x^{2}}\,x^{4}\,{\rm Tr}\left[\left(\partial_\tau A^{(n_1)}_x - \partial_x A^{(n_1)}_\tau\right)\,\bigl[A^{(n_2)}_\tau, A^{(n_3)}_x\bigr]\right] \\
		&\quad - \frac{1}{2}g^2\sum_{n_1,n_2,n_3,n_4=0}^{\infty} I^{4pt}_{n_1,n_2,n_3,n_4} \int \frac{d\tau\,dx}{x^{2}}\,x^{4}\,{\rm Tr}\left[\bigl[A^{(n_1)}_{\tau}, A^{(n_2)}_{x}\bigr]\,\bigl[A^{(n_3)}_{\tau}, A^{(n_4)}_{x}\bigr]\right]. \\
	\end{split}
\end{equation}
Here, $m_n^2 = n(n+1)$, and the coefficients of the interaction vertices are given by:
\begin{equation}
\label{eq:I3andI4}
	\begin{split}
		I^{3pt}_{n_1,n_2,n_3} &\equiv \int_{-1}^{1} dw\,\Omega_{n_1}(w)\,\Omega_{n_2}(w)\,\Omega_{n_3}(w) \\
		&= \sqrt{\frac{(2n_1 + 1)(2n_2 + 1)(2n_3 + 1)}{2}}\,\threej{n_1}{n_2}{n_3}{0}{0}{0}^2, \\
		I^{4pt}_{n_1,n_2,n_3,n_4} &\equiv \int_{-1}^{1} dw\,\Omega_{n_1}(w)\,\Omega_{n_2}(w)\,\Omega_{n_3}(w)\,\Omega_{n_4}(w) \\
		&= \sum_{n} I^{3pt}_{n_1,n_2,n}\,I^{3pt}_{n_3,n_4,n}, \\
	\end{split}
\end{equation}
where the bracket denotes the 3j symbol
\begin{equation}\label{def:3jsymbol}
	\begin{split}
		\threej{j_1}{j_2}{j_3}{0}{0}{0}&=(-1)^s\sqrt{\dfrac{s!\left(\tfrac12\right)_{s-j_1} \left(\tfrac12\right)_{s-j_2} \left(\tfrac12\right)_{s-j_3}}
		{\left(\tfrac32\right)_{s} (s-j_1)! (s-j_2)! (s-j_3)!}} \\
		& \text{if } j_i\geqslant\abs{j_k-j_l} \ \text{and}\ 2s\equiv j_1+j_2+j_3\ \text{even}, \\
		\threej{j_1}{j_2}{j_3}{0}{0}{0}&=0\quad\text{otherwise}.
	\end{split}
\end{equation} 
We list the vertex coefficients involving the zero mode $A^{(0)}$:
\begin{equation}
	\begin{split}
		I^{3pt}_{n,n,0} &= \frac{1}{\sqrt{2}}, \\
		I^{4pt}_{n_1,n_2,n_3,0} &= \frac{1}{\sqrt{2}}\,I^{3pt}_{n_1,n_2,n_3} = \frac{\sqrt{(2n_1 + 1)(2n_2 + 1)(2n_3 + 1)}}{2}\,\threej{n_1}{n_2}{n_3}{0}{0}{0}^2, \\
		I^{4pt}_{n,n,0,0} &= \frac{1}{2}. \\
	\end{split}
\end{equation}

The action \eqref{action:KKreduced} can be interpreted as a gauge field in AdS$_2$ (zero mode) coupled to an infinite tower of massive adjoint matter fields (higher KK modes). The covariant derivatives are given by:
\begin{equation}
	\begin{split}
		D_\mu A^{(n)}_\nu = \partial_\mu A^{(n)}_\nu + \frac{ig}{\sqrt{2}}\bigl[A^{(0)}_\mu, A^{(n)}_\nu\bigr],
	\end{split}
\end{equation}
where the extra factor of $\sqrt{2}$ comes from the normalization of the zero mode function $\Omega_0(z) = \tfrac{1}{\sqrt{2}}$.

Recall that we have chosen the axial gauge \eqref{AdS3:axialgauge} in AdS$_3$. The residual gauge redundancy is constrained by:
\begin{equation}\label{gauge:extra}
	\begin{split}
		\partial_z U = 0,
	\end{split}
\end{equation}
which corresponds precisely to gauge transformations in AdS$_2$:
\begin{equation}
	\begin{split}
		&A^{(0)}_\mu \rightarrow U A^{(0)}_\mu U^{-1} - \frac{i\sqrt{2}}{g} U \partial_\mu U^{-1}, \\
		&A^{(n)}_\mu \rightarrow U A^{(n)}_\mu U^{-1}\quad(n\geqslant1), \\
		&U = U(\tau, x), \quad \mu = \tau, x. \\
	\end{split}
\end{equation}
This allows us to further gauge-fix the zero mode as:
\begin{equation}\label{AdS3:extraaxialgauge}
	\begin{split}
		A^{(0)}_x = 0.
	\end{split}
\end{equation}

One can also write \eqref{action:KKreduced} in a more gauge-covariant form:
\begin{equation}\label{action:KKreducedCovariant}
	\begin{split}
		S_{\rm YM}[A] &=  \frac{1}{4}\sum_{n=0}^{\infty}\int \frac{d\tau\,dx}{x^{2}}\,{\rm Tr}\left(f^{(n)}_{\mu\nu}f^{(n)\mu\nu} + 2\,m_n^2\,A^{(n)}_{\mu}A^{(n)\mu}\right) \\
		&\quad + \frac{1}{2}ig\sum_{n_1,n_2,n_3=1}^{\infty} I^{3pt}_{n_1,n_2,n_3} \int \frac{d\tau\,dx}{x^{2}}\,{\rm Tr}\left(f^{(n_1)}_{\mu\nu}\,\bigl[A^{(n_2)\mu}, A^{(n_3)\nu}\bigr]\right) \\
		&\quad - \frac{1}{4}g^2\sum_{n_1,n_2,n_3,n_4=1}^{\infty} I^{4pt}_{n_1,n_2,n_3,n_4} \int \frac{d\tau\,dx}{x^{2}}\,{\rm Tr}\left(\bigl[A^{(n_1)}_{\mu}, A^{(n_2)}_{\nu}\bigr]\,\bigl[A^{(n_3)\mu}, A^{(n_4)\nu}\bigr]\right), \\
	\end{split}
\end{equation}
where
\begin{equation}
	\begin{split}
		f^{(0)}_{\mu\nu} &= \partial_\mu A^{(0)}_\nu - \partial_\nu A^{(0)}_\mu + \frac{ig}{\sqrt{2}}\bigl[A^{(0)}_\mu, A^{(0)}_\nu\bigr], \\
		f^{(n)}_{\mu\nu} &= D_\mu A^{(n)}_\nu - D_\nu A^{(n)}_\mu \quad (n \geqslant 1). \\
	\end{split}
\end{equation}

\subsection{Neumann boundary conditions}

The boundary of AdS$_3$ is given by
\begin{equation}
	\begin{split}
		z=\pm\infty \quad \text{or} \quad x=0.
	\end{split}
\end{equation}
The limit $z \to \pm\infty$ corresponds to approaching the two-dimensional conformal boundary. By contrast, the limit $x \to 0$ is more subtle, as it probes a one-dimensional subspace, which can equivalently be reached by first taking $z \to \pm\infty$ and then sending $x \to \infty$. Because of this subtlety, it is more convenient to impose boundary conditions in Poincar\'e half-space coordinates
\begin{equation}
	\begin{split}
		ds^2=\frac{dt^2+dy^2+dw^2}{w^2}.
	\end{split}
\end{equation}
In Poincaré coordinates, the Neumann boundary condition is given by\footnote{In generic dimensions, eq.~\eqref{AdS3:Nbc} should include an additional factor of $w^{3-d}$. However, since we are interested in $d\leqslant3$, it turns out that the inclusion of this factor does not affect the conclusion.}
\begin{equation}\label{AdS3:Nbc}
	\begin{split}
		F_{tw}\big|_{w=0}=F_{yw}\big|_{w=0}=0,
	\end{split}
\end{equation}
that is, the electric fields vanish at the boundary.


We leave the explicit relation between coordinates $(t,y,w)$ and $(\tau,x,z)$ to appendix \ref{app:AdS3}. Using the Jacobian of the coordinate transformation, one can show that
\begin{equation}\label{F:transform}
	\begin{split}
		F_{tw}=-\frac{\sinh (z)}{x}F_{\tau z}+\frac{1}{\cosh(z)}F_{\tau x},\quad F_{yw}=-\frac{\cosh (z)}{x}F_{x z}.
	\end{split}
\end{equation}
Let us first look at the boundary regime where $x\neq0$ and $z=\pm\infty$. There, the Neumann boundary condition \eqref{AdS3:Nbc} is equivalent to
\begin{equation}
	\begin{split}
		\sinh(z)\partial_zA_\tau\Big{|}_{z=\pm\infty}=\cosh(z)\partial_zA_x\Big{|}_{z=\pm\infty}=0.
	\end{split}
\end{equation}
It turns out that this condition is automatically satisfied by all the KK modes, because
\begin{equation}
	\begin{split}
		\sinh(z)\partial_z\Omega_n=\sqrt{\frac{2n+1}{2}} \frac{\sinh(z)}{\cosh^2(z)}P_n'(\tanh z)\rightarrow0\quad(z\rightarrow\pm\infty), \\
		\cosh(z)\partial_z\Omega_n=\sqrt{\frac{2n+1}{2}} \frac{1}{\cosh(z)}P_n'(\tanh z)\rightarrow0\quad(z\rightarrow\pm\infty), \\
	\end{split}
\end{equation}

We now turn to the other component of the boundary, located at \( x=0 \). As mentioned above, this corresponds to a one-dimensional subspace of the boundary that is independent of the transverse coordinate \( z \) (see appendix~\ref{app:AdS3}). Accordingly, we impose the self-consistency condition that the gauge field be independent of \( z \) at \( x=0 \). In the axial gauge \( A_z = 0 \), this condition reads
\begin{equation}\label{YM:extraBC:A}
	\begin{split}
		\partial_zA_\tau\Big{|}_{x=0}=\partial_z A_x\Big{|}_{x=0}=0\,,
	\end{split}
\end{equation}
which is equivalent to
\begin{equation}\label{YM:extraBC}
	\begin{split}
		F_{\tau z}\Big{|}_{x=0}=F_{x z}\Big{|}_{x=0}=0\,.
	\end{split}
\end{equation}

In terms of KK modes, \eqref{YM:extraBC:A} implies that
\begin{equation}
	\begin{split}
		A^{(n)}_\tau\Big{|}_{x=0}=A^{(n)}_x\Big{|}_{x=0}=0\quad(n=1,2,3,\ldots),
	\end{split}
\end{equation}
i.e. the Dirichlet boundary condition for higher KK modes.
\newpage
In summary:
\begin{itemize}
	\item The KK modes automatically meet the Neumann boundary condition \eqref{AdS3:Nbc} at $x\neq0$.
	\item The Neumann boundary condition at $x=0$ implies the Dirichlet condition on higher KK modes: $$A^{(n)}_\tau\Big{|}_{x=0}=A^{(n)}_x\Big{|}_{x=0}=0\quad(n=1,2,3,\ldots).$$
\end{itemize}

\subsection{Free propagators with the Neumann boundary condition}
Now let us compute the free gluon propagators. We use the Lie algebra basis $\{t^a\}$ and choose the following normalization in the fundamental representation
\begin{equation}
	\begin{split}
		{\rm Tr}\left(t^at^b\right)=\delta^{ab}.
	\end{split}
\end{equation}
Under this convention, the structure constants are defined by
\begin{equation}
	\begin{split}
		[t^a,t^b]=i f^{abc}t^c.
	\end{split}
\end{equation}
Then the free gluon action is given by
\begin{equation}
	\begin{split}
		S_{\rm YM}^{free}[A]&=\frac{1}{2}\sum\limits_{c=1}^{{\rm dim}G}\sum\limits_{n=0}^{\infty}\int \frac{d\tau\,dx}{x^{2}}\left[x^{4}\left(\partial_\tau A^{c,(n)}_x-\partial_x A^{c,(n)}_\tau\right)^2+x^2n(n+1)\left({A^{c,(n)}_{\tau}}^2+{A^{c,(n)}_{x}}^2\right)\right], \\
	\end{split}
\end{equation}
where $G$ is the gauge group.
\newline
\newline\noindent\textbf{Zero KK modes}
\newline For the 0-th KK modes, we have chosen the gauge $A^{(0)}_x=0$. Then its free propagator satisfies  
\begin{equation}
	\begin{split}
		{\rm e.o.m}:&\quad\partial_x\left(x^2\partial_x\right)\braket{A^{a,(0)}_\tau(\tau,x)A^{b,(0)}_\tau(\tau',x')}+\delta^{ab}\delta(\tau-\tau')\delta(x-x')=0, \\
		{\rm finiteness}:&\quad\braket{A^{a,(0)}_\tau(\tau,x)A^{b,(0)}_\tau(\tau',x')}\big{|}_{x=0}\ {\rm is\ finite}, \\
		{\rm permutation}:&\quad\braket{A^{a,(0)}_\tau(\tau,x)A^{b,(0)}_\tau(\tau',x')}=\braket{A^{b,(0)}_\tau(\tau',x')A^{a,(0)}_\tau(\tau,x)}.
	\end{split}
\end{equation}
The above equations uniquely fix the free propagators of the 0-th KK modes as follows:
\begin{equation}
	\begin{split}
		\braket{A^{a,(0)}_\tau(\tau,x)A^{b,(0)}_\tau(\tau',x')}=\begin{cases}
			\frac{\delta^{ab}\delta(\tau-\tau')}{x}, & x>x', \\
			\frac{\delta^{ab}\delta(\tau-\tau')}{x'}, & x<x'. \\
		\end{cases}
	\end{split}
\end{equation}
The free propagator of the gauge curvature is obtained by taking derivatives in $x$ and $x'$:
\begin{equation}
	\begin{split}
		\braket{F^{a,(0)}_{\tau x}(\tau,x)F^{b,(0)}_{\tau x}(\tau',x')}=-\delta^{ab}\left[\frac{\partial}{\partial x}\left(\frac{\delta(\tau-\tau')\delta(x-x')}{x}\right)+\frac{\partial}{\partial x'}\left(\frac{\delta(\tau-\tau')\delta(x-x')}{x'}\right)\right].
	\end{split}
\end{equation}
Therefore, the free propagator of the gauge curvature is localized in the bulk and there is no boundary d.o.f. corresponding to it. In other words, there is no gluon in AdS$_2$ as expected.
\newline
\newline\textbf{Higher KK modes}
\newline For higher KK modes ($n\geqslant1$), the problem is reduced to solving the free propagator of the Proca field:
\begin{equation}\label{AdS3:KKaction}
	\begin{split}
		S_{\rm Proca}[A]=\frac{1}{2}\int \frac{d\tau\,dx}{x^{2}}\left[x^{4}\left(\partial_\tau A_x-\partial_x A_\tau\right)^2+x^2m^2\left({A_{\tau}}^2+{A_{x}}^2\right)\right], \\
	\end{split}
\end{equation}
with the Dirichlet boundary condition
\begin{equation}\label{AdS3:KKbc}
	\begin{split}
		A_\tau\Big{|}_{x=0}=A_x\Big{|}_{x=0}=0.
	\end{split}
\end{equation}
Define the propagators
\begin{equation}\label{AdS3:KKpropagator}
	\begin{split}
		\Pi_{\mu\nu}(X_1,X_2):=\braket{A_\mu(X_1)A_\nu(X_2)},
	\end{split}
\end{equation}
where $X_1=(\tau_1,x_1)$ and $X_2=(\tau_2,x_2)$ are the points in AdS$_2$. They satisfy the following equations of motion
\begin{equation}\label{AdS3:KKeom}
	\begin{split}
		&\partial_{x_1}\left[x_1^2(\partial_{\tau_1}\Pi_{x\tau}-\partial_{x_1}\Pi_{\tau\tau})\right]+m^2\Pi_{\tau\tau}=\delta(\tau_1-\tau_2)\delta(x_1-x_2), \\
		&\partial_{x_1}\left[x_1^2(\partial_{\tau_1}\Pi_{xx}-\partial_{x_1}\Pi_{\tau x})\right]+m^2\Pi_{\tau x}=0, \\
		&-x_1^2\partial^2_{\tau_1}\Pi_{x\tau}+x_1^2\partial_{\tau_1}\partial_{x_1}\Pi_{\tau\tau}+m^2 \Pi_{x\tau}=0, \\
		&-x_1^2\partial^2_{\tau_1}\Pi_{xx}+x_1^2\partial_{\tau_1}\partial_{x_1}\Pi_{\tau x}+m^2 \Pi_{x x}=\delta(\tau_1-\tau_2)\delta(x_1-x_2). \\
	\end{split}
\end{equation}
The solution to \eqref{AdS3:KKeom} with Dirichlet boundary condition \eqref{AdS3:KKbc} is given by \cite{Costa:2014kfa}
\begin{equation}
	\begin{split}
		\Pi_{\mu\nu}(X_1,X_2)=-\frac{\partial^2 u}{\partial X_1^\mu\partial X_2^\nu}g_0(u)+\frac{\partial u}{\partial X_1^\mu}\frac{\partial u}{\partial X_2^\nu}g_1(u),
	\end{split}
\end{equation}
where the AdS$_2$-invariant $u$ is given by
\begin{equation}
	\begin{split}
		u(X_1,X_2)=\frac{(\tau_1-\tau_2)^2+(x_1-x_2)^2}{2x_1x_2},
	\end{split}
\end{equation}
The form factors $g_0(u)$ and $g_1(u)$ are given by
\begin{equation}
	\begin{split}
		g_0(u)&=(\Delta-1)F_1(u)+\frac{1+u}{u}F_2(u), \\
		g_1(u)&=\frac{(1+u)(\Delta-1)}{u(2+u)}F_1(u)+\frac{1+(1+u)^2}{u^2(2+u)}F_2(u), \\
	\end{split}
\end{equation}
where
\begin{equation}
	\begin{split}
		F_1(u)&=\mathcal{N}(\Delta)(2u)^{-\Delta}\hyperF{\Delta}{\Delta}{2\Delta}{-\frac{2}{u}}, \\
		F_2(u)&=\mathcal{N}(\Delta)(2u)^{-\Delta}\hyperF{\Delta}{\Delta+1}{2\Delta}{-\frac{2}{u}}, \\
		\mathcal{N}(\Delta)&=\frac{\Gamma(\Delta-1)}{2\pi^{1/2}\Gamma(\Delta+\frac{1}{2})}, \\
		m^2&=\Delta(\Delta-1),\quad(\Delta>1). \\
	\end{split}
\end{equation}
In summary, for Neumann boundary condition in AdS$_3$:
\begin{itemize}
	\item The bulk-bulk free propagators of the zero KK modes are given by
	\begin{equation}\label{AdS3:0propa}
		\begin{split}
			\braket{A^{a,(0)}_\tau(\tau,x)A^{b,(0)}_\tau(\tau',x')}=\begin{cases}
				\frac{\delta^{ab}\delta(\tau-\tau')}{x}, & x>x', \\
				\frac{\delta^{ab}\delta(\tau-\tau')}{x'}, & x<x'. \\
			\end{cases}
		\end{split}
	\end{equation}
	Here we have already chosen the axial gauge $A_x^{a,(0)}=0$.
	\item The bulk-bulk free propagators of the $n$-th KK modes ($n=1,2,\ldots$) are given by 
	\begin{equation}\label{AdS3:npropa}
		\begin{split}
			\braket{A_\mu^{a,(n)}(\tau,x)A_\nu^{b,(n)}(\tau',x')}&=\delta^{ab}\left[-\frac{\partial^2 u}{\partial X_1^\mu\partial X_2^\nu}g_0(n;u)+\frac{\partial u}{\partial X_1^\mu}\frac{\partial u}{\partial X_2^\nu}g_1(n;u)\right], \\
		\end{split}
	\end{equation}
	where the coordinate variables are given by
	\begin{equation*}
		\begin{split}
			X_1&=(\tau,x),\quad X_2=(\tau',x'), \quad u=\frac{(\tau-\tau')^2+(x-x')^2}{2xx'}, \\
		\end{split}
	\end{equation*}
	and the mode functions are given by
	\begin{equation*}
		\begin{split}
			g_0(n;u)&=nF_1(n;u)+\frac{1+u}{u}F_2(n;u), \\
			g_1(n;u)&=\frac{n(1+u)}{u(2+u)}F_1(n;u)+\frac{1+(1+u)^2}{u^2(2+u)}F_2(n;u), \\
			F_1(n;u)&=\mathcal{N}_n(2u)^{-n-1}\hyperF{n+1}{n+1}{2n+2}{-\frac{2}{u}}, \\
			F_2(n;u)&=\mathcal{N}_n(2u)^{-n-1}\hyperF{n+1}{n+2}{2n+2}{-\frac{2}{u}}, \\
			\mathcal{N}_n&\equiv\mathcal{N}(n+1)=\frac{\Gamma(n)}{2\pi^{1/2}\Gamma(n+\frac{3}{2})}. \\
		\end{split}
	\end{equation*}
\end{itemize}

\subsection{Boundary modes, bulk-to-boundary free propagators}
We define the boundary operators to be
\begin{equation}\label{AdS3:defbdryop}
	\begin{split}
		a_\tau^{(n)}(\tau)&:=\lim\limits_{x\rightarrow0}\left[x^{-n}A_\tau^{(n)}(\tau,x)\right], \\
		a_x^{(n)}(\tau)&:=\lim\limits_{x\rightarrow0}\left[x^{-n-1}A_x^{(n)}(\tau,x)\right]. \\
	\end{split}
\end{equation}
For higher KK modes, we have the conservation law coming from the equation of motion \eqref{AdS3:KKeom}:
\begin{equation}\label{AdS3:KKconservation}
	\begin{split}
		\partial_\tau A^{(n)}_\tau+\partial_x A^{(n)}_x=0,\quad (n\geqslant1).
	\end{split}
\end{equation}
The above equation holds when it is inserted away from other operators in the correlation function. By \eqref{AdS3:defbdryop} and \eqref{AdS3:KKconservation}, we get
\begin{equation}
	\begin{split}
		\partial_\tau a^{(n)}_\tau+(n+1)a^{(n)}_x=0\quad\Rightarrow\quad a^{(n)}_x=-\frac{1}{n+1}\partial_\tau a^{(n)}_\tau.
	\end{split}
\end{equation}
We see that $a_x^{(n)}$ is proportional to the level-1 descendant of $a_\tau^{(n)}$. Therefore, for each higher KK mode, there is only one boundary primary state corresponding to it. The description of boundary operators in terms of KK modes is equivalent to the description in terms of transverse derivatives of the field strength, as indicated in Table \ref{tab:examples}, where we dropped the index $\tau$ for compactness.

By taking the limit $x\rightarrow0$ for one of the bulk operator $A^{(n)}$ in \eqref{AdS3:npropa}, we get the bulk-boundary free propagator of the n-th KK mode:
\begin{equation}\label{AdS3:bulkbdry}
	\begin{split}
		\braket{a_\tau^{(n)}(\tau_1)A_\tau^{(n)}(\tau_2,x_2)}&=-\frac{(n+1)\Gamma(n)}{2\sqrt{\pi}\Gamma\left(n+\frac{3}{2}\right)}\frac{x_2^n\left[(\tau_1-\tau_2)^2-x_2^2\right]}{\left[(\tau_1-\tau_2)^2+x_2^2\right]^{n+2}}, \\
		\braket{a_\tau^{(n)}(\tau_1)A_x^{(n)}(\tau_2,x_2)}&=\frac{(n+1)\Gamma(n)}{\sqrt{\pi}\Gamma\left(n+\frac{3}{2}\right)}\frac{x_2^{n+1}(\tau_1-\tau_2)}{\left[(\tau_1-\tau_2)^2+x_2^2\right]^{n+2}}. \\
	\end{split}
\end{equation}
If we take the limit $x\rightarrow0$ for both operators in the bulk-bulk free propagator \eqref{AdS3:npropa}, we get the boundary two-point function of $a_\tau^{(n)}$:
\begin{equation}\label{AdS3:bdry2pt}
	\begin{split}
		\braket{a_\tau^{(n)}(\tau_1)a_\tau^{(n)}(\tau_2)}=-\frac{(n+1)\Gamma(n)}{2\sqrt{\pi}\Gamma\left(n+\frac{3}{2}\right)}(\tau_1-\tau_2)^{-2n-2}.	
	\end{split}
\end{equation}
This is consistent with the boundary scaling dimension of $a_\tau^{(n)}$:
\begin{equation}
	\begin{split}
		\Delta_n=n+1.
	\end{split}
\end{equation}
Using \eqref{AdS3:bulkbdry} and \eqref{AdS3:bdry2pt}, we can also work out the bulk-to-boundary OPE of $A^{(n)}$:
\begin{equation}\label{AdS3:bdryOPE}
	\begin{split}
		A^{(n)}_\tau(\tau,x)&=\sum_{k=0}^{\infty}\frac{(-1)^k(n+2)_k}{k!}\frac{n+2k+1}{n+k+1} x^{n+2k}\partial^{2k}_\tau a^{(n)}_\tau(\tau), \\
		A^{(n)}_x(\tau,x)&=-\sum_{k=0}^{\infty}\frac{(-1)^k(n+2)_k}{k!}\frac{1}{n+k+1} x^{n+2k+1}\partial^{2k+1}_\tau a^{(n)}_\tau(\tau), \\
	\end{split}
\end{equation}
In principle, one can compute all the free-theory correlation functions of the higher KK modes using \eqref{AdS3:bdry2pt} and \eqref{AdS3:bdryOPE}.

\subsection{Feynman rules}
From now on, we do perturbation theory in the coupling:
\begin{equation}
	\begin{split}
		g\equiv g_{\rm YM}R^{\frac{1}{2}}\,.
	\end{split}
\end{equation}

The Feynman rule for the 3-point vertex is given by 
\begin{equation}
	\label{eq:3vertex}
	\begin{aligned}
		&\vcenter{
			\hbox{
				\begin{tikzpicture}
					\begin{feynman}
						\vertex (a) at (-1,0.55) {\scriptsize\((n_1,a,\mu)\)};
						\vertex (b) at (1,0.55) {\scriptsize\((n_2, b, \nu)\)};
						\vertex (c) at (0,-1.2) {\scriptsize\((n_3, c, \rho)\)};
						\vertex (x) at (0,-0.2);  
						
						\diagram* {
							(a) -- [gluon] (x),
							(b) -- [gluon] (x),
							(c) -- [gluon] (x),
						};
					\end{feynman}
					
					\node at (0.18,-0.26) {\scriptsize\(x\)};
					\node at (-0.9,0.25) {\scriptsize\(x_1\)};
					\node at (0.9,0.25) {\scriptsize\(x_2\)};
					\node at (-0.2,-1) {\scriptsize\(x_3\)};
					
					\fill (-0.68,0.28) circle (0.05);
					\fill (0.68,0.28) circle (0.05);
					\fill (0,-0.95) circle (0.05);
					\fill (0,-0.2) circle (0.05);
					
					\draw[black] (0,0) circle [radius=1.7];
					
				\end{tikzpicture}
			}
		}  \\
		&= g\,I_{n_1,n_2,n_3}^{3pt}\,f^{abc}
		\Bigg[\frac{\partial\Pi^{n_1}_{\mu\beta}(x_1,x)}{\partial x^\alpha}
		\left(\Pi^{n_2}_{\nu\alpha}(x_2,x)\Pi^{n_3}_{\rho\beta}(x_3,x) -\alpha\leftrightarrow\beta\right) \\
		&\qquad\qquad\qquad\qquad+ (123)\leftrightarrow (231)+(123)\leftrightarrow(312) \Bigg]\,.\\		
	\end{aligned}
\end{equation}
The Feynman rule for the 4-point vertex (connecting to four bulk points) is given by
\begin{equation}
	\begin{aligned}
		&\vcenter{
			\hbox{
				\begin{tikzpicture}
					\begin{feynman}
						\vertex (a) at (-0.75,0.75) {\scriptsize\((n_1, a, \mu)\)};
						\vertex (b) at (0.75,0.75) {\scriptsize\((n_2, b, \nu)\)};
						\vertex (c) at (0.75,-0.75) {\scriptsize\((n_3, c, \rho)\)};
						\vertex (d) at (-0.75,-0.75) {\scriptsize\(( n_4, d, \sigma)\)};
						\vertex (x) at (0,0);  
						
						\diagram* {
							(a) -- [gluon] (x),
							(b) -- [gluon] (x),
							(c) -- [gluon] (x),
							(d) -- [gluon] (x),
						};
					\end{feynman}
					
					\node at (0.3,0) {\scriptsize\(x\)};
					\node at (-0.7,0.5) {\scriptsize\(x_1\)};
					\node at (0.75,0.5) {\scriptsize\(x_2\)};
					\node at (-0.7,-0.5) {\scriptsize\(x_3\)};
					\node at (0.75,-0.5) {\scriptsize\(x_4\)};
					\fill (0,0) circle (0.05);
					
					\fill (-0.5,0.5) circle (0.05);
					\fill (0.5,0.5) circle (0.05);
					\fill (-0.5,-0.5) circle (0.05);
					\fill (0.5,-0.5) circle (0.05);
					
					\draw[black] (0,0) circle [radius=1.7];
				\end{tikzpicture}
			}
		} \\ &=-g^2\,I_{n_1,n_2,n_3,n_4}^{4pt} \Bigg{[}f^{abe}f^{cde}\left(\Pi^{n_1}_{\mu\alpha}(x_1,x)\Pi^{n_2}_{\nu\beta}(x_2,x)-\alpha\leftrightarrow\beta\right)\Pi^{n_3}_{\rho\alpha}(x_3,x)\Pi^{n_4}_{\sigma\beta}(x_4,x) \\
		&\qquad\qquad\qquad\quad+f^{ace}f^{bde}\left(\Pi^{n_1}_{\mu\alpha}(x_1,x)\Pi^{n_3}_{\rho\beta}(x_3,x)-\alpha\leftrightarrow\beta\right)\Pi^{n_2}_{\nu\alpha}(x_2,x)\Pi^{n_4}_{\sigma\beta}(x_4,x) \\
		&\qquad\qquad\qquad\quad+f^{ade}f^{bce}\left(\Pi^{n_1}_{\mu\alpha}(x_1,x)\Pi^{n_4}_{\sigma\beta}(x_4,x)-\alpha\leftrightarrow\beta\right)\Pi^{n_2}_{\nu\alpha}(x_2,x)\Pi^{n_3}_{\rho\beta}(x_3,x)\Bigg{]}\,. \\
	\end{aligned}
\end{equation}
The integration over the internal vertices is given by the following rule:
\begin{equation}\label{eq:VertexInt}
	\begin{split}
		\int d\tau dx x^2\,,
	\end{split}
\end{equation}
where $x^2$ comes from the volume factor ($x^{-2}$) and from raising the indices ($x^4$). In what follows, we will only need the cubic vertex.

\subsection{Tree-level OPE coefficients for single-letter operators}\label{subsection:TreeSingleLetterOp}

We have now all of the ingredients necessary for the computation of the leading order three-point functions of single-letter operators. 
\begin{equation}\label{def:ThreePtSingleLetter}
	\begin{split}
		\braket{a_{\tau}^{(n_1)}(\tau_1)a_{\tau}^{(n_2)}(\tau_2)a_{\tau}^{(n_3)}(\tau_3)}_W&:=\frac{\braket{{\rm Tr}\left(\mathcal{P}a_{\tau}^{(n_1)}(\tau_1)a_{\tau}^{(n_2)}(\tau_2)a_{\tau}^{(n_3)}(\tau_3)e^{-ig\int_\gamma A}\right)}}{\braket{W}}, \\
		W&\equiv\text{Tr}\left(e^{-ig\int_\gamma A}\right).
	\end{split}
\end{equation}
Here $n_i=1,2,\ldots$ labels the KK modes.

In the weak coupling limit ($g=0$), the three-point function \eqref{def:ThreePtSingleLetter} vanishes because there is no way to Wick contract an odd number of free fields. When we turn on the coupling, at order $g$, the only non-vanishing diagram is
\begin{equation}\label{eq:3ptTreeDiag}
	\begin{aligned}
		&\vcenter{
			\hbox{
				\begin{tikzpicture}
					\begin{feynman}
						\vertex (a) at (-1.05,0.55) {\scriptsize\((n_1,a)\)};
						\vertex (b) at (1.05,0.55) {\scriptsize\((n_2, b)\)};
						\vertex (c) at (0,-1.2) {\scriptsize\((n_3, c)\)};
						\vertex (x) at (0,-0.2);  
						
						\diagram* {
							(a) -- [gluon] (x),
							(b) -- [gluon] (x),
							(c) -- [gluon] (x),
						};
					\end{feynman}
					
					\node at (0.18,-0.26) {\scriptsize\(x\)};
					\node at (-0.9,0.25) {\scriptsize\(\tau_1\)};
					\node at (0.9,0.25) {\scriptsize\(\tau_2\)};
					\node at (-0.2,-1) {\scriptsize\(\tau_3\)};
					
					\fill (-0.68,0.28) circle (0.03);
					\fill (0.68,0.28) circle (0.03);
					\fill (0,-0.93) circle (0.03);
					\fill (0,-0.2) circle (0.03);
					
					\draw[black] (0,-0.13) circle [radius=0.8];
				\end{tikzpicture}
			}
		}
		\ 
		\times\ \frac{\text{Tr}(t^a t^b t^c)}{\text{Tr}(1)}
	\end{aligned}
\end{equation}
where the color indices are summed over. 

Since this diagram only involves bulk-to-boundary propagators, the terms in~\eqref{eq:3vertex} are significantly simplified using~\eqref{AdS3:bulkbdry}.  
For example, the first term in~\eqref{eq:3vertex} becomes
\begin{equation}
	\begin{split}
		&\lim_{x_1,x_2,x_3\rightarrow0} \left[ x_1^{-n_1}x_2^{-n_2}x_3^{-n_3} \,
		\frac{\partial\Pi^{n_1}_{\tau\beta}(x_1,x)}{\partial x^\alpha}
		\left( \Pi^{n_2}_{\tau\alpha}(x_2,x)\, \Pi^{n_3}_{\tau\beta}(x_3,x) - \alpha \leftrightarrow \beta \right) \right] \\
		&\quad = -\frac{(n_2+1)(n_3+1) \, \Gamma(n_1+2) \, \Gamma(n_2) \, \Gamma(n_3)}{4 \pi^{3/2} \, \Gamma\!\left(n_1+\frac{3}{2}\right) \Gamma\!\left(n_2+\frac{3}{2}\right) \Gamma\!\left(n_3+\frac{3}{2}\right)} \\
		&\qquad\quad \times
		\frac{(\tau_2-\tau_3) \, x^{n_1+n_2+n_3}
			\left[(\tau -\tau_2)(\tau -\tau_3)+x^2\right]}
		{\left[(\tau -\tau_1)^2+x^2\right]^{n_1+1}
			\left[(\tau -\tau_2)^2+x^2\right]^{n_2+2}
			\left[(\tau -\tau_3)^2+x^2\right]^{n_3+2}} .
	\end{split}
\end{equation}
The second and third terms in~\eqref{eq:3vertex} follow from cyclic permutations of the labels~\(1,2,3\).

Including the vertex integration~\eqref{eq:VertexInt}, the computation reduces to integrals of the form
\begin{equation}\label{def:IntI}
	\begin{split}
		\mathcal{I}_{n_1,n_2,n_3}(\tau_1,\tau_2,\tau_3)
		&= \int d\tau\, dx \; I_{n_1,n_2,n_3}(\tau_1,\tau_2,\tau_3;\tau,x), \\
		I_{n_1,n_2,n_3}(\tau_1,\tau_2,\tau_3;\tau,x)
		&:= \frac{(\tau_2-\tau_3) \, x^{n_1+n_2+n_3+2}
			\left[(\tau -\tau_2)(\tau -\tau_3)+x^2\right]}
		{\left[(\tau -\tau_1)^2+x^2\right]^{n_1+1}
			\left[(\tau -\tau_2)^2+x^2\right]^{n_2+2}
			\left[(\tau -\tau_3)^2+x^2\right]^{n_3+2}} .
	\end{split}
\end{equation}

To proceed, we introduce the scalar three-point integrand
\begin{equation}
	S_{\Delta_1,\Delta_2,\Delta_3}(\tau_1,\tau_2,\tau_3;\tau,x)
	:= x^{-2} \prod_{j=1}^{3} \frac{x^{\Delta_j}}{\left[x^2 + (\tau-\tau_j)^2\right]^{\Delta_j}} .
\end{equation}
Then, following  \cite{Sleight:2016dba,Karateev:2017jgd}, \( I_{n_1,n_2,n_3} \) can be expressed as 
\begin{equation}\label{IntIvsIntS}
	\begin{split}
		I_{n_1,n_2,n_3}
		&= \frac{\tau_2-\tau_3}{2} \left[ S_{n_1+1,n_2+1,n_3+2} + S_{n_1+1,n_2+2,n_3+1} \right] \\
		&\quad + \frac{(\tau_2-\tau_3)^2}{4}
		\left[ \frac{1}{n_2+1} \, \partial_{\tau_2} S_{n_1+1,n_2+1,n_3+2}
		- \frac{1}{n_3+1} \, \partial_{\tau_3} S_{n_1+1,n_2+2,n_3+1} \right].
	\end{split}
\end{equation}

The integral of \( S_{\Delta_1,\Delta_2,\Delta_3} \) corresponds to the tree-level Witten diagram for the scalar three-point function in AdS$_2$/CFT$_1$, whose result is known~\cite{Freedman:1998tz}:
\begin{equation}\label{IntS:result}
	\begin{split}
		&\int d\tau\, dx \; S_{\Delta_1,\Delta_2,\Delta_3}(\tau_1,\tau_2,\tau_3;\tau,x)
		\equiv \int\frac{d\tau\, dx}{x^2}
		\prod_{j=1}^{3} \frac{x^{\Delta_j}}{\left[x^2 + (\tau-\tau_j)^2\right]^{\Delta_j}} \\
		&\quad = \frac{B_{\Delta_1,\Delta_2,\Delta_3}}
		{|\tau_1-\tau_2|^{\Delta_{123}} \,
			|\tau_2-\tau_3|^{\Delta_{231}} \,
			|\tau_1-\tau_3|^{\Delta_{132}}},
	\end{split}
\end{equation}
where \( \Delta_{ijk} \equiv \Delta_i + \Delta_j - \Delta_k \) and the coefficient \( B_{\Delta_1,\Delta_2,\Delta_3} \) is
\begin{equation}\label{def:B}
	B_{\Delta_1,\Delta_2,\Delta_3}
	= \frac{\sqrt{\pi} \, \Gamma\left(\tfrac{\Delta_1+\Delta_2+\Delta_3-1}{2}\right)\Gamma(\tfrac{1}{2}\Delta_{123})\Gamma(\tfrac{1}{2}\Delta_{231})\Gamma(\tfrac{1}{2}\Delta_{132})}{2\,\Gamma(\Delta_1)\Gamma(\Delta_2)\Gamma(\Delta_3)}.
\end{equation}
Combining~\eqref{def:IntI}, \eqref{IntIvsIntS}, \eqref{IntS:result} and~\eqref{def:B}, we obtain the expression for the integral $\mathcal{I}_{n_1,n_2,n_3}$:
\begin{equation}
	\begin{split}
		\mathcal{I}_{n_1,n_2,n_3}(\tau_1,\tau_2,\tau_3)
		&=-\frac{\sqrt{\pi}\,
			\Gamma\!\left(\tfrac{1}{2}n_{123}+1\right)
			\Gamma\!\left(\tfrac{1}{2}n_{132}+1\right)
			\Gamma\!\left(\tfrac{1}{2}n_{231}+1\right)
			\Gamma\!\left(\tfrac{n_1+n_2+n_3+3}{2}\right)}
		{2\,\Gamma(n_1+1)\,\Gamma(n_2+2)\,\Gamma(n_3+2)} \\
		&\quad\times
		\frac{1}{(\tau_2-\tau_1)^{n_{123}+1}
			(\tau_3-\tau_1)^{n_{132}+1}
			(\tau_3-\tau_2)^{n_{231}+1}},
	\end{split}
\end{equation}
where \( n_{ijk} \equiv n_i+n_j-n_k \).

This, together with the prefactor in~\eqref{def:IntI}, completes the computation of the first term in the Feynman diagram~\eqref{eq:3vertex}.  
The second and third terms follow by cyclic permutation of $(1,2,3)$.  
Altogether, we obtain
\begin{equation}\label{TreeInt:final}
	\begin{split}
		&-\frac{(n_2+1)(n_3+1)\,\Gamma(n_1+2)\,\Gamma(n_2)\,\Gamma(n_3)}
		{4\pi^{3/2}\,\Gamma\!\left(n_1+\tfrac{3}{2}\right)
			\Gamma\!\left(n_2+\tfrac{3}{2}\right)
			\Gamma\!\left(n_3+\tfrac{3}{2}\right)}
		\;\mathcal{I}_{n_1,n_2,n_3}(\tau_1,\tau_2,\tau_3) \\
		&\quad + (123)\!\leftrightarrow\!(231) + (123)\!\leftrightarrow\!(312) \\
		&= \frac{\Gamma\!\left(\tfrac{1}{2}n_{123}+1\right)
			\Gamma\!\left(\tfrac{1}{2}n_{132}+1\right)
			\Gamma\!\left(\tfrac{1}{2}n_{231}+1\right)
			\Gamma\!\left(\tfrac{n_1+n_2+n_3+3}{2}\right)}
		{8\pi\,n_1n_2n_3\,
			\Gamma\!\left(n_1+\tfrac{3}{2}\right)
			\Gamma\!\left(n_2+\tfrac{3}{2}\right)
			\Gamma\!\left(n_3+\tfrac{3}{2}\right)} \\
		&\qquad\times
		\sum_{i=1}^{3} n_i(n_i+1)\,
		\frac{1}{(\tau_2-\tau_1)^{n_{123}+1}
			(\tau_3-\tau_1)^{n_{132}+1}
			(\tau_3-\tau_2)^{n_{231}+1}}.
	\end{split}
\end{equation}

The final result for~\eqref{eq:3ptTreeDiag} is~\eqref{TreeInt:final} multiplied by the factor
\begin{equation}
	g\,I_{n_1,n_2,n_3}^{3pt}\,f^{abc}
	\frac{\mathrm{Tr}(t^a t^b t^c)}{\mathrm{Tr}(1)}
	= i\,\frac{C_2(G)\,d_G}{2\,d_r}\; g\,I_{n_1,n_2,n_3}^{3pt},
\end{equation}
where $C_2(G)$ denotes the quadratic Casimir $C_2\equiv\sum_a t^a t^a$ in the adjoint representation, $d_G$ is the adjoint dimension, and $d_r$ the dimension of the chosen representation for the trace.  
For $SU(N)$ in the fundamental representation,\footnote{In our convention, $t^a$ differs from~\cite{Peskin:1995ev} by a factor $\sqrt{2}$, so $C_2$ differs by a factor of $2$.}
\begin{equation}
	\frac{C_2(G)\,d_G}{2\,d_F} = N^2-1.
\end{equation}

Thus, the tree-level three-point function~\eqref{def:ThreePtSingleLetter} reads
\begin{equation}
	\begin{split}
		&\braket{a_{\tau}^{(n_1)}(\tau_1)a_{\tau}^{(n_2)}(\tau_2)a_{\tau}^{(n_3)}(\tau_3)}_W \\
		&\quad = i\,g\,(N^2-1)\,I_{n_1,n_2,n_3}^{3pt}
		\frac{\Gamma\!\left(\tfrac{1}{2}n_{123}+1\right)
			\Gamma\!\left(\tfrac{1}{2}n_{132}+1\right)
			\Gamma\!\left(\tfrac{1}{2}n_{231}+1\right)
			\Gamma\!\left(\tfrac{n_1+n_2+n_3+3}{2}\right)}
		{8\pi\,n_1n_2n_3\,
			\Gamma\!\left(n_1+\tfrac{3}{2}\right)
			\Gamma\!\left(n_2+\tfrac{3}{2}\right)
			\Gamma\!\left(n_3+\tfrac{3}{2}\right)} \\
		&\qquad\times \sum_{i=1}^{3} n_i(n_i+1)\,
		\frac{1}{(\tau_2-\tau_1)^{n_{123}+1}
			(\tau_3-\tau_1)^{n_{132}+1}
			(\tau_3-\tau_2)^{n_{231}+1}}
		+ O(g^2).
	\end{split}
\end{equation}

This correlator is purely imaginary because the operators have not yet been normalized.  
In the current normalization, the two-point function is
\begin{equation}
	\begin{split}
		\braket{a_{\tau}^{(n_1)}(\tau_1)a_{\tau}^{(n_2)}(\tau_2)}_W
		&:= \frac{\braket{\mathrm{Tr}\!\left(\mathcal{P}\,a_{\tau}^{(n_1)}(\tau_1)\,a_{\tau}^{(n_2)}(\tau_2)\,e^{-ig\int_\gamma A}\right)}}{\braket{W}} \\
		&= -\frac{(n_1+1)\,\Gamma(n_1)}{2\sqrt{\pi}\,\Gamma\!\left(n_1+\frac{3}{2}\right)}
		\frac{1}{(\tau_1-\tau_2)^{2n_1+2}}
		\frac{N^2-1}{N}\,\delta_{n_1,n_2}
		+ O(g^2),
	\end{split}
\end{equation}
where we used~\eqref{AdS3:bdry2pt} and $C_2 = \frac{N^2-1}{N}$ for $SU(N)$ in the fundamental representation. To obtain the standard OPE coefficient, we define the canonically normalized operators,
\begin{equation}
    \braket{\tilde{a}^{(n_1)}(0)\,\tilde{a}^{(n_2)}(1)}_W = \delta_{n_1,n_2}\ .
\end{equation}
They are related to the unnormalized single-letter operators in the following way,
\begin{equation}
	\tilde{a}^{(n)} :=
	i\left[\frac{(n+1)\,\Gamma(n)}{2\sqrt{\pi}\,\Gamma\!\left(n+\frac{3}{2}\right)}
	\frac{N^2-1}{N}\right]^{-\frac12} {a}^{(n)} + O(g^2).
\end{equation}
In this normalization, the OPE coefficient can be easily extracted from the three-point function,
\begin{equation}
\label{eq:OPECaaa}
	\braket{\tilde{a}^{(n_1)}(\tau_1)\,\tilde{a}^{(n_2)}(\tau_2)\,\tilde{a}^{(n_3)}(\tau_3)}_W
	= \frac{C_{n_1,n_2,n_3}}
	{(\tau_2-\tau_1)^{n_{123}+1}
		(\tau_3-\tau_1)^{n_{132}+1}
		(\tau_3-\tau_2)^{n_{231}+1}}
	+ O(g^2),
\end{equation}
with the final answer
\begin{equation}\label{eq:pertC}
	\begin{split}
		C_{n_1,n_2,n_3}
		&= g \sqrt{\frac{N^3}{N^2 - 1}}
		\left[ \sum_{j=1}^3 n_j (1 + n_j) \right]
		\frac{
			\big( \tfrac{n_1 + n_2 + n_3}{2} \big)!\,
			\Gamma\!\left( \tfrac{n_{123} + 1}{2} \right)
			\Gamma\!\left( \tfrac{n_{132} + 1}{2} \right)
			\Gamma\!\left( \tfrac{n_{231} + 1}{2} \right)
		}{
			\pi^{2}
			\prod_{j=1}^{3}
			\sqrt{
				2 n_j\,(n_j + 1)\, n_j!\,
				\big( \tfrac{1}{2} \big)_{n_j}
			}
		}
		+ O(g^2).
	\end{split}
\end{equation}

Now we take $N$ large. If we choose to define the confinement scale  as $\Lambda_\text{QCD}=g_\text{YM}^2 N$, then we can express the answer in terms of $\lambda$, defined in \eqref{def:lambda}, which is the dimensionless 't Hooft coupling in our model. 


\begin{equation}
\label{eq:finalC}
	\begin{split}
		C_{n_1,n_2,n_3} =\left[ \sum_{j=1}^3 n_j (1 + n_j) \right]
			\frac{
				\big( \tfrac{n_1 + n_2 + n_3}{2} \big)!\,
				\Gamma\!\left( \tfrac{n_{123} + 1}{2} \right)
				\Gamma\!\left( \tfrac{n_{132} + 1}{2} \right)
				\Gamma\!\left( \tfrac{n_{231} + 1}{2} \right)
			}{
				\pi^{2}
				\prod_{j=1}^{3}
				\sqrt{
					2 n_j\,(n_j + 1)\, n_j!\,
					\big( \tfrac{1}{2} \big)_{n_j}
				}
			}\lambda^{1/2}+O(\lambda) \,,
	\end{split}
\end{equation}
which is the main result of this section. Note that, by the property of the $3j$ symbol in~\eqref{eq:I3andI4}, the tree-level OPE coefficient is non-zero only when $n_1$, $n_2$, and $n_3$ satisfy the triangle inequality and their sum is an even number.

\section{Nonlinearly Realized Conformal Symmetry}\label{sec:Nonlin}

In flat space, the EST is strongly constrained by nonlinearly realized Poincaré symmetry.  
For example, the soft behavior of higher-point S-matrix elements is related to lower-point elements via soft theorems~\cite{Cooper:2014noa,Dubovsky:2015zey}.  
It is natural to expect that similar constraints also arise for an effective string in AdS.  
In this setting, the analog of a soft on-shell Goldstone boson is a displacement operator integrated over the Wilson line, with a kernel that implements a broken conformal transformation.  
Such constraints were derived and studied recently in~\cite{Gabai:2025zcs,NadavUnpublished}.  
One of the key results of these works is that, in certain cases, these constraints relate contributions to correlation functions that appear at different orders in perturbation theory. 

In this section, we apply the same technique to our case: $SU(N)$ Yang--Mills theory in AdS$_3$ with an infinite fundamental Wilson loop.  
We recall that we work in the large-$N$ limit with small dimensionless 't~Hooft coupling $\lambda = g_\text{YM}^2 N R $.\footnote{The constraints also hold at large~$\lambda$, but without mixing between different perturbative orders. They were tested for the expression~\eqref{eq:01thStrong} in~\cite{Gabai:2025zcs}.} 

As discussed in Section~\ref{subsec:WilsonLoopDefect}, the Wilson line on the boundary defines a conformal line defect containing the displacement operator~$\mathbb{D}$, whose scaling dimension is protected, $\Delta_{\mathbb{D}} = 2$. Our discussion below assumes this property holds for all values of the coupling~$\lambda$.  
Under this assumption, the constraints from nonlinearly realized conformal symmetry will allow us to determine not only an infinite set of tree-level OPE coefficients---a particular subset of those given in~\eqref{eq:pertC}---but also the leading-order anomalous dimensions of all the single-letter operators introduced in section~\ref{subsection:TreeSingleLetterOp}.   
In principle, conformal data involving multi-letter operators can be constrained in a similar manner, and determined up to a small set of constants; we leave this analysis to future work.



Recall that for AdS$_3$, a conformal defect line living on the boundary admits only a single displacement operator because there is only one transverse direction.  
In this case, the constraints from nonlinearly realized conformal symmetry take a particularly simple form.  
We begin by setting up some conventions.  

Let $\mathcal{O}$ be any CFT$_1$ primary operator. We define
\begin{equation}\label{def:FODDO}
	\mathcal{F}^{\mathcal{O} \mathbb{D} \mathbb{D} \mathcal{O}}(t) \equiv 
	\frac{\langle\mathcal{O}(\tau_1)\,\mathbb{D}(\tau_2)\,\mathbb{D}(\tau_3)\,\mathcal{O}(\tau_4) \rangle_W}
	{\langle \mathbb{D}(\tau_2) \mathbb{D}(\tau_3)\rangle_W \, \langle\mathcal{O}(\tau_1)\mathcal{O}(\tau_4)\rangle_W}
	- [{\rm GFF}]\,,
\end{equation}
and
\begin{equation}\label{def:FDODO}
	\mathcal{F}^{\mathbb{D} \mathcal{O} \mathbb{D} \mathcal{O}}(t) \equiv 
	\frac{1}{(1+t)^4}
	\frac{\langle \mathbb{D}(\tau_1)\,\mathcal{O}(\tau_2)\,\mathbb{D}(\tau_3)\,\mathcal{O}(\tau_4) \rangle_W}
	{\langle \mathbb{D}(\tau_1) \mathbb{D}(\tau_3)\rangle_W \, \langle\mathcal{O}(\tau_2)\mathcal{O}(\tau_4)\rangle_W}
	- [{\rm GFF}]\,.
\end{equation}
Here $t$ is the conformal cross ratio
\begin{equation}
	t \equiv \frac{\tau_{21}\,\tau_{43}}{\tau_{41}\,\tau_{32}}\,,
\end{equation}
with $\tau_{ij} \equiv \tau_i - \tau_j$.  
The symbol $[{\rm GFF}]$ denotes the sum of pairwise products of non-vanishing two-point functions, i.e.\ the generalized free field (disconnected) contribution. When $\mathcal{O}$ and $\mathbb{D}$ are distinct primary operators, the $[{\rm GFF}]$ terms in \eqref{def:FODDO} and \eqref{def:FDODO} are
\begin{equation}\label{eq:GFFex1}
	\mathcal{F}_{\rm GFF}^{\mathcal{O}\mathbb{D}\mathbb{D}\mathcal{O}}(t) = 1\,, 
	\qquad 
	\mathcal{F}_{\rm GFF}^{\mathbb{D}\mathcal{O}\mathbb{D}\mathcal{O}}(t) = \frac{1}{(1+t)^4}\,.
\end{equation}
When $\mathcal{O} = \mathbb{D}$, they coincide and are given by
\begin{equation}\label{eq:GFFex2}
	\mathcal{F}_{\rm GFF}^{\mathbb{D}\mathbb{D}\mathbb{D}\mathbb{D}}(t) 
	= 1 + t^{-4} + \left(\frac{1}{1+t}\right)^4\,.
\end{equation}

In these conventions, the crossing relations take the form
\begin{equation}
	\mathcal{F}^{\mathbb{D}\mathbb{D}\mathcal{O}\mathcal{O}}(t) 
	= t^{-4}\,\mathcal{F}^{\mathcal{O}\mathbb{D}\mathbb{D}\mathcal{O}}\!\left(\frac{1}{t}\right), 
	\qquad
	\mathcal{F}^{\mathbb{D}\mathcal{O}\mathbb{D}\mathcal{O}}(t)  
	= t^{-4}\,\mathcal{F}^{\mathbb{D}\mathcal{O}\mathbb{D}\mathcal{O}}\!\left(\frac{1}{t}\right).
\end{equation}

Now we are ready to present the integrated constraints \cite{Gabai:2025zcs,NadavUnpublished} (see a simplified derivation in appendix \ref{app:nonlin}). 
There are two homogeneous integral constraints:
\begin{equation}
	\label{homo1}
	\int_0^\infty \! dt \,\big[\,t^{2} \,\mathcal{F}^{\mathbb{D}\mathcal{O}\mathbb{D}\mathcal{O}}(t) 
	+ \big(1+2t+2t^{2}\big)\mathcal{F}^{\mathcal{O}\mathbb{D}\mathbb{D}\mathcal{O}}(t)\big] = 0\,,
\end{equation}
\begin{equation}
	\label{homo2}
	\int_0^\infty \! dt \,\big[ -t \,\mathcal{F}^{\mathbb{D}\mathcal{O}\mathbb{D}\mathcal{O}}(t) 
	+ 2(t+t^{2})\mathcal{F}^{\mathcal{O}\mathbb{D}\mathbb{D}\mathcal{O}}(t) \big] = 0\,,
\end{equation}
and one inhomogeneous constraint:
\begin{equation}
	\label{inhomo}
	\begin{split}
		\int_0^\infty \! dt \,\big[ (t^{2}-1)\log t \,\mathcal{F}^{\mathbb{D}\mathcal{O}\mathbb{D}\mathcal{O}}(t) 
		&- 2(1+2t)\log\!\left(\frac{t}{1+t}\right)\mathcal{F}^{\mathcal{O}\mathbb{D}\mathbb{D}\mathcal{O}}(t) \big] 
		= \frac{2\Delta_{\mathcal{O}}}{\Lambda}\,.
	\end{split}
\end{equation}
Here, $\Lambda$ is the normalization of the displacement two-point function:
\begin{equation}
	\langle {\mathbb{D}}(x) {\mathbb{D}}(0)\rangle_W = \frac{\Lambda}{x^{4}}\,,
\end{equation}
which depends non-trivially on the coupling. 
To leading order, the displacement operator is given by $i g_{\mathrm{YM}} F_{\tau 1}$ (see \eqref{eq:dispWeak}), 
and its two-point function, computed in \eqref{2pfDisp}, is
\begin{equation}
	\Lambda = \frac{2\lambda}{\pi} + O(\lambda^{3/2})\,.
\end{equation}

Before proceeding to use these constraints, we must address a subtlety: 
when an operator of sufficiently low dimension is exchanged in the four-point function, 
one or more of the integrals in \eqref{homo1}–\eqref{inhomo} diverge. 
In such cases, the integrals should be regularized by analytically continuing the exchanged operator dimension from a sufficiently large value. 
The condition \eqref{eq:genericDefect} ensures that this analytic continuation yields a finite result.

\subsection{Weak coupling expansion: enhanced contributions}
\label{sec:enhancements}

As we have assumed, the boundary Wilson line satisfies the constraints above for any finite value of the coupling $\lambda$. This allows us to do perturbation theory at small coupling and impose the constraints order by order. Notably, the integrals appearing in \eqref{homo1}–\eqref{inhomo} generally do not commute with the perturbative expansion of the four-point correlators \cite{Gabai:2025zcs,Cavaglia:2022yvv}. As a result, perturbative data from higher orders can mix into the lower-order expansions of the constraints. Whenever a contribution is promoted to a lower perturbative order in this way, we refer to it as an \emph{enhanced contribution}. In this subsection we work through one example of such an enhancement in detail.

Consider the four-point function of displacement operators, which are proportional to the first KK mode in the language of section \ref{sec:dimreduction}. The first constraint \eqref{homo1} reads in this case
\begin{equation}\label{Homo1:DDDD}
	\int_0^\infty \! dt \,\big( 1 + 2t + 3t^2 \big)\, \mathcal{F}^{\mathbb{D}\mathbb{D}\mathbb{D}\mathbb{D}}(t) = 0\,.
\end{equation}
The displacement–displacement OPE contains the second KK mode, whose scaling dimension is
\begin{equation}
	\Delta_{2} = 3 + \gamma_{2}\,,
\end{equation}
where we introduce the notation $\gamma_n$ for the anomalous dimension of the operator $a_{\tau}^{(n)}$:
\begin{equation}
	\Delta_n = n + 1 + \gamma_n\,.
\end{equation}

Since \eqref{Homo1:DDDD} is homogeneous in $\mathbb{D}$, we can rescale $\mathbb{D}$ and replace them with the first KK mode $a_{\tau}^{(1)}$. We now zoom in on the contribution from the exchange of $a_{\tau}^{(2)}$ near $t = 0$:
\begin{equation}
	C_{1,1,2}^2 \!\int_0^\delta \! dt \,\big( 1 + 2t + 3t^2 \big) \big( t^{\Delta_2 - 4} + \dots \big)
	= C_{1,1,2}^2 \left[ \frac{\delta^{\Delta_2 - 3}}{\Delta_2 - 3} + \frac{2\,\delta^{\Delta_2 - 2}}{\Delta_2 - 2} + \frac{3\,\delta^{\Delta_2 - 1}}{\Delta_2 - 1} \right] + \dots
\end{equation}
Here, $C_{1,1,2}$ is the OPE coefficient for three KK modes as defined in \eqref{eq:OPECaaa}. Expanding the right-hand side in perturbation theory gives
\begin{equation}\label{DDDD:enhance1}
	C_{1,1,2}^2 \left[ \frac{\delta^{\Delta_2 - 3}}{\Delta_2 - 3} + \frac{2\,\delta^{\Delta_2 - 2}}{\Delta_2 - 2} + \frac{3\,\delta^{\Delta_2 - 1}}{\Delta_2 - 1} \right] 
	= \frac{C_{1,1,2}^2}{\gamma_2} + O(\lambda)\,.
\end{equation}
Since $C_{1,1,2} = O(\lambda^{1/2})$ (as shown in \eqref{eq:OPECaaa}) and $\gamma_2 = O(\lambda)$ (as will be shown later), their ratio scales as $O(1)$. Thus, this exchange contributes at order $O(1)$ to \eqref{homo1} in the $\lambda \to 0$ limit.

A similar analysis for the region $t \to \infty$ yields another $O(1)$ contribution:
\begin{equation}\label{DDDD:enhance2}
	\begin{split}
		C_{1,1,2}^2 \!\int_\delta^\infty \! dt \,\big( 1 + 2t + 3t^2 \big) \big( t^{-\Delta_2} + \dots \big) 
		&= 3\,C_{1,1,2}^2 \left[ \frac{\delta^{\,3 - \Delta_2}}{\Delta_2 - 3} + \dots \right] + \dots \\
		&= \frac{3\,C_{1,1,2}^2}{\gamma_2} + O(\lambda)\,,
	\end{split}
\end{equation}
where the factor of $3$ originates from the $3 t^2$ term in the prefactor.

We therefore see that even though $a_\tau^{(2)} \sim \partial_1 F_{\tau 1}$ is exchanged at order $O(\lambda)$ in the displacement four-point function, it contributes to \eqref{homo1} at order $O(1)$. For the homogeneous constraints, any enhancement is determined by the inverse scaling of the leading nontrivial anomalous dimension of either the external or exchanged operator. For the inhomogeneous constraint, the enhancement is doubled due to the logarithm. For instance,
\begin{equation}
	C_{1,2,1}^2 \!\int_0^\delta \! dt \,(t^2 - 1)\,\log t \, t^{\Delta_2} 
	= C_{1,2,1}^2 \left[ \frac{1}{(\Delta_2 - 3)^2} + O\big( (\Delta_2 - 3)^0 \big) \right]\,.
\end{equation}

This enhancement mechanism is quite general, and we will exploit it below to determine an infinite set of conformal data to leading nontrivial order.

\subsection{Solving the leading order constraints: first KK mode}

We begin by determining the couplings of the displacement operator (the first KK mode) to the second KK mode, $a^{(2)}_\tau \sim \partial_1 F_{\tau 1}$. 
At zeroth order, the full four-displacement correlator is
\begin{equation}
	1 + t^{-4}\,.
\end{equation}
Subtracting the GFF contribution gives the connected piece
\begin{equation}\label{Homo1Integrand:DDDD}
	\mathcal{F}^{\mathbb{D}\mathbb{D}\mathbb{D}\mathbb{D}}(t) 
	= -\left( \frac{1}{1+t} \right)^{4} + O(\lambda)\,.
\end{equation}
When $\mathcal{O} = \mathbb{D}$, the two homogeneous constraints are in fact equivalent.  
Multiplying \eqref{Homo1Integrand:DDDD} by the homogeneous kernel \eqref{homo1}, the first term integrates to  
\begin{equation}
	C^{\mathrm{finite}}_{\mathrm{Homo}} := 
	-\int_0^\infty \! dt\, \big( 1 + 2t + 3t^2 \big) \left( \frac{1}{1+t} \right)^{4} 
	= -\frac{5}{3}\,.
\end{equation}
If there were no mixing between perturbative orders, this result would contradict \eqref{homo1}.  
However, as explained in section \ref{sec:enhancements}, part of the $O(\lambda)$ piece in \eqref{Homo1Integrand:DDDD} is \emph{enhanced} to order $O(1)$ upon integration.  
This enhancement leads to an additional contribution,
\begin{equation}
	C^{\mathrm{enhanced}}_{\mathrm{Homo}} :=
	\lim_{\lambda \to 0} \int_0^\infty \! dt\, \big( 1 + 2t + 3t^2 \big)
	\left[ \mathcal{F}^{\mathbb{D}\mathbb{D}\mathbb{D}\mathbb{D}}(t) 
	+ \left( \frac{1}{1+t} \right)^{4} \right].
\end{equation}
To satisfy \eqref{homo1}, these two pieces must cancel:
\begin{equation}
	C^{\mathrm{finite}}_{\mathrm{Homo}} + C^{\mathrm{enhanced}}_{\mathrm{Homo}} = 0\,.
\end{equation}

We now compute $C^{\mathrm{enhanced}}_{\mathrm{Homo}}$.  
By carefully analyzing the integral \eqref{Homo1:DDDD} near $t = 0$ and $t = \infty$, one finds that only operators with dimensions close to $\Delta = 1,2,3$ can contribute.  
In our case, at $\lambda = 0$ we have $\Delta_1 = 2$ (the displacement operator) and $\Delta_2 = 3$ (the second KK mode).  
The reflection symmetry forces $C_{1,1,1} \equiv 0$ for all couplings, so the only possible contribution comes from the second KK mode, whose enhanced effect was already computed in \eqref{DDDD:enhance1} and \eqref{DDDD:enhance2}.  
The total is
\begin{equation}\label{DDDDresult:homo}
	C^{\mathrm{enhanced}}_{\mathrm{Homo}}=\lim_{\lambda \to 0} \frac{4\, C_{1,1,2}^2}{\gamma_{2}} .
\end{equation}
The cancellation condition therefore implies
\begin{equation}
	\lim_{\lambda \to 0} \frac{4\, C_{1,1,2}^2}{\gamma_{2}} = \frac{5}{3}\,.
\end{equation}

In fact, using \eqref{eq:OPECaaa} we can determine $\gamma_{2}$ at order $O(\lambda)$ directly.  
However, to illustrate the power of the constraints, we will \emph{pretend} that $C_{1,1,2}$ is unknown and re-derive it from the constraints.

Now we consider the inhomogeneous constraint \eqref{inhomo} for the displacement four-point function:
\begin{equation}\label{inhomo:DDDD}
	\begin{split}
		\int_0^\infty \! dt \,
		\left[(t^{2}-1)\log t \, - 2(1+2t)\log\!\left(\frac{t}{1+t}\right) \right]
		\mathcal{F}^{\mathbb{D}\mathbb{D}\mathbb{D}\mathbb{D}}(t)
		= \frac{4}{\Lambda}\,.
	\end{split}
\end{equation}
The $O(\lambda^0)$ part of $\mathcal{F}^{\mathbb{D}\mathbb{D}\mathbb{D}\mathbb{D}}$ yields
\begin{equation}
	C^{\mathrm{finite}}_{\mathrm{InHomo}} := 
	-\int_0^\infty \! dt\, 
	\left[(t^{2}-1)\log t \, - 2(1+2t)\log\!\left(\frac{t}{1+t}\right) \right] 
	\left( \frac{1}{1+t} \right)^{4}
	= -\frac{25}{9}\,.
\end{equation}
However, $\Lambda = O(\lambda)$ at small coupling, so the right-hand side of \eqref{inhomo:DDDD} diverges as $\lambda^{-1}$ when $\lambda \to 0$.  
Therefore, an enhanced contribution must cancel this divergence.  
By an analysis parallel to that of the homogeneous constraint, the only enhancement arises from the second KK mode, giving
\begin{equation}
	C^{\mathrm{enhanced}}_{\mathrm{InHomo}} 
	= \frac{4\, C_{1,1,2}^2}{\gamma_{2}^2} + O(\lambda^0)\,.
\end{equation}
Here, the factor of $4$ comes from the sum of the contributions near $t = 0$ (which gives $3$) and $t \to \infty$ (which gives $1$).  
The cancellation of divergences leads to
\begin{equation}\label{DDDDresult:inhomo}
	\frac{4\, C_{1,1,2}^2}{\gamma_{2}^2}
	= \frac{4}{\Lambda} + O(\lambda^{0})\,.
\end{equation}

Recalling that $\Lambda = \tfrac{2\lambda}{\pi} + O(\lambda^{2})$, the combined system \eqref{DDDDresult:homo} and \eqref{DDDDresult:inhomo} is sufficient to determine
\begin{equation}
	\label{eq:initCondNonlin}
	C_{1,1,2}^2 = \frac{25}{72\pi}\,\lambda + O(\lambda^{2})\,, 
	\qquad 
	\gamma_{2} = \frac{5}{6\pi}\,\lambda + O(\lambda^{2})\,.
\end{equation}
In particular, the OPE coefficient agrees with \eqref{eq:OPECaaa}.

\subsection{Solving the leading order constraints: higher KK modes} 
Consider the finite coupling version of $a_\tau^{(n)} \sim \partial_1^{(n-1)} F_{\tau 1}$, rescaled to be canonically normalized. This is (classically) a dimension $1+n$ operator. In the 't Hooft limit, the four-point functions in \eqref{def:FODDO} and \eqref{def:FDODO} are given by
\begin{equation}
	\mathcal{F}^{\mathbb{D}\mathcal{O}\mathbb{D}\mathcal{O}}(t) = -\frac{1}{(1+t)^4} +O(\lambda) \quad,\quad \mathcal{F}^{\mathcal{O}\mathbb{D}\mathbb{D}\mathcal{O}}(t) = O(\lambda)\ .
\end{equation}
Here we have chosen $\mathcal{O}=a_\tau^{(n)}$ with $n\geqslant2$.

Starting from the first homogeneous constraint \eqref{homo1}, we have the finite piece,
\begin{equation}
	C^{\rm finite}_{\rm Homo} = -1/3\ .
\end{equation}
In principle, the enhancement can now come with any operator of dimension lower or equal to $2+n$. However, we know from the comment below \eqref{eq:finalC} that only operators with $m=n \pm 1$ will couple to this four point function at the next order. Hence, we only consider enhancements from such operators. We get,
\begin{equation}
	C^{\rm enhanced}_{\rm Homo} = \frac{2(1+n+2n^2) C_{1,n,n-1}^2}{n (1+2n) (\gamma_{n-1} - \gamma_{n})} + \frac{2 C_{1,n,n+1}^2}{\gamma_{n+1} - \gamma_{n}} + \frac{2C_{1,1,2}C_{n,n,2}}{\gamma_2}
\end{equation}
giving,
\begin{equation}
\label{eq:1stconstWeak}
	\frac{2(1+n+2n^2) C_{1,n,n-1}^2}{n (1+2n) (\gamma_{n-1} - \gamma_{n})} + \frac{2 C_{1,n,n+1}^2}{\gamma_{n+1} - \gamma_{n}} + \frac{2C_{1,1,2}C_{n,n,2}}{\gamma_2} = 1/3\ .
\end{equation}

Similarly, the second homogeneous constraint gives,
\begin{equation}
\label{eq:2ndconstWeak}
	\frac{2 C_{1,n,n-1}^2}{n (\gamma_{n-1} - \gamma_{n})} + \frac{2C_{1,1,2}C_{n,n,2}}{\gamma_2} = -1/6\ .
\end{equation}

Finally, from the inhomogeneous constraint we get, 
\begin{equation}
\label{eq:3rdconstWeak}
	\frac{4 (1+n)(1-2n) C_{1,n,n-1}^2}{n(1+2n) (\gamma_{n-1} - \gamma_{n})^2} + \frac{4 C_{1,n,n+1}^2}{(\gamma_{n+1} - \gamma_{n})^2} = \frac{2(n+1)}{\Lambda}\ .
\end{equation}

\paragraph{Solution}
By plugging \eqref{eq:2ndconstWeak} into \eqref{eq:1stconstWeak}, we get,
\begin{equation}
	C_{1,1,2}C_{n+1,n+1,2} = \frac{5(1+n)\Lambda + 72n (2n-1)C_{1,1,2}C_{n,n,2} }{72(n+1)(2n+1)} + O(\lambda^2)\ .
\end{equation}
Combined with the initial condition \eqref{eq:initCondNonlin}, the solution to this recursion is,
\begin{equation}
	C_{1,1,2}C_{n,n,2} = \frac{5 \Lambda \left(n^2+n+3\right)}{144 n (2 n-1)}+ O(\lambda^2)\ .
\end{equation}
Plugging this back into \eqref{eq:2ndconstWeak}, we find,
\begin{equation}
	\frac{C^2_{1,n-1,n}}{\gamma_n-\gamma_{n-1}} = \frac{n^2+1}{8 n-4}+ O(\lambda)\ .
\end{equation}
Finally, from \eqref{eq:3rdconstWeak} we get,
\begin{equation}
	\frac{C^2_{1,n-1,n}}{(\gamma_n-\gamma_{n-1})^2} = \frac{(1-n^2)n}{(2 -4   n)\Lambda}+ O(\lambda^0)\ .
\end{equation}
Plugging in the explicit value \eqref{2pfDisp}, we conclude,
\begin{eqnarray}
	C_{1,1,2}C_{n,n,2} &=& \frac{5  \left(n^2+n+3\right)}{72 n (2 n-1)\pi}\lambda + O(\lambda^2)\ ,\\
	C^2_{1,n-1,n} &=& \frac{  \left(n^2+1\right)^2}{4 n (2 n-1) \left(n^2-1\right)\pi}\lambda+ O(\lambda)\\
	\gamma_n &=& \frac{1}{2\pi} \left(\frac{2}{n+1}+2 (\psi(n)+\gamma )-1\right)\lambda+ O(\lambda^2)\ .
\end{eqnarray}
Here, $\psi$ is the digamma function. Note that the top two rows are in agreement with the perturbative computation \eqref{eq:pertC}. However, the direct computation of anomalous dimensions seems more complicated since it involves loop diagrams. All anomalous dimensions are positive -- consistent with a monotonic growth towards strong coupling (see figure \ref{fig:Inter}). It is also interesting to note that they behave as $\log(n)$ at large $n$.

As mentioned above, it is also possible to determine all multi-letter leading anomalous dimensions up to a small set of constants. We leave this to future work.

\section{Conclusions}


In this paper, we have carried out an analysis of a two-dimensional theory that describes a long confining string in AdS space. We focused on the large-$N$ pure Yang–Mills theory in three dimensions; however, we expect the qualitative features of our analysis to remain valid for a broad class of confining gauge theories. The string preserves an $SO(1,2)$ subgroup of the original AdS isometries, and can thus be regarded as a conformal defect on the boundary. This point of view allows us to identify a set of observables, namely correlation functions of operators living on this defect, which are unambiguously defined at any AdS radius. In the bulk, the theory admits two perturbative descriptions, depending on whether the AdS scale is longer or shorter than the confinement scale. In the former case, the appropriate description is in terms of effective string theory (EST), while in the latter it is in terms of weakly coupled Yang–Mills theory. In either case, we can identify a special operator of protected dimension $\Delta = 2$, known as the displacement operator. In the EST, it is identified with the Goldstone mode (branon), whereas in the Yang–Mills description it corresponds to a specific component of the boundary field strength, or equivalently to the first KK mode of the bulk gauge field. In either of the weakly coupled regimes, all other operators in the theory can be built from various combinations of the displacement and its derivatives, making it the primary building block of the theory. At the moment the absence of other particles on the EST side can be inferred from the lattice results, but eventually we hope to derive it from matching to the gauge theory description. On top of this, correlation functions involving displacement satisfy certain constraints that follow from nonlinearly realized AdS isometries. These constraints are true non-perturbatively and also serve as a powerful tool in perturbative calculations.

From the gauge theory point of view, the string can be thought of as a gigantic meson. When the AdS radius is small, all partons inside this meson have energies of at least order $1/R_\text{AdS}\gg \Lambda_\text{QCD}$, and are thus well-described by weakly coupled gluons. At large radius it is natural to assume that the partons become the Goldstones. Essentially, this is the manifestation of confinement in this setup -- the gluon field got replaced by string-like excitations. A stronger hypothesis that we are proposing is that the transition between the two descriptions happens continuously, and moreover, can be observed perturbatively in either description. This is, at first sight, in contrast with the idea that confinement is a fully non-perturbative effect. Our main goal is to test this hypothesis. If it turns out to be correct, it is likely to also hold in other regimes of YM and QCD, for example in the study of flat space flux tubes reviewed in section \ref{sec:flat}. A similar relation should also hold in high-energy scattering in QCD, once confinement effects become important, at least for sufficiently large $N$.

Our current results are not precise enough to decisively prove or disprove the hypothesis, but they are certainly consistent with it, and also reveal some subtleties. The most trivial way in which the hypothesis could have worked would be if the spectra of operators on the two sides was very close, as occurs for weakly coupled RG flows between two CFTs. We know this cannot be literally true, since the number of primary operators with the same dimension is significantly larger on the small radius side. This difference becomes increasingly significant as we go to higher dimensions. In fact, we expect operators of dimension $\Delta$ at small radius to acquire an anomalous dimension of order $\Delta^2$ once we interpolate to large radius, in order to be consistent with \eqref{sqrtformula}, which appears to be very non-perturbative. We expect, however, that this effect may be accounted for non-perturbatively in the spirit of equation \eqref{TTbarAdS}, at least for a large number of operators. That is, the large change in dimensions corresponds to the semiclassical effect of string tension, which is, in some sense, integrable. It should be possible to understand this phenomenon better by using the conformal formulation of the worldsheet theory \eqref{eq:PS} adjusted to AdS. In this formulation, the square-root relation is expected to arise from solving the leading-order constraints, and can thus be separated from the rest of the worldsheet dynamics. A similar effect was observed in the relation between different worldsheet gauges in the context of integrable critical strings in AdS \cite{Baggio:2018gct,Frolov:2019nrr}\footnote{We thank Fiona Seibold for discussions on this point.}. On the small radius side, the same effect should be captured by the linear potential created by the zero KK mode of the gauge field. Again, this effect may be described semiclassically because in 2D YM theory the linear potential arises already at tree level.

Another important difference between the gauge theory and the worldsheet description of the defect is the existence of infinitely many single-letter operators $\d_1^n F_{\tau 1} $. One possibility is that these operators interpolate to the operators of the form $X^{n+1}$. We expect to be able to check this explicitly in perturbation theory, at least for the low-lying operators shown in figure \ref{fig:Inter}.

A more refined version of the hypothesis is that, once the large but ``simple'' effect of the string tension is accounted for non-perturbatively, the remaining change in dimensions is more similar to that of a perturbative RG flow. The next step in our program is to test this hypothesis by performing higher-order calculations on both sides in 3D YM theory \cite{paper2}. Needless to say, everything we discussed so far straightforwardly generalizes to 4D YM theory. An important new ingredient is the worldsheet axion present in the EST description. It naturally maps to the operator $F_{ij}$ on the gauge theory side, and we again expect to be able to compute some of the axion’s properties starting from the gauge theory side. It is particularly interesting to see whether the value of the leading axion coupling can be related to the one-loop beta function of YM at weak coupling, as conjectured in \cite{Dubovsky:2018vde}. Related to this, we expect additional features to appear in our construction in 4D, due to the fact that YM theory is conformal at tree level. 

Even though four-dimensional confining theories are definitely our ultimate goal, many of the interesting phenomena we discussed can also be studied in lower dimensions, in particular in 2D YM with adjoint matter. These theories can also be placed in AdS$_2$ and become weakly coupled at small radius. There has been a lot of progress in understanding these theories in flat space \cite{Dubovsky:2018dlk,Donahue:2019adv,Cherman:2019hbq, Komargodski:2020mxz,Dempsey:2021xpf,Dempsey:2022uie,Popov:2022vud,Dempsey:2023fvm}, including a small circle expansion with periodic boundary conditions \cite{Dempsey:2024ofo}, which bears a similarity to the small AdS radius expansion. Of particular interest is the supersymmetric version of the theory \cite{Kutasov:1993gq}. Its recent covariant formulation \cite{Klebanov:2025mbu} suggests that it may be possible to preserve supersymmetry in AdS as well, in which case a confining string filling AdS$_2$ would host a Goldstino. This scenario would then also lead to a weakly coupled description at large radius, this time in terms of a fermion with protected dimension, and would again allow us to test how the two descriptions interpolate as the radius is varied. This simpler model also exhibits the square-root relation between dimensions \eqref{sqrtformula} and is likely the simplest setup in which to understand it. It is likely that, in the worldsheet description, the change in the density of operators is related to the formation of zig-zags, which effectively alter the statistics of the worldsheet particles \cite{Dubovsky:2018dlk,Donahue:2019adv}.

In addition to perturbative calculations, one can apply the numerical 1D conformal bootstrap \cite{Paulos:2016fap,Paulos:2019fkw} to our model. We expect this approach to be especially productive if the dimensions of low-lying operators can be constrained with sufficient precision using our perturbative results.\footnote{We thank Miguel Paulos and Philine van Vliet for discussions on this point.} An example of an extremely successful application of this procedure is the theory of a 1/2-BPS Wilson line in $\mathcal{N}=4$ SYM \cite{Rey:1998ik,Maldacena:1998im}. In fact, we would like to argue that this system and a confining string in AdS have many more similarities than may appear at first sight. Let us briefly summarize this perspective. Consider a straight Wilson line in  $\mathcal{N}=4$ theory at large-$N$ and with the 't Hooft coupling $\lambda_{\mathcal{N}=4}$. It corresponds to a conformal defect, dual to a fundamental string in AdS$_5\times$S$^5$ that stretches radially and spans an AdS$_2$ slice inside it \cite{Giombi:2017cqn}. At small $\lambda_{\mathcal{N}=4}$ this defect is very similar to the Wilson line in pure YM, the only difference being additional adjoint fields that can be inserted on the $\mathcal{N}=4$ Wilson line. At strong coupling the theory is also qualitatively similar to the flux tube in AdS. In fact, the leading-order action for the Goldstones is the same as \eqref{NGleading}, with the string length related to the AdS radius by $\ell_s=R_{\text{AdS}_5} (\lambda_{\mathcal{N}=4})^{-\frac{1}{4}}$, a relation qualitatively analogous to \eqref{def:lambda}. At this order, the only important difference is the presence of an additional displacement operator, which allows for the nonlinear realization of the full four-dimensional conformal group. In this regard, holography does to the  $\mathcal{N}=4$ Wilson line the same thing that confinement does to a QCD or YM Wilson line on the boundary of AdS -- at strong coupling, it allows the gauge theory degrees of freedom to be replaced by the worldsheet Goldstone fields. Of course, the holographic correspondence is a much deeper statement which includes the emergence of gravity in the bulk; however, at large $N$, gravity does not appear to play an important role as far as the Wilson line is concerned.

A crucial distinction of the $\mathcal{N}=4$ Wilson line defect is that, unlike the confining string, it is integrable \cite{Drukker:2012de,Correa:2012hh}. Presumably due to its unusual UV-completion, the worldsheet theory provides a counter-example to the no-go theorems for integrability in AdS$_2$ recently derived in \cite{Antunes:2025iaw}. Integrability allows us to compute the spectrum of defect operators with incredible precision \cite{Gromov:2015dfa,Grabner:2020nis}, and, combined with the numerical bootstrap, has also led to a determination of the OPE coefficients  \cite{Cavaglia:2021bnz,Cavaglia:2022qpg,Cavaglia:2024dkk}. For a large number of low-lying operators, one can observe how operators interpolate between the weak and strong coupling regimes. Note that the asymptotic operator counting remains the same in the $\mathcal{N}=4$ theory, and consequently, most dimensions should satisfy the ``square-root'' relation \eqref{sqrtformula}. Another effect is the disappearance of an infinite number of single-letter operators like $\partial_i^n F_{\tau 1}$, where $i$ denotes a generic transverse index. They likely transition to operators of the form $(X^i)^{n+1}$ at strong coupling. We expect these non-perturbative phenomena to be common to both confining and holographic theories, and thus having an exactly solvable example of these phenomena is extremely valuable. Due to integrability, one observes level crossings between dimensions of operators from the same symmetry sectors. We do not expect these level crossings to occur for confining strings, as particle production on the worldsheet will lead to mixing between operators corresponding to different numbers of worldsheet excitations. An interesting question is how large these particle production effects are. 

It would also be interesting to study supersymmetric confining theories in AdS. Even though we do not expect corresponding flux tubes to be integrable, supersymmetry certainly provides additional constraints. The 4D $\mathcal{N}=2$ YM theories on AdS were very recently studied in \cite{Bason:2025zpy}.

In this paper we focused on long static strings in large-$N$ pure Yang-Mills theory, since in our opinion they are the simplest objects characteristic of confining theories that are amenable to study with current techniques. 
It would be natural to generalize our AdS approach to other observables and theories. In particular, we can consider rotating closed strings, or open strings with fundamental quarks at the end points. The leading order description of such objects was already discussed in \cite{Alday:2007mf}. As mentioned in the introduction, pions in a theory with massless or light quarks also correspond to (approximately) protected operators, and hence the AdS approach can be readily generalized to them. More generally, it could be that the small-radius expansion converges well for certain QCD observables that usually do not admit any perturbative description. In fact one can even attempt a calculation of the real world proton mass, although we do not expect it to be the simplest calculation. The question of the convergence properties of the perturbative expansion in AdS is likely related to non-perturbative objects such as instantons, center vortices, and renormalons. In fact, the original motivation of \cite{Callan:1989em} was precisely this question. Since at small radius the theory is fully perturbative, it should be possible to systematically include these effects; however, to the best of our knowledge, it has not been done yet.


\section*{Acknowledgements}
We thank Ofer Aharony, António Antunes, Lorenzo Di Pietro, Sergei Dubovsky, Kolya Gromov, Julius Julius, Shota Komatsu, Petr Kravchuk, Mehrdad Mirbabayi, Bendeguz Offertaler, Miguel Paulos, Joao Penedones, Orr Sela, Amit Sever,  Steve Shenker, Philine van Vliet for useful discussions. 
The work of VG and JQ is supported by Simons Foundation grant 994310 (Simons Collaboration on Confinement and QCD Strings).
\appendix

\section{State Counting in the Weak Coupling Limit}
\label{app:WeakCounting}
This appendix is aimed at counting the total number of $SO(1,2)$ primary states at dimension $\Delta$ in the weak coupling limit. Recall that in unitary 1D CFTs, the irreducible representations of the conformal group $SO(1,2)$ are rather simple: one state per level,
$$\ket{\Delta}\ \text{ (primary)},\ P\ket{\Delta}\ \text{(level-1 descendant)},\ P^2\ket{\Delta}\ \text{(level-2 descendant)},\ldots,$$
except for the trivial representation which does not contain any descendants. Our basic strategy is to first count
$$\mathcal{N}_\Delta:=\text{number of states at dimension }\Delta.$$
Here $\mathcal{N}_\Delta$ includes both primaries and descendants. Then the total number of primaries, denoted as $\mathcal{N}^\text{p}_\Delta$, is given by
\begin{equation}\label{counting:StatesToPrimaries}
    \mathcal{N}^\text{p}_\Delta=\begin{cases}
        \mathcal{N}_\Delta-\mathcal{N}_{\Delta-1}, &\Delta>1, \\
        \mathcal{N}_\Delta, & \Delta\leqslant 1. \\
    \end{cases}
\end{equation}
Here we used the unitarity of 1D CFT, which implies that there are no states with negative dimension and the dimension-0 operators have no descendants.

In principle, we can use whatever way to count the states. For pure gauge theory in AdS$_3$, we find it clean to count states in terms of KK modes in AdS$_2$ (introduced in section \ref{sec:KKmodes}). This is because in AdS$_2$, each field will correspond to only one irreducible representation of $SO(1,2)$.

For gauge theory in AdS$_3$ with Neumann boundary condition, the gauge field is decomposed into the following field content.
\begin{itemize}
    \item(0-th KK mode) $A^{(0)}$, gauge field in AdS$_2$ with Neumann boundary condition. This mode does not have propagating degrees of freedom in the bulk, so it does not correspond to any field at the boundary.
    \item (Higher KK modes) $A^{(n)}$ ($n=1,2,\ldots$), Proca fields with mass $m_n=\sqrt{n(n+1)}$ and Dirichlet boundary condition. Each of them corresponds to a boundary primary field $a_\tau^{(n)}$ with dimension $\Delta=n+1$.
\end{itemize}
We emphasize that all these vector fields are matrix-valued. Each boundary state corresponds to an operator of the form
\begin{equation}\label{counting:defO}
    \mathcal{O}_{k_1,n_1;k_2,n_2;\ldots}=\left(\partial_\tau^{k_1}a_\tau^{(n_1)}\right)\left(\partial_\tau^{k_2}a_\tau^{(n_2)}\right)\left(\partial_\tau^{k_3}a_\tau^{(n_3)}\right)\ldots,
\end{equation}
and a general correlation function on a Wilson line at zero coupling is of the form
\begin{equation}
    \braket{{\rm Tr}\left[\mathcal{O}_1\mathcal{O}_2\ldots\right]},
\end{equation}
where the trace is over the representation of the gauge group.

To count states for each dimension, we define the character of the Hilbert space:
\begin{equation}\label{def:chi}
    \chi(q):={\rm tr}\left[q^{D}\right]=\sum_{\Delta}N_\Delta\, q^\Delta,
\end{equation}
where $D$ is the dilatation operator of the 1D CFT, and the trace is over the whole 1D CFT Hilbert space. 

According to the representation of $SO(1,2)$, the n-th KK mode contributes a factor of 
$$q^{n+1}+q^{n+2}+q^{n+3}+\ldots=\frac{q^{n+1}}{1-q}.$$
Then the calculation of $\chi(q)$ is as follows:
\begin{equation}
    \begin{split}
        \chi(q)&=\sum\limits_{\left\{N_1,N_2,\ldots\right\}}\frac{\left(\sum\limits_{n=1}^{\infty}N_n\right)!}{\prod\limits_{n=1}^{\infty}N_n!}\prod\limits_{n=1}^{\infty}\left(\frac{q^{n+1}}{1-q}\right)^{N_n}=\frac{(1-q)^2}{1-2q}. \\
    \end{split}
\end{equation}
Here, in the first equality, the number $N_n$ denotes the number of $n$-th KK modes appearing in \eqref{counting:defO}, and the combinatorial number takes into account the ordering of $\partial^k a_\tau^{(n)}$. The second equality is a straightforward computation. 

Now the number of dimension-$\Delta$ states is simply the coefficient of $q^\Delta$ in the character:
\begin{equation}
    \chi(q)=1+\sum_{n=0}^{\infty}2^n q^{n+2},
\end{equation}
from which we can read the number of states:
\begin{equation}
    \mathcal{N}_{\Delta}=\begin{cases}
        1, & \Delta=0, \\
        2^{\Delta-2}, & \Delta=2,3,4,\ldots, \\
        0, & \text{otherwise}. \\
    \end{cases}
\end{equation}
Then by \eqref{counting:StatesToPrimaries}, we get the counting for primaries:
\begin{equation}
    \mathcal{N}^\text{p}_{\Delta}=\begin{cases}
        1, & \Delta=0,2 \\
        2^{\Delta-3}, & \Delta=3,4,5,\ldots, \\
        0, & \text{otherwise}. \\
    \end{cases}
\end{equation}

\section{Nonlinearly Realized AdS Isometries}\label{app:NLRAdSIso}

Euclidean AdS$_{d+1}$ can be realized as a hypersurface embedded in $\mathbb{R}^{1,d+1}$. Given the coordinates $(\tau, x, z, \Omega_{d-2})$ for AdS$_{d+1}$, the embedding  is described by:
\begin{equation}\label{Embedding:AdSLorentz}
	\begin{split}
		X^0 &= \frac{1 + \tau^2 + x^2}{2x} \cosh z, \\
		X^1 &= \frac{1 - \tau^2 - x^2}{2x} \cosh z, \\
		X^2 &= \frac{\tau}{x} \cosh z, \\
		X^M &= \Omega^{M-2} \sinh z, \quad M = 3, 4, \ldots, d+1,
	\end{split}
\end{equation}
where $\Omega$ denotes a unit vector on the $(d-2)$-sphere, i.e., a point on $S^{d-2}$ in $\mathbb{R}^{d-1}$:
\begin{equation}
	\begin{split}
		(\Omega^1)^2 + (\Omega^2)^2 + \ldots + (\Omega^{d-1})^2 = 1.
	\end{split}
\end{equation}

Using the embedding coordinates, the generators of the continuous isometry group of AdS$_{d+1}$ are
\begin{equation}\label{generators:AdSTarget}
	\begin{split}
		L_{MN} &= X_M \frac{\partial}{\partial X^N} - X_N \frac{\partial}{\partial X^M}, \\
		X_M &= \eta_{MN} X^N, \quad \eta = \text{diag}(-1, 1, \ldots, 1).
	\end{split}
\end{equation}

In the presence of a long string described in section \ref{subsec:WilsonLoopDefect}, the preserved generators are:
\begin{itemize}
	\item $L_{01}$, $L_{02}$, and $L_{12}$, which together generate the spacetime symmetry group of AdS$_2$, namely $SO(1,2)$;
	\item $L_{ij}$ with $i, j = 3, 4, \ldots, d+1$, which leave the boundary Wilson line invariant. These generate the internal rotation group $SO(d-1)$, interpreted as a global symmetry group from the AdS$_2$ point of view.
\end{itemize}

The Goldstone modes correspond to the broken generators:
\begin{equation}\label{gene:broken}
	\begin{split}
		L_{ai}, \quad a = 0, 1, 2,\quad i = 3, 4, \ldots, d+1.
	\end{split}
\end{equation}
Below, for convenience, without specification we will always use letters $a,b\ldots$ for indices $0,1,2$ and $i,j,k\ldots$ for transverse indices $3,\ldots,d+1$. The presence of the single transverse index $i$ indicates that the Goldstone modes transform as a vector under the $SO(d-1)$ global symmetry. However, the index $a$ does not necessarily imply that the Goldstone modes transform as a vector under the AdS$_2$ spacetime symmetry, since these indices often appear as part of derivative structures and do not label independent dynamical fields. As common for spontaneously broken space-time symmetries there are fewer Goldstones than there are broken generators, see for example \cite{Bonifacio:2018zex} for details. As we will see below, for a long string in AdS$_{d+1}$, the Goldstone modes living on the AdS$_2$ worldsheet are in fact scalar fields from the AdS$_2$ point of view.

Now let us derive the action of the generators \eqref{generators:AdSTarget} on the worldsheet fields $z$ and $\Omega_{d-1}$, under the gauge fixing condition \eqref{gaugefix}.

When there is no gauge fixing on the worldsheet, the infinitesimal transformations on the fields $(\tau,x,z,\Omega_{d-1})$ are simply via
\begin{equation}\label{LorentzTransf}
	\begin{split}
		L_{MN}:\quad\delta X^L(\sigma)=\epsilon\left(X_M(\sigma)\delta^{L}_{N}-X_N(\sigma)\delta^{L}_{M}\right),
	\end{split}
\end{equation}
or, in terms of group action,
\begin{equation}
	\begin{split}
		X^{'L}(\sigma)=\Lambda^L_{M}X^M(\sigma),\quad \Lambda\in SO(1,d+1),
	\end{split}
\end{equation}
where $\sigma=(\sigma^0,\sigma^1)$ is the worldsheet coordinates. Then the infinitesimal transformations on $\tau(\sigma)$, $x(\sigma)$, $z(\sigma)$ and $\Omega_{d-1}(\sigma))$ can be derived from \eqref{Embedding:AdSLorentz} and \eqref{LorentzTransf}. However, when we gauge fix the worldsheet coordinates as \eqref{gaugefix}, we need to redefine $\sigma$ to make it consistent with the gauge. The new coordinate $\sigma'\equiv(\tau',x')$ can be computated by inverting \eqref{Embedding:AdSLorentz} with $X'$:
\begin{equation}\label{InvEmbedding:AdSLorentz}
	\begin{split}
		\tau&=\frac{X^2}{X^0+X^1},\quad x=\frac{1}{X^0+X^1}\sqrt{(X^0)^2-(X^1)^2-(X^2)^2}, \\
		\cosh z&=\sqrt{(X^0)^2-(X^1)^2-(X^2)^2}, \\
		\Omega^{i}&=\frac{X^i}{\sqrt{\sum\limits_{j}(X^j)^2}}. \\
	\end{split}
\end{equation} 

Let us first look at the action generated by $L_{ab}$, the AdS$_2$ isometry group $SO(1,2)$. In the first step (no gauge fixing), $X^a$'s get changed while $X^i$'s remain the same. Therefore, after the second step (recover the gauge condition), we get the final transformation rule under $SO(1,2)$
\begin{equation}
	\begin{split}
		X^{i}(\sigma)\rightarrow X^{i}(\Lambda^{-1}\cdot\sigma),
	\end{split}
\end{equation}
where $\Lambda^{-1}\cdot\sigma$ denotes the inverse of $\Lambda\in SO(1,2)$ acting on the point $\sigma$ in AdS$_2$. In terms of $z$ and $\Omega$ it is given by
\begin{equation}
	\begin{split}
		z(\sigma)\rightarrow z(\Lambda^{-1}\cdot\sigma),\quad \Omega^{i}(\sigma)\rightarrow\Omega^{i}(\Lambda^{-1}\sigma).
	\end{split}
\end{equation}
The above transformation rules imply that $z$ and $\Omega^i$ act as scalars of $SO(1,2)$.

Then let us look at the action generated by $L_{ij}$, the $SO(d-1)$ rotation. In the first step, $X^i$'s get rotated while $X^a$'s remain the same. Therefore, according to \eqref{InvEmbedding:AdSLorentz}, there is no need to redefine the worldsheet coordinates (no second step): gauge \eqref{gaugefix} is preserved under $SO(d-1)$. This leads to the final transformation rule under $SO(d-1)$
\begin{equation}
	\begin{split}
		X^{i}(\sigma)\rightarrow \Lambda^{i}_{j}X^{j}(\sigma).
	\end{split}
\end{equation}
In terms of $z$ and $\Omega$ it is given by
\begin{equation}
	\begin{split}
		z(\sigma)\rightarrow z(\sigma),\quad \Omega^i(\sigma)\rightarrow \Lambda^{i}_{j}\Omega^{j}(\sigma),
	\end{split}
\end{equation}
which means that $SO(d-1)$ acts as a global symmetry group and under which $X^i$ (or $\Omega^i$) transforms under the vector representation. Furthermore, $SO(d-1)$ is an internal symmetry for the worldsheet theory in the sense that it commutes with $SO(1,2)$.

Now let us look at the spontaneously broken part of the action, generated by $L_{ai}$. In the first step, the infinitesimal transformation is given by
\begin{equation}
	\begin{split}
		\delta X^b(\sigma)&=-\epsilon X_i(\sigma)\delta^{b}_{a}, \\
		\delta X^j(
		\sigma)&=\epsilon X_a(\sigma)\delta^{j}_{i}. \\
	\end{split}
\end{equation}
Then in the second step, we redefine $\sigma\equiv(\tau,x)$ according to \eqref{InvEmbedding:AdSLorentz}:
\begin{equation}
	\begin{split}
		\delta\tau&\equiv\delta\left(\frac{X^2}{X^0+X^1}\right) =\epsilon x\Omega^i\tanh z\left[\tau (\delta^{0}_{a}+\delta^{1}_a)-\delta^{2}_{a}\right], \\
		\delta x&\equiv\delta\left(\frac{1}{X^0+X^1}\sqrt{(X^0)^2-(X^1)^2-(X^2)^2}\right) =\epsilon x\Omega^i\tanh z\left(\tilde{x}_a+x(\delta^0_a+\delta^1_a)\right). \\
	\end{split}
\end{equation}
Here we introduced the AdS$_2$ notation $X^a\equiv \tilde{x}^a\cosh z$, i.e.
\begin{equation}
	\begin{split}
		\tilde{x}_a=\eta_{ab}\tilde{x}^b,\quad \tilde{x}^a=\left(\frac{1+\tau^2+x^2}{2x},\frac{1-\tau^2-x^2}{2x},\frac{\tau}{x}\right)\in\text{AdS}_2. 
	\end{split}
\end{equation}
Combining the first and the second steps we get the final transformation
\begin{equation}\label{transf:combine}
	\begin{split}
		X^b(\sigma)&\rightarrow X^b(\sigma)-\epsilon X_i(\sigma)\delta^{b}_{a}-\delta\tau \partial_\tau X^b(\sigma)-\delta x \partial_x X^b(\sigma),  \\
		X^j(\sigma)&\rightarrow X^j(\sigma)+\epsilon X_a(\sigma)\delta^{j}_{i}-\delta\tau \partial_\tau X^j(\sigma)-\delta x \partial_x X^j(\sigma).  \\
	\end{split}
\end{equation}
Using \eqref{transf:combine} we can compute how $z$ and $\Omega^i$ transform under gauge \eqref{gaugefix}. The final result is 
\begin{equation}
	\begin{split}
		\delta z(\sigma)&=\epsilon\Omega^i\big{[}\tilde{x}_a-x\tanh z(\partial_\tau z)(\tau(\delta^0_a+\delta^1_a)-\delta^2_a) \\
		&\qquad -x\tanh z(\partial_x z)(\tilde{x}_a+x(\delta^0_a+\delta^1_a))\big{]}, \\
		\delta\Omega^j(\sigma)&=\epsilon\big{[}\coth z\, \tilde{x}_a(\delta^j_i-\Omega^j\Omega_i)-x\Omega_i\tanh z(\partial_\tau)\Omega^j(\tau(\delta^0_a+\delta^1_a)-\delta^2_a) \\
		&\qquad -x\Omega_i\tanh z(\partial_x\Omega^j)(\tilde{x}_a+x(\delta^0_a+\delta^1_a))\big{]}. \\
	\end{split}
\end{equation}
In the case of AdS$_3$, $\Omega=\pm1$ is discrete, so we have $\delta\Omega^i=0$ and
\begin{equation}
	\begin{split}
		\delta z(\sigma)=\begin{cases}
			-\epsilon\left[\frac{1+\tau^2+x^2}{2x}+x\tau\tanh z\partial_\tau x+\frac{x^2-\tau^2-1}{2}\tanh z\partial_x z\right] & \text{for}\ L_{03} , \\
			\epsilon\left[\frac{1-\tau^2-x^2}{2x}-x\tau\tanh z\partial_\tau x-x^2\tanh z\partial_x z\right] & \text{for}\ L_{13} , \\
			\epsilon\left[\frac{\tau}{x}+x\tanh z\partial_\tau z-\tau\tanh z\partial_x z\right] & \text{for}\ L_{23} . \\
		\end{cases}
	\end{split}
\end{equation}

\section{Coordinate Systems for AdS$_3$}\label{app:AdS3}

In this appendix, we discuss two coordinate systems for Euclidean AdS$_3$:
\begin{itemize}
	\item \textbf{AdS$_2$-sliced coordinates} $(\tau,x,z)$, with metric
	\begin{equation}
		ds^2=\cosh^2 z\,\frac{d\tau^2+dx^2}{x^2}+dz^2\,.
	\end{equation}
	\item \textbf{Poincar\'e half--space coordinates} $(t,y,w)$, with metric
	\begin{equation}
		ds^2=\frac{dt^2+dy^2+dw^2}{w^2}\,.
	\end{equation}
\end{itemize}

Both coordinate systems may be embedded into $\mathbb{R}^{1,3}$, where Euclidean AdS$_3$ is realized as the hyperboloid
\begin{equation}
	(X^0)^2-(X^1)^2-(X^2)^2-(X^3)^2=1,
	\qquad X^0>0\,.
\end{equation}
The corresponding embedding coordinates are given by
\begin{equation}
	\begin{split}
		\text{AdS$_2$ slicing:}&\quad
		\begin{cases}
			X^0 = \cosh z\,\dfrac{1+\tau^2+x^2}{2x}\,, \\
			X^1 = \cosh z\,\dfrac{\tau}{x}\,, \\
			X^2 = \cosh z\,\dfrac{1-\tau^2-x^2}{2x}\,, \\
			X^3 = \sinh z\,,
		\end{cases} \\[1ex]
		\text{Poincar\'e:}&\quad
		\begin{cases}
			X^0 = \dfrac{1+t^2+y^2+w^2}{2w}\,, \\
			X^1 = \dfrac{t}{w}\,, \\
			X^2 = \dfrac{1-t^2-y^2-w^2}{2w}\,, \\
			X^3 = \dfrac{y}{w}\,.
		\end{cases}
	\end{split}
\end{equation}

By comparing the two parameterizations, one finds that the coordinates are related by
\begin{equation}\label{eq:coordinateTransf}
	t=\tau,\qquad
	y=x\tanh z,\qquad
	w=\frac{x}{\cosh z}\,.
\end{equation}
The associated Jacobian matrix reads
\begin{equation}
	\frac{\partial(t,y,w)}{\partial(\tau,x,z)}=
	\begin{pmatrix}
		1 & 0 & 0 \\[0.5ex]
		0 & \tanh z & \dfrac{x}{\cosh^2 z} \\[1ex]
		0 & \dfrac{1}{\cosh z} & -\dfrac{x\tanh z}{\cosh z}
	\end{pmatrix},
\end{equation}
which will be used to determine how gauge fields transform between the two coordinate systems. Here we list the resulting components of the gauge curvature:
\begin{equation}
	\begin{split}
		F_{ty} &= \tanh z\, F_{\tau x} + \frac{1}{x}\, F_{\tau z}\,, \\
		F_{tw} &= -\frac{\sinh z}{x}\, F_{\tau z} + \frac{1}{\cosh z}\, F_{\tau x}\,, \\
		F_{yw} &= -\frac{\cosh z}{x}\, F_{x z}\,.
	\end{split}
\end{equation}

Poincar\'e coordinates are most commonly employed in the context of the AdS/CFT correspondence, since the conformal boundary at $w=0$ is naturally identified with the flat space on which the CFT resides. Boundary conditions are therefore most conveniently imposed in this coordinate system.

In AdS$_2$-sliced coordinates, the limit $w\to0$ corresponds to
\begin{equation}
	z\to\pm\infty,
	\qquad \tau \ \text{and}\ x \ \text{fixed}.
\end{equation}
Standard boundary conditions may thus be formulated in terms of the asymptotic behavior of fields as $z\to\pm\infty$.

However, in the AdS$_2$ slicing one may also consider the alternative limit
\begin{equation}
	x\to0,
	\qquad \tau \ \text{and}\ z \ \text{fixed}.
\end{equation}
Using \eqref{eq:coordinateTransf}, one finds that this limit always corresponds to the same boundary point in Poincar\'e coordinates,
\begin{equation}
	(t,y,w)=(\tau,0,0),
\end{equation}
independent of the transverse coordinate $z$. This property explains why AdS$_2$-sliced coordinates are particularly well suited for studying flux tubes: the transverse direction does not induce geometric fluctuations at the boundary (see also figure~\ref{fig:ads_slicing}). 

When considering bulk fields in AdS using these coordinates, the $x\to0$ limit must be treated with care. Requiring that fields be independent of the transverse coordinate at $x=0$ leads to additional boundary conditions. For example, for a scalar field $\phi$ in AdS$_3$, one must impose
\begin{equation}
	\partial_z\phi(x=0,\tau,z)=0\,.
\end{equation}
For Yang--Mills theory in AdS$_3$, the analogous condition applies to the gauge field and leads to the constraint \eqref{YM:extraBC:A}.

For the same reason, AdS$_2$-sliced coordinates have been utilized in other settings where the relevant physics is associated with a fixed boundary point. As a special case of AdS$_{d+1}$ foliated by AdS$_d$, they have been used to describe interface geometries in AdS \cite{Bak:2003jk}. When AdS$_{d+1}$ is instead foliated by AdS$_2$, these coordinates have been used to describe strings with fixed endpoints on the boundary \cite{Giombi:2017cqn}, which provides direct motivation for our choice of coordinates in this work.

\section{Derivation of Integrated Constraints}
\label{app:nonlin}
In this section, we give a simplified derivation of constraints on conformal defect lines\footnote{We assume the line has a displacement operator. This is guaranteed for theories with a stress tensor, but is also true in the context of this paper} in two-dimensional conformal theories \cite{Gabai:2025zcs,NadavUnpublished}. The constraints are associated with a symmetry that is broken by the presence of the line, and hence are nonlinearly realized. The global conformal symmetry is implemented by the generators $L_{n}$ and $\bar{L}_{n}$ with $n=-1,0,1$. In a theory with a stress tensor, these are,
\begin{equation}
    L_{n} = \oint \frac{dz}{2\pi i}z^{n+1} T(z) \quad,\quad \bar{L}_{n} = \oint \frac{d\bar{z}}{2\pi i}\bar{z}^{n+1} \bar{T}(\bar{z})\ .
\end{equation}
We set the defect to lie along the real line. The theory of interest doesn't have a stress tensor. However, it is still useful to consider what happens in theories that do. The defect being conformal is equivalent to,
\begin{equation}
\label{eq:Tunbroken}
    T(x+i \epsilon) - \bar{T}(x+i \epsilon) - T(x-i \epsilon) + \bar{T}(x-i \epsilon) = 0\ ,
\end{equation}
while the displacement is defined as,
\begin{equation}
\label{eq:Tbroken}
    T(x+i \epsilon) + \bar{T}(x+i \epsilon) - T(x-i \epsilon) - \bar{T}(x-i \epsilon) = 2 \pi i \mathbb{D}(x)\ .
\end{equation}
Consequently, we say that the generators,
\begin{equation}
\label{eq:unbroken}
    L_n + \bar{L}_n\ ,
\end{equation}
are preserved while,
\begin{equation}
\label{eq:broken}
    L_n - \bar{L}_n\ ,
\end{equation}
are broken by the presence of the line. The conclusion is still true in theories without a stress tensor. Note the change of sign in the two expressions \eqref{eq:unbroken} and \eqref{eq:broken} compared to \eqref{eq:Tunbroken} and \eqref{eq:Tbroken}, due to the reversal of the contour in the anti-holomorphic case. It is more common to derive identities resulting from the preserved symmetries. While some physical consequences have been derived from the broken generators \cite{Billo:2016cpy}, they remained underutilized until very recently \cite{Gabai:2023lax, Gabai:2025zcs, NadavUnpublished}.

In addition, we assume that the theory and the defect are sufficiently generic. More precisely, we assume,
\begin{equation}
\label{eq:genericDefect}
    \begin{split}
        &\text{For any $\Delta_*$ fixed, there are no two primary operators living on the defect such that} \\
    &\text{their dimensions are (a) below $\Delta_*$ and (b) integer spaced,} \\
    &\text{except for the displacement operator $\mathbb{D}$ and the identity operator.}
    \end{split}
\end{equation}
This assumption is clearly violated in the strict weak- or strong-coupling limit. The strategy we employ is to derive constraints that hold at \textit{generic finite} coupling and only a posteriori expand them in one of the perturbative limits. In some cases, it will be clear that the perturbative limit does not commute with the integral appearing in the constraint.

\subsection{Homogeneous constraints}
We now study the effect of a broken conformal transformation \eqref{eq:broken} in the presence of a straight line and a number of local scalar primary operators. Later in the analysis, we will take an OPE limit of all of these operators to the line, in order to get constraints that are purely within the one-dimensional theory on the line. 

\begin{figure}
    \centering
    \includegraphics[width=1.\linewidth]{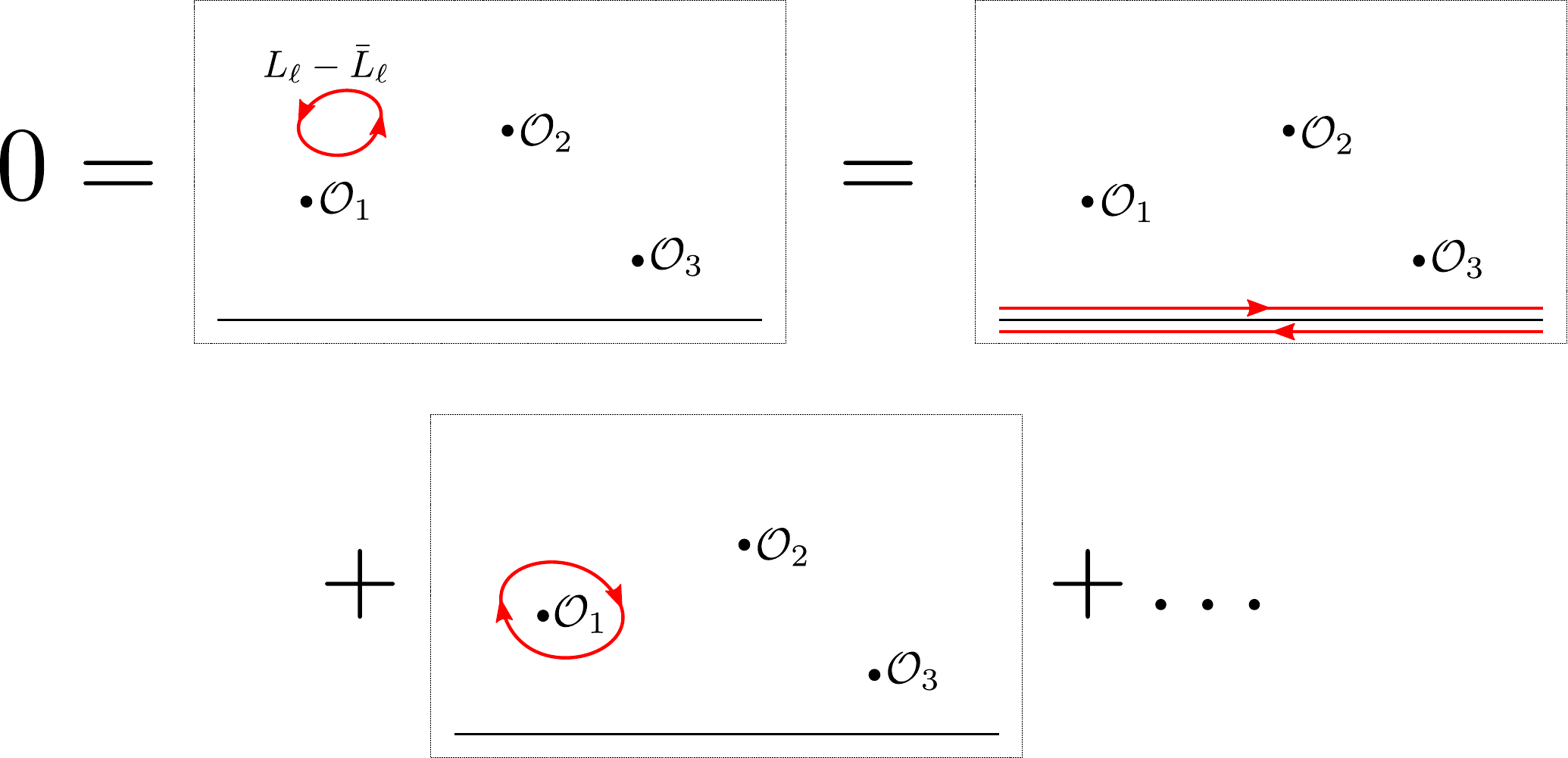}
    \caption{The starting point of the analysis leading to the homogeneous identities. The term in the RHS of the first row is evaluated as the integral of the displacement operators. The second row consists of the sum of actions on each of the local operators.}
    \label{fig:basic}
\end{figure}

We start from a small shrinkable circular contour that evaluates to zero, and then blow it up (see figure \ref{fig:basic}),
\begin{equation}
\label{eq:transHomo}
   \sum_j \left(\hat{\mathcal{D}}^{(\ell)}_{j} - \hat{\bar{\mathcal{D}}}^{(\ell)}_{j}\right) \Big{\langle}\prod_k \mathcal{O}_k(z_k)\Big{\rangle} + \int\limits_{-\infty}^{\infty} dx\, x^{\ell+1}\, \Big{\langle}\mathcal{L}\{\mathbb{D}(x)\}\prod_j \mathcal{O}_j(z_j)\Big{\rangle} = 0\ .
\end{equation}
Here the first term comes from the contour encircling the operator insertions, while the second term comes from the contour wrapping the real line. $\mathcal{L}\{\}$ stands for the insertion of the line operator, and the operators inside the brackets are defect operators on top of it, in path ordering. We will see in the rest of the section that it is only the second term that eventually contributes. The differential operators are defined as,
\begin{equation}
\begin{split}
    \hat{\mathcal{D}}^{(\ell)}_{j} = -z_j^{\ell+1}\partial_{z_j} - (\ell+1) z_j^\ell h_j\ ,\\
    \hat{\bar{\mathcal{D}}}^{(\ell)}_{j} = -\bar{z}_j^{\ell+1}\partial_{\bar{z}_j} - (\ell+1) \bar{z}_j^\ell h_j\ .
\end{split}
\end{equation}
\eqref{eq:transHomo} is a valid Ward identity as is. However, to make it more useful we would like to take the operators to be on the line. To do that, we use the bulk to defect OPE,
\begin{equation}
\label{eq:bBOpe}
    \mathcal{O}(x+iy) = \sum_{\widetilde{\mathcal{O}}}b_{\mathcal{O} \widetilde{\mathcal{O}}}\left[y^{\Delta_{\widetilde{\mathcal{O}}}-\Delta_{\mathcal{O}}}\widetilde{\mathcal{O}}(x) + O(y^{\Delta_{\widetilde{\mathcal{O}}}-\Delta_{\mathcal{O}}+2})\right]
\end{equation}
Here, the sum is over all defect primary operators that can appear on the line. Here and in what follows, we use $\mathcal{O}$ and $\widetilde{\mathcal{O}}$ to denote generic bulk and defect operators, respectively. We choose one such operator, $\widetilde{\mathcal{O}}_j$, for each insertion s.t. $b_{\mathcal{O}_j \widetilde{\mathcal{O}}_j} \neq 0$. Expanding \eqref{eq:transHomo} in all $y_j$ simultaneously, we pick out the coefficient of,
\begin{equation}
\label{eq:lineLim}
    \prod_j y_j^{\Delta_{\widetilde{\mathcal{O}}_j}-\Delta_{\mathcal{O}_j}}\ ,
\end{equation}
which has to vanish independently. The potential contributions can be either from $\widetilde{\mathcal{O}}_j$ itself, or from a level $m$ conformal descendant (or primary if $m=0$) of a defect primary operator with dimensions $\Delta_{\widetilde{\mathcal{O}}_j}-m$, with the same quantum number under the reflection that preserves the line as $\widetilde{\mathcal{O}}_j$. According to assumptions \eqref{eq:genericDefect}, such a primary operator does not exist, and only the $\widetilde{\mathcal{O}}_j$ can contribute. As an intermediate step, we evaluate,
\begin{equation}
\begin{split}
\label{eq:bBOpeDer}
    \partial_{z}\mathcal{O}_j(x+iy) &\sim b_{\mathcal{O}_j\widetilde{\mathcal{\mathcal{O}}}_j}\, y^{\Delta_{\widetilde{\mathcal{O}}_j}-\Delta_{\mathcal{O}_j}-1} \left(\frac{-i}{2} (\Delta_{\widetilde{\mathcal{O}}_j}-\Delta_{\mathcal{O}_j}) + y\, \partial_x\right)\widetilde{\mathcal{O}}_j(x) + \dots \\
    \partial_{\bar{z}}\mathcal{O}_j(x+iy) &\sim b_{\mathcal{O}_j\widetilde{\mathcal{\mathcal{O}}}_j}\, y^{\Delta_{\widetilde{\mathcal{O}}_j}-\Delta_{\mathcal{O}_j}-1} \left(\frac{i}{2} (\Delta_{\widetilde{\mathcal{O}}_j}-\Delta_{\mathcal{O}_j}) + y\, \partial_x\right)\widetilde{\mathcal{O}}_j(x)+\dots
\end{split}
\end{equation}
Here, and in the rest of this section, the ellipses stand for any power that cannot mix with $\Delta_{\widetilde{\mathcal{O}}_j}-\Delta_{\mathcal{O}_j}$. Namely, they stand for both powers that are separated from $\Delta_{\widetilde{\mathcal{O}}_j}-\Delta_{\mathcal{O}_j}$ by a non-integer amount, and powers that are larger than $\Delta_{\widetilde{\mathcal{O}}_j}-\Delta_{\mathcal{O}_j}$ by an integer. By combining \eqref{eq:bBOpe} and \eqref{eq:bBOpeDer}, we find,
\begin{equation}
\begin{split}
    (z^{\ell+1}\partial_{z}\mathcal{O}_j(x+iy)-\bar{z}^{\ell+1}\partial_{\bar{z}}\mathcal{O}_j(x+iy)) &\sim \\
    &\hspace{-70pt} - i x^{\ell+1} y^{\Delta_{\widetilde{\mathcal{O}}_j}-\Delta_{\mathcal{O}_j}-1}(\Delta_{\widetilde{\mathcal{O}}_j}-\Delta_{\mathcal{O}_j})+ {O}(y^{\Delta_{\widetilde{\mathcal{O}}_j}-\Delta_{\mathcal{O}_j}+1}) + \dots 
\end{split}
\end{equation}
and,
\begin{equation}
\begin{split}
    (z^{\ell}\mathcal{O}_j(x+iy)-\bar{z}^{\ell}\mathcal{O}_j(x+iy)) &\sim {O}\left(y^{\Delta_{\widetilde{\mathcal{O}}_j}-\Delta_{\mathcal{O}_j}+2}\right) + \dots 
\end{split}
\end{equation}
We see that the power $\Delta_{\widetilde{\mathcal{O}}_j}-\Delta_{\mathcal{O}_j}$ makes no appearance. Hence, there are no contributions of order $O(y^{\Delta_{\widetilde{\mathcal{O}}_j} - \Delta_{\mathcal{O}_j}})$ from the first term of \eqref{eq:transHomo}. 

Let us turn to analyze the second term of \eqref{eq:transHomo}. There is one type of potentially singular behavior in the bulk to defect OPE. That is, when $x$ is close to one of the $x_j$'s. To regularize that, we introduce a cutoff $|x-x_j| > \epsilon$. Now, we are left with two types of contributions. The first type is, 
\begin{equation}
\label{eq:singularContDInt}
    \int_{x_j-\epsilon}^{x_j+\epsilon} dx\, (x-x_j)^{\alpha} \langle\mathcal{L}\{\mathbb{D}(x) \prod_{k\neq j} \widetilde{\mathcal{O}}_k(x_k)\}\mathcal{O}_j(x_j + i y_j)\rangle\ ,
\end{equation}
where we have already taken all bulk-to-line limits \eqref{eq:lineLim}, except for that of the $j$-th operator. Also, we re-wrote $x^{\ell+1}$ as a sum of powers of $(x-x_j)^\alpha$ (with $\alpha=0,1,\ldots\ell+1$). As we take $y_j \to 0$, by dimensional analysis, this is proportional to, 
\begin{equation}
    \sum_{\widetilde{\mathcal{O}}} \int d\mu^{\widetilde{\mathcal{O}}}_a \, y_j^{\Delta_{\widetilde{\mathcal{O}}} - \Delta_{\mathcal{O}_j} + \alpha - a -1} \epsilon^{a}  \langle\mathcal{L}\{ \widetilde{\mathcal{O}}(x_j) \prod_{k\neq j} \widetilde{\mathcal{O}}_k(x_k)\}\rangle\ ,
\end{equation}
where $a$ is integrated over the real line with some measure $\mu_a^{\widetilde{\mathcal{O}}}$. Any contribution with a nonzero power of $\epsilon$ cannot mix with the finite contribution and must cancel separately. Hence, we only need to keep track of contributions with $a=0$. Since we are picking out only the power,
\begin{equation}
    y_j^{\Delta_{\widetilde{\mathcal{O}}_j}- \Delta_{\mathcal{O}_j} }\ ,
\end{equation}
the only operator that can contribute is an operator of dimension,
\begin{equation}
    \Delta_{\widetilde{\mathcal{O}}} = \Delta_{\widetilde{\mathcal{O}}_j} + 1 - \alpha\ ,
\end{equation}
and the opposite reflection quantum number than $\widetilde{\mathcal{O}}$. Again, by assumption \ref{eq:genericDefect}, this cannot happen. We conclude that the first type of contributions \eqref{eq:singularContDInt} vanishes. We are left with the simple identity,
\begin{equation}
\label{eq:const1PV}
    \left[\,\,\int\limits_{|x-x_j| > \epsilon} dx\, x^{\ell+1} \langle\mathcal{L}\{\mathbb{D}(x)\prod_j \widetilde{\mathcal{O}}_j(x_j)\}\rangle\right]_{\rm finite} = 0\ ,
\end{equation}

To get from this expression to the one in the main text, there are a few necessary steps:
\begin{enumerate}
    \item We set $\widetilde{\mathcal{O}}_1 = \mathbb{D}$
    \item We note by explicit integration that the constraint is satisfied by the GFF correlators \eqref{eq:GFFex1} and \eqref{eq:GFFex2}. As a result, it must be satisfied by the connected correlator (or GFF subtracted correlator)
    \item We split the integral into the different orderings (in $1$-dimensions these correspond to independent analytic functions)
    \item The finite piece of the integral in different regularizations is the same up to contributions from counter terms that have the right quantum numbers to appear without the need of an additional scale (like the cutoff). In our case the absence of potential counter terms appearing in the OPE of any $2$ operators is guaranteed by assumption \eqref{eq:genericDefect}. As a result, the finite part in the point splitting regularization in \eqref{eq:const1PV} is the same as the finite part computed by analytic continuation in the dimension of the low lying exchanged operator appearing in the OPE as in section \ref{sec:Nonlin} 
\end{enumerate}
After these steps, we end up with \eqref{homo1} and \eqref{homo2}.

\subsection{Inhomogeneous constraint}
Let us now move on to the derivation of the inhomogeneous constraint. We lay out an alternative derivation to \cite{Gabai:2025zcs}, following the approach of \cite{NadavUnpublished,MiguelUnpublished}.

\begin{figure}
    \centering
    \includegraphics[width=1.\linewidth]{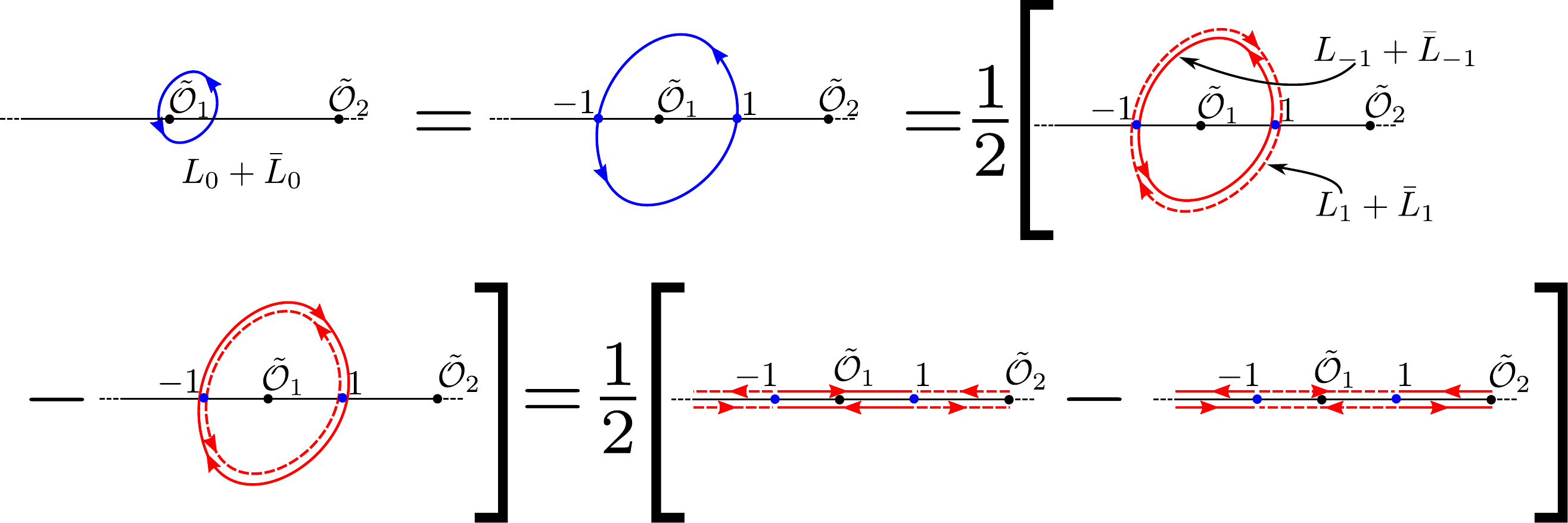}
    \caption{The identity that leads to the inhomogeneous constraint \eqref{inhomo}. The blue line corresponds to an unbroken generator, while the red lines correspond to the broken generators. The main step is the second equality, where we replace an unbroken generator by a commutator of $2$ broken generators.}
    \label{fig:inhom}
\end{figure}

Consider now the following setup. We again place the defect along the real line, but add $2$ operators on the line at $0$ and $\infty$. Now, we act on one of the operators with the generator $L_{0} + \bar{L}_0$. We evaluate this in $2$ ways, as in figure \ref{fig:inhom}.
First, we act on the operator with the transformation, giving,
\begin{equation}
\label{eq:inhomoway1}
     \mathcal{I} = -\left(x_1\partial_{x_1} + \Delta_{\widetilde{\mathcal{O}}_1}\right)\langle  \mathcal{L}\{\widetilde{\mathcal{O}}_1(0)\widetilde{\mathcal{O}}_2(\infty)\}\rangle = \Delta_{\widetilde{\mathcal{O}}_1}\langle  \mathcal{L}\{\widetilde{\mathcal{O}}_1(0)\widetilde{\mathcal{O}}_2(\infty)\}\rangle\ .
\end{equation}
Second, we blow up the contour such that it still envelopes the origin, but also crosses the real line in two points, say $-1$ and $1$. Now, we use the identity,
\begin{equation}
	L_{0} + \bar{L}_0 = \frac{1}{2}[L_{1}-\bar{L}_{1},L_{-1}-\bar{L}_{-1}]\ ,
\end{equation}
to exchange the unbroken generator with the commutator of the broken ones. Next, in each of the two terms in the commutator, we blow up the external generator to $\infty$ and shrink the internal one to $0$. As we deform each of the contours, there are two possible effects. Namely, they can either transform one of the defect operators on the line, or act on the line itself giving rise to an integrated displacement operator. However, due to the argument carefully presented in the previous subsection, any contribution where either of the broken generators acts on an operator vanishes. Hence, we are left with the integrated contribution,
\begin{equation}
\label{eq:inhomoWay2}
	\begin{split}
		\mathcal{I} =\frac{1}{2}\left[\int\limits_{-\infty}^{-1} + \int\limits^{\infty}_{1}\right]dy \left[\int\limits_{-1}^{0} + \int\limits^{1}_{0}\right]du\, (y^2-u^2)\, \langle\mathcal{L}\{\mathbb{D}(y)\mathbb{D}(u) \widetilde{\mathcal{O}}_1(0)\widetilde{\mathcal{O}}_2(\infty)\}\rangle \ .
	\end{split}
\end{equation}
In principle, we should work with a point splitting regularization as in \eqref{eq:const1PV}. With that regularization, we can verify that the integral \eqref{eq:inhomoWay2} exactly vanishes for the GFF correlators \eqref{eq:GFFex1} and \eqref{eq:GFFex2}. Hence, we can safely subtract a multiple of the GFF correlator to set the identity exchange to $0$. We will not be careful about the regularization in the rest of this appendix to simplify the presentation, since it does not introduce any new subtleties that were not dealt with in the homogeneous case. Now, we study each of the $4$ contributions to the integral separately. For example, the last one gives,
\begin{equation}
	\begin{split}
		-\frac{\Lambda}{2}\int\limits^{\infty}_{1}dy \int\limits^{1}_{0} du\frac{u^2-y^2}{(y-u)^4}\, \mathcal{F}^{\widetilde{\mathcal{O}}\mathbb{D}\mathbb{D}\widetilde{\mathcal{O}}}\left(\frac{u}{y-u}\right) \ ,
	\end{split}
\end{equation}
where the notation for the four-point function is defined in \eqref{def:FODDO}, and $\Lambda$ is the two point function of displacement. It's physical significance is explained above equation \eqref{2pfDisp}. By a change of variables $y=\frac{1+t}{t} u$ we get,
\begin{equation}
	\begin{split}
		-\frac{\Lambda}{2}\int\limits_{0}^{\infty}dt \int\limits^{t/(1+t)}_{1} du\,\frac{1+2t}{u}\, \mathcal{F}^{\widetilde{\mathcal{O}}\mathbb{D}\mathbb{D}\widetilde{\mathcal{O}}}\left(t\right) = -\frac{\Lambda}{2}\int\limits_{0}^{\infty}dt\,(1+2t)\log\left(\frac{t}{1+t}\right)\, \mathcal{F}^{\widetilde{\mathcal{O}}\mathbb{D}\mathbb{D}\widetilde{\mathcal{O}}}\left(t\right)\ .
	\end{split}
\end{equation}
The first contribution to \eqref{eq:inhomoWay2} gives the same, while the other two contributions combine to give,
\begin{equation}
	\begin{split}
		\frac{\Lambda}{2}\int\limits_{0}^{\infty}dt\,(t^2-1)\log\left(t\right)\, \mathcal{F}^{\mathbb{D}\widetilde{\mathcal{O}}\mathbb{D}\widetilde{\mathcal{O}}}\left(t\right)\ .
	\end{split}
\end{equation}
we get,
\begin{equation}
    \mathcal{I} = \frac{\Lambda}{2}\int\limits_{0}^{\infty}dt\,(t^2-1)\log\left(t\right)\, \mathcal{F}^{\mathbb{D}\widetilde{\mathcal{O}}\mathbb{D}\widetilde{\mathcal{O}}}\left(t\right) -\Lambda\int\limits_{0}^{\infty}dt\,(1+2t)\log\left(\frac{t}{1+t}\right)\, \mathcal{F}^{\widetilde{\mathcal{O}}\mathbb{D}\mathbb{D}\widetilde{\mathcal{O}}}\left(t\right)
\end{equation}
By comparing it to \eqref{eq:inhomoway1}, we find \eqref{inhomo}.

\small
	
\bibliography{string_ads}
\bibliographystyle{utphys}
	
\end{document}